\documentclass[a4paper,11pt]{article}
\usepackage[english]{babel}
\usepackage{jheppub}
\pdfoutput=1
\usepackage[T1]{fontenc}
\usepackage{bbold}
\usepackage{float}
\usepackage{amsmath}
\usepackage{empheq}
\usepackage{booktabs}
\usepackage{color}
\usepackage[utf8]{inputenc}
\usepackage{xspace}
\usepackage{scalerel}
\usepackage[most]{tcolorbox}
\usepackage{multirow}
\usepackage{scalefnt}
\usepackage{bold-extra}
\usepackage[shortlabels]{enumitem}
\usepackage[tikz]{bclogo}
\usepackage{subcaption}
\usepackage{tikz-feynman}
\usepackage{cancel}
\usepackage{verbatim}
\usetikzlibrary{arrows,shapes}
\usepackage{afterpage}
\usepackage{graphicx,grffile}

\newcommand\F{${\rm F}$}
\newcommand\FJ{${\rm FJ}$}
\newcommand\FJJ{${\rm FJJ}$}

\newcommand\PhiB{\Phi_{\scriptscriptstyle \rm F}}

\newcommand\ww{$W^+W^-$}
\newcommand\zz{$ZZ$}

\newcommand{\flav}{\ell}

\newcommand{\flavB}{\flav_{\scriptscriptstyle \rm F}}

\newcommand{\flavBJ}{\flav_{\scriptscriptstyle \rm FJ}}

\newcommand{\projflav}{\flavB\leftarrow\flavBJ}

\newcommand{\CA}{C_{\mathrm{A}}}

\newcommand{\nf}{N_f}

\newcommand\PhiBJ{\Phi_{\scriptscriptstyle \rm FJ}}

\newcommand{\as}{\alpha_s}
\newcommand{\aew}{\alpha_{\text{\scalefont{0.77}EW}}}

\newcommand{\pt}{{p_{\text{\scalefont{0.77}T}}}}
\newcommand{\GZ}{{\Gamma_Z}}
\newcommand{\GW}{{\Gamma_W}}
\newcommand{\thW}{{\theta_W}}

\newcommand{\qt}{{q_{\text{\scalefont{0.77}T}}}}

\newcommand{\ptrad}{{p_{\text{\scalefont{0.77}T,rad}}}}

\newcommand{\ptww}{{p_{\text{\scalefont{0.77}T,$WW$}}}}
\newcommand{\ptwp}{{p_{\text{\scalefont{0.77}T,$W^+$}}}}
\newcommand{\ptwm}{{p_{\text{\scalefont{0.77}T,$W^-$}}}}
\newcommand{\mtww}{{m_{\text{\scalefont{0.77}T,$WW$}}}}
\newcommand{\mtwwexp}{{m_{\text{\scalefont{0.77}T,$WW$}}^{\rm exp}}}

\newcommand{\ptll}{{p_{\text{\scalefont{0.77}T,}\ell\ell}}}
\newcommand{\ptj}{{p_{\text{\scalefont{0.77}T,$j$}}}}
\newcommand{\ptjone}{{p_{\text{\scalefont{0.77}T,$j_1$}}}}
\newcommand{\ptjoneveto}{{p_{\text{\scalefont{0.77}T,$j_1$}}^{\rm veto}}}

\newcommand{\ptl}{{p_{\text{\scalefont{0.77}T,$\ell$}}}}
\newcommand{\ptlone}{{p_{\text{\scalefont{0.77}T,$\ell_1$}}}}

\newcommand{\ptmiss}{{p_{\text{\scalefont{0.77}T,miss}}}}
\newcommand{\ptmissrel}{{p_{\text{\scalefont{0.77}T,miss,rel}}}}

\newcommand{\ywp}{{y_{\text{\scalefont{0.77}$W^+$}}}}

\newcommand{\yll}{{y_{\text{\scalefont{0.77}$\ell\ell$}}}}
\newcommand{\dphill}{{\Delta\phi_{\text{\scalefont{0.77}$\ell\ell$}}}}
\newcommand{\yww}{{y_{\text{\scalefont{0.77}$WW$}}}}

\newcommand{\dyww}{{\Delta y_{\text{\scalefont{0.77}$W^-,W^+$}}}}

\newcommand{\mz}{{m_Z}}
\newcommand{\mw}{{m_W}}
\newcommand{\mww}{{m_{\text{\scalefont{0.77}WW}}}}

\newcommand{\mll}{{m_{\text{\scalefont{0.77}$\ell\ell$}}}}
\newcommand{\etal}{{\eta_{\text{\scalefont{0.77}$\ell$}}}}

\newcommand{\etaj}{{\eta_{\text{\scalefont{0.77}j}}}}

\newcommand{\drej}{{\Delta R_{\text{\scalefont{0.77}$e j$}}}}

\newcommand{\muF}{{\mu_{\text{\scalefont{0.77}F}}}}
\newcommand{\muR}{{\mu_{\text{\scalefont{0.77}R}}}}

\newcommand{\Q}{{Q_{\text{\scalefont{0.77}$0$}}}}

\newcommand{\noun}[1]{{\scshape #1}}

\newcommand{\POWHEG}{\noun{Powheg}}

\newcommand{\POWHEGBOX}{\noun{Powheg-Box}}
\newcommand{\POWHEGBOXRES}{\noun{Powheg-Box-Res}}
\newcommand{\POWHEGBOXVTWO}{\noun{Powheg-Box-V2}}

\newcommand{\minlo}{{\noun{MiNLO$^{\prime}$}}}
\newcommand{\minnlo}{{\noun{MiNNLO$_{\rm PS}$}}}
\newcommand{\Matrix}{{\noun{Matrix}}}
\newcommand{\OpenLoops}{{\noun{OpenLoops}}}
\newcommand{\PYTHIA}[1]{\noun{Pythia{#1}}}

\newcommand{\setupinclusive}{{\tt setup-inclusive}}
\newcommand{\setupone}{{\tt fiducial-1-JV}}
\newcommand{\setuptwo}{{\tt fiducial-2-JV}}
\newcommand{\setuponenoJV}{{\tt fiducial-1-noJV}}
\newcommand{\setuptwonoJV}{{\tt fiducial-2-noJV}}

\newcommand{\abarmu}[1]{\frac{\as(#1)}{2\pi}}

\newcommand{\citere}[1]{ref.\,\cite{#1}}

\newcommand{\citeres}[1]{refs.\,\cite{#1}}
\newcommand{\citeresc}[1]{Refs.\,\cite{#1}}
\newcommand{\eqn}[1]{eq.\,(\ref{#1})}
\newcommand{\eqnc}[1]{Eq.\,(\ref{#1})}

\newcommand{\fig}[1]{figure\,\ref{#1}}

\newcommand{\tab}[1]{table\,\ref{#1}}
\newcommand{\sct}[1]{section~\ref{#1}}

\newcommand{\LambdaPWG}{\Lambda_{\rm pwg}}

\newcommand{\M}{ {\rm M} }
\newcommand{\C}{ \mathcal{C} }
\newcommand{\FF}{ \mathcal{F} }
\newcommand{\sm}{ \tilde{s} }
\newcommand{\tm}{ \tilde{t} }

\newcommand{\sphid}[1]{}

\usepackage{xcolor}
\newcommand{\mathd}{\mathrm{d}}
\newcommand{\tmop}[1]{\ensuremath{\operatorname{#1}}}

\newtcolorbox{empheqboxed}{colback=white!35, 
 colframe=black,
 width=\textwidth,
 sharpish corners,
 top=-2mm, % default value 2mm
 bottom=0pt
}

\title{{\boldmath{$W^+W^-$} production at NNLO+PS with M{\scalefont{0.77}I}NNLO\boldmath{$_{\text{PS}}$}}}

\author[]{Daniele Lombardi,}
\author[]{Marius Wiesemann}
\author[]{and Giulia Zanderighi}

\emailAdd{lombardi@mpp.mpg.de}
\emailAdd{wieseman@mpp.mpg.de}
\emailAdd{zanderi@mpp.mpg.de}

\affiliation[]{Max-Planck-Institut f\"ur Physik, F\"ohringer Ring 6,
  80805 M\"unchen, Germany}

\abstract{ We consider $W^+W^-$ production in hadronic collisions and
  present the computation of next-to-next-to-leading order accurate
  predictions consistently matched to parton showers (NNLO+PS) using
  the \minnlo{} method. Spin correlations, interferences and off-shell
  effects are included by calculating the full process $pp \to e^+\nu_e \mu^-\bar{\nu}_\mu$.
  This is the first NNLO+PS calculation for
  $W^+W^-$ production that does not require an a-posteriori
  multi-differential reweighting.  The evaluation time of the two-loop
  contribution has been reduced by more than one order of magnitude
  through a four-dimensional cubic spline interpolation.  We find good
  agreement with the inclusive and fiducial cross sections measured by ATLAS and CMS. 
  Both NNLO corrections and matching to parton showers are important
  for an accurate simulation of the \ww{} signal, and their matching
  provides the best description of fully exclusive \ww{} events to date.}


\preprint{MPP-2021-27}

\begin{document}

\maketitle

\section{Introduction}
\label{sec:intro}

Precision phenomenology has evolved to one of the cornerstones of 
todays physics programme at the Large Hadron Collider (LHC). Without 
clear hints for new physics, the precise measurement of production rates 
and distributions of Standard Model (SM) processes provides a 
valuable path towards the observation of deviations from the SM picture. The production 
of vector-boson pairs is among the most important LHC signatures in that respect.
Those processes are crucial to constrain or measure anomalous interactions among 
SM particles, such as anomalous couplings among three vector bosons (triple-gauge 
couplings), as any small deviation from the expected rates or shapes of distributions 
could be a signal of new physics.

\ww{} production has the largest cross section among the massive diboson processes and it 
provides direct access to triple-gauge couplings, which appear already in the leading 
perturbative contribution to the cross section. The measurement of this process at the LHC
is a direct probe of the gauge symmetry structure of electroweak (EW) interactions and 
of the mechanism of EW symmetry breaking in the SM. 
Moreover, \ww{} final states are an irreducible background to Higgs measurements in the 
$H\to W^+W^-$ decay channel and to direct searches for BSM particles decaying into two 
leptons, missing energy, and/or jets. 
The \ww{} cross section has been measured at both the Tevatron \cite{Aaltonen:2009aa,Abazov:2009ys,Abazov:2011cb} and the LHC (at 7 TeV \cite{Aad:2012oea,ATLAS:2012mec,Chatrchyan:2013yaa,Aad:2014mda}, 8 TeV \cite{Aad:2016wpd,Chatrchyan:2013oev,Khachatryan:2015sga,Aaboud:2017cgf} and 13 TeV \cite{Aaboud:2019nkz,Aaboud:2017qkn,CMS:2016vww,Sirunyan:2020jtq}).
The high sensitivity to anomalous triple-gauge couplings has been exploited 
in various indirect BSM searches \cite{Abazov:2009ys,Aaltonen:2009aa,ATLAS:2012mec,Aad:2012oea,Chatrchyan:2013yaa,Chatrchyan:2013fya,Aad:2014mda,CMS:2015uda,Khachatryan:2015sga,Aad:2016wpd,Aaboud:2017cgf,Sirunyan:2017bey,Aad:2021dse} and the irreducible \ww{} background has been extensively studied in the context of $H\to W^+W^-$ decays in \citeres{Binoth:2005ua,Campbell:2011cu,Aad:2012me,Aad:2013wqa,Chatrchyan:2013iaa,Campbell:2013wga,ATLAS:2014aga,Khachatryan:2014kca,Aad:2015xua,Aad:2015rwa,Aad:2015ona,Aad:2016lvc,Caola:2016trd}.

The theoretical description of fiducial cross sections and kinematic
distributions has been greatly improved by the calculation of
next-to-next-to-leading order (NNLO) corrections in QCD perturbation
theory, which have become the standard for $2\to 1$ and $2\to 2$
colour-singlet production
\cite{Ferrera:2011bk,Ferrera:2014lca,Ferrera:2017zex,Campbell:2016jau,Harlander:2003ai,Harlander:2010cz,Harlander:2011fx,Buehler:2012cu,Marzani:2008az,Harlander:2009mq,Harlander:2009my,Pak:2009dg,Neumann:2014nha,deFlorian:2013jea,deFlorian:2016uhr,Grazzini:2018bsd,Catani:2011qz,Campbell:2016yrh,Grazzini:2013bna,Grazzini:2015nwa,Campbell:2017aul,Gehrmann:2020oec,Cascioli:2014yka,Grazzini:2015hta,Heinrich:2017bvg,Kallweit:2018nyv,Gehrmann:2014fva,Grazzini:2016ctr,Grazzini:2016swo,Grazzini:2017ckn,Baglio:2012np,Li:2016nrr,deFlorian:2019app}.
With $\gamma\gamma\gamma$ production even the first $2\to 3$ LHC
process was recently pushed to NNLO
accuracy~\cite{Chawdhry:2019bji,Kallweit:2020gcp}.  In comparison to
LHC measurements NNLO corrections are crucial for a more accurate and
precise description of data.  On the other hand, the validity of
fixed-order calculations is challenged in kinematical regimes
sensitive to soft and collinear radiation through the appearance of
large logarithmic contributions. In such regimes an all-order
description is mandatory to obtain physically meaningful predictions.  The
analytic resummation of large logarithmic contributions is usually
restricted to a single observable or at most two observables, see
e.g.\ \citere{Kallweit:2020gva} for the recent
next-to-next-to-next-to-logarithmic (N$^3$LL) result of the \ww{}
transverse-momentum ($\ptww$) spectrum and the joint resummation of
logarithms in $\ptww$ and in the transverse momentum of the leading
jet ($\ptjone$) at next-to-next-to-logarithmic (NNLL) accuracy.  By
contrast, parton showers are based on a numerical resummation approach
with limited logarithmic accuracy, but they include all-order effects
in all regions of phase space at the same time. Moreover, the fully
exclusive description of the final state enables a full-fledged
hadron-level simulation that is indispensable for experimental
analyses.

In order to meet the experimental demands for having both high-precision predictions
and exclusive hadron-level events, an enormous effort is made by the theory
community to include higher-order corrections in parton showers.
Almost two decades ago the matching of next-to-leading order (NLO) QCD predictions and parton showers (NLO+PS) was formulated in seminal publications \cite{Frixione:2002ik,Nason:2004rx,Frixione:2007vw}.
More recently, the first NNLO+PS approaches have been developed for colour-singlet 
processes \cite{Hamilton:2012rf,Alioli:2013hqa,Hoeche:2014aia,Monni:2019whf,Monni:2020nks},
and the \minnlo{} approach of  \citeres{Monni:2019whf,Monni:2020nks} was very recently extended to heavy-quark pair production \cite{Mazzitelli:2020jio}.
The methods of \citeres{Hamilton:2012rf,Monni:2019whf,Monni:2020nks}
originate from the \minlo{} procedure~\cite{Hamilton:2012np,Hamilton:2012rf}, which upgrades 
a NLO calculation for colour singlet plus jet production to become NLO accurate for both zero-jet and one-jet observables by exploiting features of the all-order structure of the transverse momentum resummation formula. 
In \citere{Frederix:2015fyz}, a numerical extension of the \minlo{} procedure to
higher jet multiplicities was presented and applied to Higgs production in association 
with up to two jets. Most NNLO+PS applications have been done for 
simple $2\to1$ LHC processes or $1\to 2$ decays so far, such as Higgs-boson production
\cite{Hamilton:2013fea,Hoche:2014dla,Monni:2019whf,Monni:2020nks}, 
Drell-Yan (DY) production
\cite{Hoeche:2014aia,Karlberg:2014qua,Alioli:2015toa,Monni:2019whf,Monni:2020nks},
Higgsstrahlung \cite{Astill:2016hpa,Astill:2018ivh,Alioli:2019qzz},
which is still a $2\to1$ process with respect to QCD corrections, and 
the $H\to b \bar b$ decay~\cite{Bizon:2019tfo,Alioli:2020fzf}.
There are a few notable exceptions where NNLO+PS matching was achieved for more
involved colour-singlet processes, namely $W^+W^-$ \cite{Re:2018vac}, $Z\gamma$ \cite{Lombardi:2020wju}, $\gamma\gamma$ \cite{Alioli:2020qrd} and $ZZ$ \cite{Alioli:2021egp} production. 
Moreover, with top-quark pair production the very first NNLO+PS calculation for a 
coloured initial and final state has been presented in \citere{Mazzitelli:2020jio}. 

In the case of $W^+W^-$ production at the LHC, substantial advancements 
have been made in the theoretical description of the process in terms of 
both fixed-order and all-order calculations. $W$-boson pairs are produced
in quark annihilation at LO, which was calculated several decades ago for on-shell $W$ bosons~\cite{Brown:1978mq}. NLO QCD corrections were obtained in the on-shell 
approximation first~\cite{Ohnemus:1991kk,Frixione:1993yp}, and in \citeres{Campbell:1999ah,Dixon:1999di,Dixon:1998py,Campbell:2011bn}
the leptonic $W$ decays with off-shell effects and spin correlations were accounted for.
Also, NLO EW corrections are known both for on-shell $W$ bosons~\cite{Bierweiler:2012kw,Baglio:2013toa,Billoni:2013aba} and including their off-shell treatment~\cite{Biedermann:2016guo,Kallweit:2017khh,Kallweit:2019zez}. The simplest ${\cal O}(\as^2)$ contribution 
is the loop-induced gluon fusion channel. Being separately finite and 
enhanced by the large gluon luminosities, its LO cross section is known already for a long time \cite{Glover:1988rg,Dicus:1987dj,Matsuura:1991pj,Zecher:1994kb,Binoth:2008pr,Campbell:2011bn,Kauer:2013qba,Cascioli:2013gfa,Campbell:2013una,Ellis:2014yca,Kauer:2015dma}.
The full NNLO QCD corrections were first obtained
for the inclusive cross section in the on-shell approximation~\cite{Gehrmann:2014fva}, 
while the fully differential NNLO calculation for off-shell $W$ bosons 
was presented in \citere{Grazzini:2016ctr}, using the $q\bar{q}\to VV'$ two-loop helicity amplitudes \cite{Gehrmann:2014bfa,Caola:2014iua,Gehrmann:2015ora}. Recently, NNLO corrections 
were studied for polarized \ww{} production \cite{Poncelet:2021jmj}.
Also NLO QCD corrections to the loop-induced gluon fusion contribution, which are formally of ${\cal O}(\as^3)$, were evaluated using the $gg\to VV'$ two-loop helicity amplitudes of \citeres{Caola:2015ila,vonManteuffel:2015msa}: first in an approximation without quark initial states \cite{Caola:2015rqy} and later including all relevant contributions \cite{Grazzini:2020stb}.
To date the most advanced fixed-order prediction for \ww{} production combines all of those contributions
and is available in the \Matrix{} framework~\cite{Grazzini:2017mhc}:
the combination of NNLO QCD \cite{Grazzini:2016ctr} and NLO EW predictions has been achieved in \citere{Kallweit:2019zez} using \Matrix{} and \OpenLoops{}~\cite{Cascioli:2011va, Buccioni:2017yxi, Buccioni:2019sur}.
Approximate N$^{3}$LO predictions (labelled as nNNLO) have been calculated by combining 
the NNLO quark-initiated cross section with the NLO gluon-initiated cross section in 
\citere{Grazzini:2020stb}, 
where the nNNLO cross section has also been combined with NLO EW corrections.

All-order predictions for the \ww{} process have been obtained for various observables
using state-of-the-art resummation techniques: threshold resummation at NLO+NNLL 
was presented in \citere{Dawson:2013lya}, $b$-space resummation was used to obtain the NNLO+NNLL transverse momentum spectrum of the \ww{} pair \cite{Grazzini:2015wpa} and the NNLO+NNLL jet-vetoed cross section was 
computed in \citere{Dawson:2016ysj}. 
More recently, the {\sc Matrix+RadISH} framework was introduced \cite{Kallweit:2020gva,Wiesemann:2020gbm,MatrixRadishurl}, which combines  
NNLO calculations in \Matrix{} with high-accuracy resummation through the \textsc{RadISH} formalism \cite{Monni:2016ktx,Bizon:2017rah,Monni:2019yyr}.
For all $2\to1$ and $2\to2$ colour-singlet processes the {\sc Matrix+RadISH} code makes NNLO+N$^3$LL predictions for the transverse momentum of the colour singlet, NNLO+NNLL predictions 
for the transverse momentum of the leading jet, as well as their joint resummation at NNLO+NNLL publicly available.
In particular, \citere{Kallweit:2020gva} has applied this resummation framework as an example to  \ww{} production, presenting
state-of-the-art predictions for the $\ptww{}$ spectrum, the $\ptjone{}$ spectrum, 
the jet-vetoed cross section and the $\ptww{}$ spectrum with a jet veto.
Indeed, one important aspect of the theoretical description of \ww{} production 
is the correct modelling of the jet veto
(see \citeres{Jaiswal:2014yba,Meade:2014fca,Becher:2014aya,Monni:2014zra,Dawson:2016ysj,Kallweit:2020gva} for example),
which is applied by the experimental analyses to suppress backgrounds involving top-quarks ($t\bar{t}$ and $tW$).
A strict veto against jets in the final state increases the
sensitivity to higher-order QCD effects due to potentially large
logarithms of the ratio of the small jet-veto scale over the large invariant
mass of the system. Such terms challenge the reliability of
fixed-order predictions and induce large uncertainties in theory
predictions that are typically not covered by scale-variation
procedures, especially when extrapolating cross-sections measured
in the fiducial region to the total phase space. 
In particular, the tension with NLO+PS predictions observed in earlier
\ww{} measurements \cite{ATLAS:2014xea,Chatrchyan:2013oev} challenged the validity of lower-order
Monte Carlo predictions for \ww{} production \cite{Monni:2014zra}.
Only through the calculation of NNLO corrections
\cite{Gehrmann:2014fva,Grazzini:2016ctr} this tension could be
released, and their combination with all-order resummation confirmed
that the jet-vetoed \ww{} cross section is under good theoretical
control \cite{Dawson:2016ysj,Kallweit:2020gva}.  Moreover, it was
shown that resummation effects are eventually required to obtain
reliable predictions in the tails of some kinematical distributions,
for instance in the invariant mass distribution of the \ww{}
pair~\cite{Arpino:2019fmo} when a jet-veto is imposed.
These issues show the relevance of fully flexible, hadron-level Monte
Carlo predictions with state-of-the-art perturbative precision for the
\ww{} production process, which is achieved by the the combination of
NNLO corrections with parton-shower simulations. 

Several Monte Carlo simulations for \ww{} production were performed in
the past years: NLO+PS predictions were presented in {\sc MC@NLO}
\cite{Frixione:2002ik}, {\sc Herwig}
\cite{Hamilton:2010mb,Bellm:2016cks,Bellm:2015jjp}, {\sc Sherpa}
\cite{Hoche:2010pf} and \POWHEGBOX{}
\cite{Nason:2013ydw,Melia:2011tj}. More recently, NLO+PS events with
zero-jet and one-jet multiplicities have been merged in the {\sc
  MEPS@NLO} approach \cite{Gehrmann:2012yg,Hoeche:2012yf} within {\sc
  OpenLoops+Sherpa} \cite{Cascioli:2013gfa}, in the {\sc FxFx} scheme
\cite{Frederix:2012ps} within {\sc MadGraph5\_aMC@NLO}
\cite{Alwall:2014hca}, and using the \minlo{} procedure
\cite{Hamilton:2012np,Hamilton:2012rf} within \POWHEGBOX{}
\cite{Nason:2004rx,Frixione:2007vw,Alioli:2010xd} through the {\tt
  WWJ-MiNLO} generator \cite{Hamilton:2016bfu}.  The latter
calculation was even upgraded to a full-fledged NNLO+PS generator
\cite{Re:2018vac} (referred to as NNLOPS in the following) 
using numerically highly demanding multi-dimensional
reweighting in the Born phase space to the NNLO cross section from
\Matrix{} \cite{Grazzini:2016ctr,Grazzini:2017mhc}.
More recently, the combination of NLO QCD and NLO EW corrections matched to 
parton showers was studied \cite{Brauer:2020kfv,Chiesa:2020ttl}.

In this paper, we obtain NNLO+PS predictions for \ww{} production using
the \minnlo{} method. For the first time NNLO QCD corrections are 
directly included during the generation of \ww{} events, without any
post-processing or reweighting being required. In fact,
this is also the first time a NNLO \ww{} calculation 
independent of a slicing cutoff is performed
(cf.\ \citeres{Gehrmann:2014fva,Grazzini:2016ctr}).
To this end, we have applied the recently developed \minnlo{}
method \cite{Monni:2019whf,Monni:2020nks} and its extension to $2 \to 2$ reactions presented in \citere{Lombardi:2020wju}.
At variance with the NNLOPS calculation of \citere{Re:2018vac},
our new \minnlo{} generator does not include any of the approximations or limitations related to the reweighting approach used in \citere{Re:2018vac}.
In particular, \citere{Re:2018vac} had to resort to a number of
features of the $W$-boson decays, such as the fact that the full angular
dependence of each vector-boson decay can be parametrized through eight
spherical harmonic functions \cite{Collins:1977iv} and the fact that QCD corrections are
largely independent of the off-shellness of the vector bosons, in order to simplify 
the parametrization of the nine dimensional $W^+W^-\to e^+ \nu_e\mu^- \bar\nu_\mu$ 
Born phase space. Moreover, the discretization of the residual variables in the parametrization of 
the Born phase space for the reweighting limits
the numerical accuracy in phase-space regions sensitive to
coarse bins. Not rarely, such regions can be 
relevant for BSM searches, especially when situated in the 
tails of kinematic distributions.
Without those limitations, our new \minnlo{} calculation provides the most flexible and most general 
simulation of \ww{} signal events with NNLO accuracy at the LHC.
For the two-loop contribution, we use the helicity amplitudes
for the production of a pair of off-shell vector bosons
\cite{Gehrmann:2015ora} from the public code {\tt VVAMP}~\cite{hepforge:VVamp}
and exploit their implementation 
for all $q\bar{q} \, \rightarrow \,4 \,\text{leptons}$ processes
in the \Matrix{} framework~\cite{Grazzini:2017mhc,Matrixurl}.  The
evaluation of these two-loop amplitudes turns out to be the major bottleneck
in our calculation. In order to deal with this we substantially speed up
the evaluation time by using a
four-dimensional cubic spline interpolation procedure of the two-loop
coefficients entering the helicity amplitudes.

In the present calculation we consider all topologies that lead to two opposite-charge leptons and two 
neutrinos in the final state ($\ell^+\nu_\ell\,\ell'^-\bar{\nu}_{\ell'}$) with off-shell effects, interferences, and spin correlations.
As a basis we exploit the \ww{}+jet generator 
of \citere{Hamilton:2016bfu} and include NNLO QCD corrections to \ww{} production 
through the \minnlo{} method. The ensuing \minnlo{} generator is implemented and will be made publicly available within the \POWHEGBOXRES{} framework \cite{Nason:2004rx,Frixione:2007vw,Alioli:2010xd,Jezo:2015aia}, which provides a general interface to 
parton showers. This is necessary for a complete and realistic event
simulation. Especially, non-perturbative QCD effects using hadronization and
underlying event models, as well as multiple photon emissions through 
a QED shower can be included. Those can induce sizable corrections 
in jet-binned cross sections, on the lepton momenta (especially invariant mass distributions/line shapes), 
and other more exclusive observables measured at the LHC.
In our calculation and throughout this paper we omit the loop-induced gluon-fusion contribution, as 
it is already known to higher-order in QCD \cite{Caola:2015rqy,Grazzini:2020stb} and can be evaluated with known tools at LO+PS, such as the {\tt gg2ww} generator \cite{Binoth:2006mf,Kauer:2012hd} used by ATLAS and CMS.
In fact, also a NLO+PS generator was presented for this process recently \cite{Alioli:2021wpn} in the 
\POWHEGBOXRES{} framework.
Finally, we define \ww{} signal events free of top-quark contamination by exploiting the four-flavour scheme with massive bottom quarks and drop all contributions with final-state bottom quarks. \citeresc{Gehrmann:2014fva,Grazzini:2016ctr} have shown for both total and fiducial rates at NNLO that this approach agrees
within $\sim$1-2\% with an alternative procedure to obtain top-free \ww{} predictions. The latter one is defined in the five-flavour scheme and exploits the resonance structure of top-quark contributions to extract the part of the cross section independent of the top-quark width.

This manuscript is organized as follows: in \sct{sec:calculation} we
provide all details about our calculation and implementation. In
particular, we introduce the process and its resonance structures
(\sct{sec:process}), describe the \minnlo{} formulae (\sct{sec:nnlo})
and the practical implementation in \POWHEGBOXRES{}+\Matrix{} (\sct{sec:implementation}). 
We also discuss in detail how we obtain the full two-loop contributions 
by interpolating the basic
two-loop coefficients entering the helicity-amplitudes and how we
validated this procedure (\sct{sec:interpolator}).
 In \sct{sec:phenomenology}, after describing the setup and the set of
 fiducial cuts used in the analysis (\sct{sec:setup}), we 
 present phenomenological results for \minnlo{} and compare them
 against 
 \minlo{}, NNLOPS, NNLO, analytic resummation, and data
 for both integrated cross sections (\sct{sec:crosssection})
 and differential observables (\sct{sec:distributions}).
 We conclude and summarize in \sct{sec:summary}.

\section{Outline of the calculation}
\label{sec:calculation}

\subsection{Description of the process}
\label{sec:process}

\begin{figure}[t]
  \begin{center}
    \begin{subfigure}[b]{0.5\linewidth}
      \centering
      \begin{tikzpicture}
        \begin{feynman}
          \vertex (a1) {\(u\)};
          \vertex[below=1.6cm of a1] (a2){\(\bar u\)};
          \vertex[right=2cm of a1] (a3);
          \vertex[right=2cm of a2] (a4);
          \vertex[right=1.5cm of a3] (a5);
          \vertex[right=1cm of a5] (a20);
          \vertex[above=0.3cm of a20] (a7){\(\ell^+\)} ;
          \vertex[below=0.3cm of a20] (a8){\(\nu_\ell\)};
          \vertex[right=1.5cm of a4] (a6);
          \vertex[right=1cm of a6] (a21);
          \vertex[above=0.3cm of a21] (a9){\(\ell'^-\)} ;
          \vertex[below=0.3cm of a21] (a10){\(\bar{\nu}_{\ell'}\)};

          \diagram* {
            {[edges=fermion]
              (a1)--(a3)--[edge label'=\(d\)](a4)--(a2),
              (a7)--(a5)--(a8),
              (a10)--(a6)--(a9),
            },
            (a3) -- [boson, edge label=\(W^+\)] (a5),
            (a4) -- [boson, edge label=\(W^-\)] (a6),
          };

        \end{feynman}
      \end{tikzpicture}
      \caption{}
        \label{subfig:res1}
    \end{subfigure}%                                                                                                                                                                                        
\begin{subfigure}[b]{.5\linewidth}
\centering
  \begin{tikzpicture}
    \begin{feynman}
      \vertex (a1) {\(q\)};
      \vertex[below=1.6cm of a1] (a2){\(\bar q\)};
      \vertex[below=0.8cm of a1] (a3);
      \vertex[right=1.5cm of a3] (a4);
      \vertex[right=1.3cm of a4] (a5);
      \vertex[dot,fill=black] (d) at (a5){};
      \vertex[right=0.8cm of a5] (a6);
      \vertex[below=0.8cm of a6](a8);
      \vertex[above=0.8cm of a6](a7);

      \vertex[right=1cm of a7] (a9);
      \vertex[above=0.25cm of a9] (a11){\(\ell^+\)} ;
      \vertex[below=0.25cm of a9] (a12){\(\nu_\ell\)};

      \vertex[right=1cm of a8] (a10);
      \vertex[above=0.25cm of a10] (a13){\(\ell'^-\)} ;
      \vertex[below=0.25cm of a10] (a14){\(\bar{\nu}_{\ell'}\)};
      \diagram* {
        {[edges=fermion]
          (a1)--(a4)--(a2),
          (a11)--(a7)--(a12),
          (a14)--(a8)--(a13),
        },
        (a4) -- [boson, edge label'=\(Z/\gamma^*\)] (a5),
        (a5) -- [boson,edge label=\(W^+\),inner sep=0pt] (a7),
        (a5) -- [boson,edge label'=\(W^-\), inner sep = 0pt] (a8),
      };
      
    \end{feynman}
  \end{tikzpicture}
  \caption{}
        \label{subfig:res2}
\end{subfigure}%                                                                                                                                                                                            

\begin{subfigure}[b]{1\linewidth}
      \centering
\begin{tikzpicture}
  \begin{feynman}
    \vertex (a1) {\(q\)};
    \vertex[below=1.6cm of a1] (a2){\(\bar q\)};
    \vertex[below=0.8cm of a1] (a3);
    \vertex[right=1.5cm of a3] (a4);

    \vertex[right=1.3cm of a4] (a5);

    \vertex[right=0.8cm of a5](a6);
    \vertex[above=0.3cm of a6](a7);

    \vertex[right=0.8cm of a7](a8);
    \vertex[above=0.3cm of a8](a9);

    \vertex[right=0.8cm of a9](a10);
    \vertex[above=0.2cm of a10](a11){\(\ell'^-\)};
    \vertex[below=0.2cm of a10](a12){\(\bar{\nu}_{\ell'}\)};
    \vertex[below=0.7cm of a10](a13){\(\nu_\ell\)};
    \vertex[below=1.2cm of a10](a14){\(\ell^+\)};

     \diagram* {
       {[edges=fermion]
         (a1)--(a4)--(a2),
         (a14)--(a5)--[edge label=\(\ell^-\),inner sep=0pt,near end](a7)--(a13),
         (a12)--(a9)--(a11),
       },
       (a4) -- [boson, edge label=\(Z/\gamma^*\)] (a5),
       (a7) -- [boson, edge label=\(W^-\),inner sep=0pt,near end] (a9),
       };

  \end{feynman}
\end{tikzpicture}
\caption{}
        \label{subfig:res3}
\end{subfigure}
\end{center}
  \caption{\label{DiagramsWW} Sample LO diagrams in the different-flavour channel $\ell^+\nu_\ell \ell'^{-}\bar\nu_{\ell'}$ for (a) $t$-channel \ww{} production, (b) $s$-channel $Z/\gamma^\star\to W^+W^-$ production, and (c) DY-type production.
  }
\end{figure}
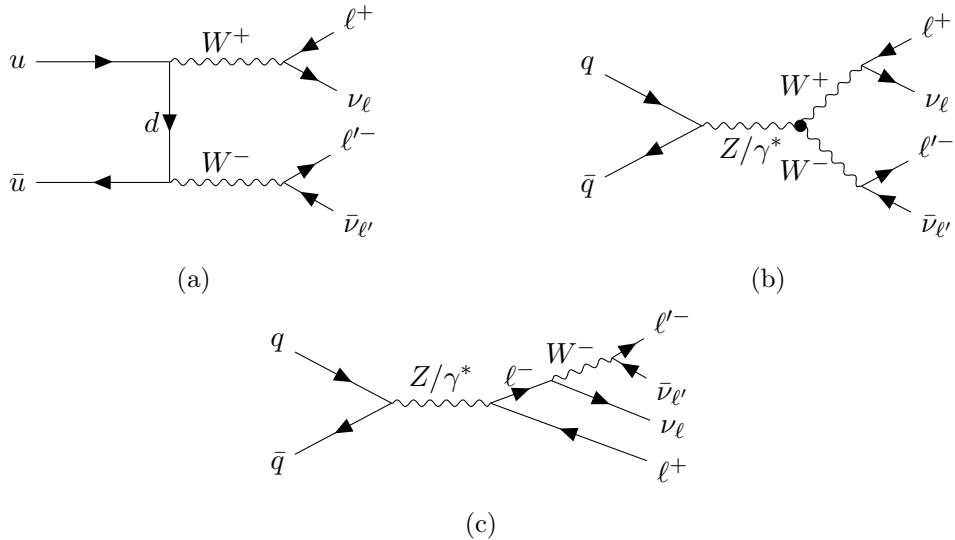

We study the process
\begin{align}
p p  \to \ell^+\nu_\ell\,\ell^{\prime -}{\bar\nu}_{\ell^\prime} + X\,,
\label{eq:process}
\end{align}
for any combination of massless leptons $\ell,\ell^\prime \in\{e,\mu,\tau\}$ with different flavours
$\ell\neq\ell^\prime$.
For simplicity and without loss of generality we consider only the process \mbox{$p p  \to e^+\nu_e \mu^{-}{\bar\nu}_{\mu} + X$} here, which we will refer to as \ww{} production in the following. 
By including all resonant and non-resonant topologies leading to this process, off-shell effects, interferences 
and spin correlations are taken into account. Sample LO diagrams are shown in \fig{DiagramsWW}, 
including
\begin{enumerate}[(a)]
\item double-resonant $t$-channel \ww{} production,
\item double-resonant $s$-channel $Z/\gamma^\star \to W^+W^-$ topologies via a triple-gauge coupling,  
with either the \ww{} pair, or the $Z$ boson and one $W$ boson being resonant,
\item double-resonant DY-type production, where both the $Z$ boson and the $W$ boson can become 
simultaneously resonant.
\end{enumerate}

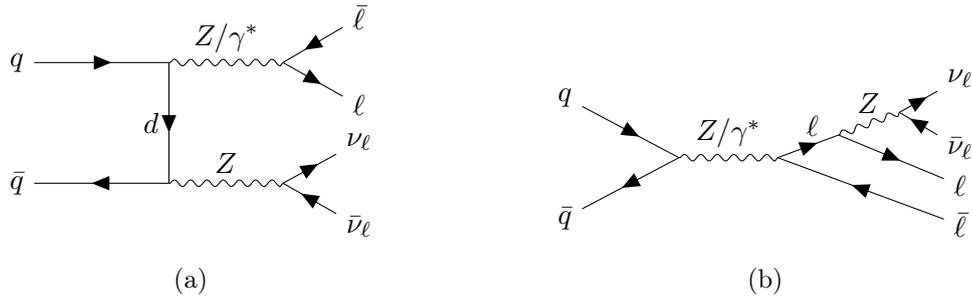
\begin{figure}[t]
  \begin{center}
\begin{subfigure}[b]{0.5\linewidth}
      \centering
      \begin{tikzpicture}
        \begin{feynman}
          \vertex (a1) {\(q\)};
          \vertex[below=1.6cm of a1] (a2){\(\bar q\)};
          \vertex[right=2cm of a1] (a3);
          \vertex[right=2cm of a2] (a4);
          \vertex[right=1.5cm of a3] (a5);
          \vertex[right=1cm of a5] (a20);
          \vertex[above=0.3cm of a20] (a7){\(\bar{\ell}\)} ;
          \vertex[below=0.3cm of a20] (a8){\(\ell\)};
          \vertex[right=1.5cm of a4] (a6);
          \vertex[right=1cm of a6] (a21);
          \vertex[above=0.3cm of a21] (a9){\(\nu_\ell\)} ;
          \vertex[below=0.3cm of a21] (a10){\(\bar{\nu}_\ell\)};

          \diagram* {
            {[edges=fermion]
              (a1)--(a3)--[edge label'=\(d\)](a4)--(a2),
              (a7)--(a5)--(a8),
              (a10)--(a6)--(a9),
            },
            (a3) -- [boson, edge label=\(Z/\gamma^*\)] (a5),
            (a4) -- [boson, edge label=\(Z\)] (a6),
          };

        \end{feynman}
      \end{tikzpicture}
      \caption{}
        \label{subfig:zzres1}
    \end{subfigure}%                                                                                                                                                                                       
\begin{subfigure}[b]{0.5\linewidth}
      \centering
\begin{tikzpicture}
  \begin{feynman}
    \vertex (a1) {\(q\)};
    \vertex[below=1.6cm of a1] (a2){\(\bar q\)};
    \vertex[below=0.8cm of a1] (a3);
    \vertex[right=1.5cm of a3] (a4);

    \vertex[right=1.3cm of a4] (a5);

    \vertex[right=0.8cm of a5](a6);
    \vertex[above=0.3cm of a6](a7);

    \vertex[right=0.8cm of a7](a8);
    \vertex[above=0.3cm of a8](a9);

    \vertex[right=0.8cm of a9](a10);
    \vertex[above=0.2cm of a10](a11){\(\nu_\ell\)};
    \vertex[below=0.2cm of a10](a12){\(\bar{\nu}_\ell\)};
    \vertex[below=0.7cm of a10](a13){\(\ell\)};
    \vertex[below=1.2cm of a10](a14){\(\bar{\ell}\)};

     \diagram* {
       {[edges=fermion]
         (a1)--(a4)--(a2),
         (a14)--(a5)--[edge label=\(\ell\),inner sep=1.5pt,near end](a7)--(a13),
         (a12)--(a9)--(a11),
       },
       (a4) -- [boson, edge label=\(Z/\gamma^*\)] (a5),
       (a7) -- [boson, edge label=\(Z\),inner sep=1.5pt,near end] (a9),
       };

  \end{feynman}
\end{tikzpicture}
\caption{}
        \label{subfig:zzres2}
\end{subfigure}
\end{center}
  \caption{\label{DiagramsZZ} Sample additional LO diagrams appearing in the same-flavour channel $\ell^+\nu_\ell \ell^{-}\bar\nu_{\ell}$ for
    (a) $t$-channel $ZZ$ production, and (b) DY-type production.
  }
\end{figure}

The corresponding production of opposite-charge same-flavour 
leptons $pp\to\ell^+\nu_\ell \ell^{-}\bar\nu_{\ell}+X$ involves the
same type of \ww{} diagrams as shown in \fig{DiagramsWW}, but also
additional \zz{} diagrams as shown in \fig{DiagramsZZ}. By focusing on
the different-flavour case ($\ell\neq\ell^\prime$) we avoid the
complications originating from the mixing of the \ww{} and \zz{}
topologies. In fact, as shown in
\citeres{Melia:2011tj,Kallweit:2018nyv,Kallweit:2017khh}, \ww{} and \zz{} interference
effects can be largely neglected and, to a very good approximation,
predictions for the two processes can be added incoherently.

An important aspect of \ww{} production is that its cross section is
subject to a severe contamination from top-quark contributions. Not
only does this affect \ww{} measurements at the LHC, which usually
employ a jet veto, a $b$-jet veto, or both to suppress top-quark
backgrounds, it also renders the theoretical definition of the \ww{}
cross section cumbersome. Indeed, resonant top-quark contributions enter radiative
corrections to \ww{} production through interference with
real-emission diagrams involving two bottom quarks in the final state.
Those interference terms are numerically so large that they easily
provide the dominant contribution to the cross section.  Specifically,
in the inclusive phase space genuine \ww{} contributions are more than
one order of magnitude smaller. Therefore, the consistent removal of
the top-quark contamination is mandatory to define a top-free \ww{}
cross section.  To this end, we exploit the four-flavour scheme (4FS),
where bottom quarks are treated as being massive, do not enter in the
initial state and diagrams with real bottom-quark radiation are
separately finite. This allows us to drop all contributions with
final-state bottom quarks, thereby cancelling the top-quark
contamination and obtaining top-free \ww{} results.  We note that
there exists an alternative approach to define a top-free \ww{} cross
section that can be used in the five-flavour scheme (5FS).  However,
this approach is much less practical as it requires the repeated
evaluation of the cross section (and distributions) with increasingly
small values of the top-quark width $\Gamma_t$ to extract the top-free
\ww{} cross section as the contribution that is not enhanced by
$1/\Gamma_t$. Indeed, it was shown in \citere{Gehrmann:2014fva} at the
inclusive level and in \citere{Grazzini:2016ctr} for the
fully-differential case that the 4FS and the 5FS definition of the
\ww{} cross section agree at the level of $\sim$1-2\%. For the sake of
simplicity, the easier 4FS approach is employed throughout this paper.

\subsection{The \minnlo{} method}
\label{sec:nnlo}
We employ the \minnlo{} method to build a NNLO+PS generator for \ww{} production. The method
was introduced in \citere{Monni:2019whf}, optimized for $2\to 1$ scattering processes in \citere{Monni:2020nks}, and generalized to 
$2\to 2$ colour-singlet scattering processes in \citere{Lombardi:2020wju}.
In the following we recall the basic ideas and essential ingredients of \minnlo{},
adapting the notation of \citeres{Monni:2019whf,Monni:2020nks,Lombardi:2020wju}.

\minnlo{} formulates a NNLO calculation fully differential in the phase space 
$\PhiB$ of a produced colour singlet \F{} with invariant mass $Q$, in such a 
way that it can be subsequently matched to a parton shower. 
It starts from a differential description of colour singlet plus jet (\FJ{}) production in the \POWHEG{} approach \cite{Nason:2004rx,Frixione:2007vw,Alioli:2010xd} 
\begin{align}
\frac{\mathd\sigma}{\mathd\PhiBJ}={\bar B}(\PhiBJ) \times
\bigg\{\Delta_{\rm pwg} (\LambdaPWG) + \int\mathd \Phi_{\tmop{rad}} 
  \Delta_{\rm pwg} (\ptrad)  \frac{R (\PhiBJ{}, \Phi_{\tmop{rad}})}{B
  (\PhiBJ{})}\bigg\}\,,
\label{eq:master}
\end{align}
and it achieves NNLO accuracy for \F{} production by modifying the content of the ${\bar B}(\PhiBJ)$ function.
With $\PhiBJ$ we have denoted the \FJ{} phase space, 
$\Delta_{\rm pwg}$ is the \POWHEG{} Sudakov form factor, $\Phi_{\tmop{rad}} $ ($\ptrad$) is the phase space 
(transverse momentum) of the second-hardest radiation, and $B$ and $R$ denote the squared tree-level matrix elements for \FJ{} and \FJJ{} production, respectively.
The central ingredient of the \minnlo{} method is the modified ${\bar B}(\PhiBJ)$ function, which describes the \F{} process at NNLO 
and the \FJ{} process at NLO, including both zero and one QCD emissions, respectively. The content of the curly brackets
generates the second QCD emission according to the \POWHEG{} mechanism, with a default \POWHEG{} cutoff of $\LambdaPWG=0.89$\,GeV.
Additional radiation that contributes at $\mathcal{O}(\as^3(Q))$ and beyond to all orders in perturbation theory
is added by the parton shower.

The \minnlo{} ${\bar B}(\PhiBJ)$ function can be expressed as follows \cite{Monni:2019whf,Monni:2020nks,Lombardi:2020wju}
\begin{align}
\label{eq:Bbar}
{\bar B}(\PhiBJ)&\equiv \sum_{\flavBJ} \Bigg\{\exp[-\tilde{S}_{\projflav}(\pt)] \bigg\{\abarmu{\pt}\left[\frac{\mathd\sigma_{\scriptscriptstyle\rm FJ}}{\mathd\PhiBJ}\right]^{(1)}_{\flavBJ} \left(1+\abarmu{\pt} [\tilde{S}_{\projflav}(\pt)]^{(1)}\right)\notag
  \\
&+ \left(\abarmu{\pt}\right)^2\left[\frac{\mathd\sigma_{\scriptscriptstyle\rm FJ}}{\mathd\PhiBJ}\right]^{(2)}_{\flavBJ}\bigg\} + \bigg\{\sum_{\flavB} \exp[-\tilde{S}_{\flavB}(\pt)]\,\mathcal{D}_{\flavB}(\pt)\bigg\}\, F^{\tmop{corr}}_{\flavBJ}(\PhiBJ)\Bigg\}\,,
\end{align}
where $\pt$ refers to the transverse momentum of the color singlet.
The overall sum runs over all flavour 
structures $\flavBJ$ of \FJ{} production, while $\flavB$ denotes the flavour 
structures of the Born process $pp\to\text{\F}$.
With $\projflav$ we denote a projection of the flavour structures, which
is trivial in the case of \ww{} production,
since the Born is always $q\bar{q}$ initiated.
All quantities with index $\flavB$ have to be evaluated in the Born kinematics
$\PhiB$, which requires a suitable projection $\PhiBJ\to\PhiB$ as introduced in appendix~A of \citere{Monni:2019whf}.
The notation 
$[X]^{(i)}$ is used for the $i$-th term in the perturbative expansion of a quantity $X$.
$\tilde{S}_{\flavB}(\pt)$ represents the Sudakov form factor and
$\mathd\sigma_{\scriptscriptstyle\rm FJ}$ is the differential fixed-order cross section, 
as defined in eqs.\,(2.9) and (2.11) of \citere{Monni:2019whf}, respectively.
The last term in \eqn{eq:Bbar} is the central contribution added by
the \minnlo{} method to achieve NNLO accuracy. The precise
definition and derivation of $\mathcal{D}_{\flavB}(\pt)$ is discussed below.
The factor $F^{\tmop{corr}}(\PhiBJ)$ encodes the dependence of the
Born-like NNLO corrections upon the full $\PhiBJ$ phase space, as discussed 
in detail in section 3 of \citere{Monni:2019whf} and section 3.3 of \citere{Lombardi:2020wju}.

A few comments are in order: a crucial feature of the \minnlo{} method is 
that the renormalisation and factorisation scales are evaluated as 
$\muR\sim\muF\sim \pt$. As a consequence, each term 
contributes to the total cross section with scales
$\muR\sim\muF\sim Q$ according to the following power counting formula:
\begin{equation}
\label{eq:counting}
    \int_{\Lambda}^{Q} \mathd \pt \frac{1}{\pt} \as^m(\pt) \log^n\frac{Q}{\pt}
\exp(-\tilde{S}(\pt))    \approx {\cal O}\left(\as^{m-\frac{n+1}{2}}(Q)\right)\,.
\end{equation}
This implies that, when including terms up to second order in
$\as(\pt)$ in \eqn{eq:Bbar}, upon integration over $\pt$, the cross
section is NLO accurate, as observed first in \citere{Hamilton:2012rf}.
By deriving also all (singular) contributions in \eqn{eq:Bbar} at
third order in $\as(\pt)$, 
NNLO accuracy is achieved after integration over $\pt$ \cite{Monni:2019whf}.
Indeed, $\mathcal{D}_{\flavB}(\pt)$ consistently adds the relevant 
singular $\as^3(\pt)$ corrections, while regular contributions at this order 
can be safely omitted as a consequence of the counting in \eqn{eq:counting}.
In fact, two results for $\mathcal{D}_{\flavB}(\pt)$ have been 
derived \cite{Monni:2019whf,Monni:2020nks} that differ only by 
terms of ${\cal O}(\as^4)$ and higher.
Their derivation stems from the analytic formulation of the NNLO cross section 
differential in $\pt$ and $\PhiB$:
\begin{align}
\label{eq:start}
  \frac{\mathd\sigma}{\mathd\PhiB\mathd \pt} &= \frac{\mathd}{\mathd \pt}
     \Bigg\{\sum_{\flavB}\exp[-\tilde{S}_{\flavB}(\pt)] {\cal L}_{\flavB}(\pt)\Bigg\} +
                                               R_f(\pt) \\
                                               &= \sum_{\flavB}\exp[-\tilde{S}_{\flavB}(\pt)] \, D_{\flavB}(\pt) +
                                               R_f(\pt) \nonumber\,,
\end{align}
where $R_f$ includes only non-singular contributions at small $\pt$, and 
\begin{equation}
\label{eq:Dterms}
  D_{\flavB}(\pt)  \equiv -  \frac{\mathd \tilde{S}_{\flavB}(\pt)}{\mathd \pt} {\cal L}_{\flavB}(\pt)+\frac{\mathd {\cal L_{\flavB}}(\pt)}{\mathd \pt}\,.
\end{equation}
The luminosity factor ${\cal L}_{\flavB}(\pt)$ contains the parton
densities, the squared hard-virtual matrix elements for \F{}
production up to two loops as well as the NNLO collinear coefficient
functions, and its expression is given in eq.\,(3.5) of
\citere{Lombardi:2020wju}. 

As discussed in detail in \citere{Monni:2019whf}, by choosing a suitable 
resummation scheme ($\muR\sim\muF\sim \pt$) and matching scheme
(factoring out $\tilde{S}_{\flavB}(\pt)$ from $R_f$ as well), and by making \eqn{eq:start}
accurate to third order in $\as(\pt)$, the relevant corrections to achieve NNLO accuracy upon integration over 
$\pt$ are derived. 
In the original \minnlo{} formulation of \citere{Monni:2019whf} the expansion 
was truncated beyond third order in $\as(\pt)$, so that $\mathcal{D}_{\flavB}(\pt)$ 
would be derived as
\begin{align}
\label{eq:Dold}
\mathcal{D}_{\flavB}(\pt) \equiv \left(\abarmu{\pt}\right)^3 [D_{\flavB}(\pt)]^{(3)}+\mathcal{O}(\as^4)\,,
\end{align}
which breaks the total derivative of the starting formula in \eqn{eq:start}.
Instead, \citere{Monni:2020nks} suggested a new prescription that
preserves the total derivative by keeping into 
account additional terms beyond accuracy, so that we use
\begin{align}
\label{eq:Dnew}
\mathcal{D}_{\flavB}(\pt) \equiv D_{\flavB}(\pt) -\abarmu{\pt} [D_{\flavB}(\pt)]^{(1)} -\left(\abarmu{\pt}\right)^2 [D_{\flavB}(\pt)]^{(2)}\,,
\end{align}
as our default choice throughout this paper. 
The relevant expressions for its evaluation, including the ones of 
the $[D_{\flavB}(\pt)]^{(i)}$ coefficients, 
are reported in appendix~C and D of \citere{Monni:2019whf} and in appendix~A of \citere{Monni:2020nks}, where the flavour dependence can be simply included 
through the replacements $H^{(1)}\rightarrow H^{(1)}_{\flavB}$, $H^{(2)}\rightarrow H^{(2)}_{\flavB}$, and $\tilde B^{(2)}\rightarrow \tilde B^{(2)}_{\flavB}$.

We further note that the flavour dependence of $\tilde{S}_{\flavB}(\pt)$ and
${\cal L}_{\flavB}$ originates entirely from the hard-virtual coefficient function 
$H_{\flavB}$, which  for a general $2\to 2$ hadronic process depends on both the flavour and the Born phase space.
This dependence propagates to the Sudakov form factor through the 
replacement $\tilde{B}_{\flavB}^{(2)} = B^{(2)} + 2\zeta_3 (A^{(1)})^2 + 2 \pi \beta_0\,H^{(1)}_{\flavB}$
in eq.\,(4.26) of \citere{Monni:2019whf}, where $\beta_0 = \frac{11 \CA{} - 2 \nf}{12\pi}$.
Moreover, $H^{(1)}_{\flavB}$ and  $H^{(2)}_{\flavB}$ are unambiguously defined in section (3.3) of \citere{Lombardi:2020wju}.

\subsection{Practical implementation in \POWHEGBOXRES{}+\Matrix{}}
\label{sec:implementation}

As a starting point, we exploit the \ww{}+jet generator developed in 
\citere{Hamilton:2016bfu} for \POWHEGBOXVTWO{} \cite{Alioli:2010xd} and 
integrated it into the \POWHEGBOXRES{} framework \cite{Jezo:2015aia}. To this end, we
 had 
to adapt the  \POWHEGBOXRES{} code to automatically 
find all relevant resonance 
histories for \ww{}+jet production. 
This was required, because the automatic generation of resonance histories
is not fully functional for processes with a jet in the final state.
As described in detail in \citere{Jezo:2015aia} and recalled 
in section 2.2 of \citere{Lombardi:2020wju}, the correct implementation of all 
resonance histories is necessary to take advantage of the efficient phase-space 
sampling within \POWHEGBOXRES{}. We have then upgraded the $W^-W^+$+jet
generator to include NNLO accuracy for \ww{} production by means of the \minnlo{} method.
This has been achieved by making use of the general \minnlo{} implementation for colour singlet production 
developed in \citere{Lombardi:2020wju} and adapting it consistently to the 4FS.

As far as the physical amplitudes are concerned, all tree-level real and double-real
matrix elements (i.e.\ for $\ell^+\nu_\ell \ell'^{-}\bar\nu_{\ell'}$+1,2-jet production)
are evaluated through the \POWHEGBOX{} interface to 
\noun{Madgraph 4}~\cite{Alwall:2007st} developed in \citere{Campbell:2012am}. 
The $\ell^+\nu_\ell \ell'^{-}\bar\nu_{\ell'}$+jet one-loop amplitude is obtained from \noun{GoSam 2.0}~\cite{Cullen:2014yla},
neglecting one-loop fermion box diagrams, which have been shown to give a negligibly 
contribution, but slow down the code substantially (cf.\ \citere{Hamilton:2016bfu}).\footnote{Note that there is an option in the Makefile of our code to include the one-loop fermion box diagrams.}
The Born-level and one-loop amplitudes for $\ell^+\nu_\ell \ell'^{-}\bar\nu_{\ell'}$ production 
have been extracted from \noun{MCFM} \cite{Campbell:2019dru}. 
The (one-loop and) two-loop $q\bar{q}\to \ell^+\nu_\ell \ell'^{-}\bar\nu_{\ell'}$ helicity amplitudes that were derived in \citere{Gehrmann:2015ora}
are obtained through their implementation in \Matrix{}
by suitably adapting the interface created in \citere{Lombardi:2020wju}.
Those amplitudes are known only in the massless approximation, but the 
effect of including massive quark loops is expected to be negligible 
because of the smallness of closed fermion-loop contributions.
For a fast evaluation of the two-loop amplitudes, we have generated interpolation 
grids, as discussed in detail in the next section.

The calculation of  $\mathcal{D}_{\flavB}(\pt)$ in \eqn{eq:Dnew} involves
the evaluation of several convolutions with the parton distribution functions (PDFs), 
which are performed through \noun{hoppet}~\cite{Salam:2008qg}.
Moreover, the collinear coefficient functions require the computation of 
polylogarithms, for which we employ the \noun{hplog} package~\cite{Gehrmann:2001pz}.

Finally, we report some of the most relevant (non-standard) settings
we have used to produce \ww{} events.  We refer the reader to
\citere{Monni:2020nks} for a detailed discussion on these settings.
In particular, to avoid spurious contributions from higher-order
logarithmic terms at large $\pt$ we consistently introduce modified
logarithms with the choice of $p=6$, as defined in eq.\,(10) of
\citere{Monni:2020nks}.  At small $\pt$, we use the standard \minnlo{}
scale setting in eq.\,(14) of \citere{Monni:2020nks}, while we activate
the option {\tt largeptscales 1} to set the scales entering the NLO
\ww{}+jet cross section at large $\pt$ as in eq.\,(19) of
\citere{Monni:2020nks}.  We use those scale settings with the
parameter $\Q =0$\,GeV, and instead regularize the Landau singularity
by freezing the strong coupling and the PDFs for scales below
$0.8$\,GeV.  We turn on the \POWHEGBOX{} option \texttt{doublefsr 1}, which
was introduced and discussed in detail in \citere{Nason:2013uba}.
As far as the parton-shower settings are  concerned, we have used the 
standard ones (also for the recoil scheme).

\subsection{Fast evaluation of the two-loop amplitude}
\label{sec:interpolator}
As discussed before, the two-loop helicity amplitudes
for the production of a pair of off-shell vector bosons
were computed in \citere{Gehrmann:2015ora} 
and the relevant coefficients functions to construct the amplitudes
can be obtained from the publicly available code {\tt
  VVAMP}~\cite{hepforge:VVamp}. Using those results all $q\bar{q} \, \rightarrow \,4 \,\text{leptons}$ amplitudes have been implemented in the \Matrix{}
framework~\cite{Grazzini:2017mhc,Matrixurl}. To exploit this
implementation for our calculation, we have compiled
\Matrix{} as a {\tt C++} library and linked it to our \minnlo{}
generator using the interface created in \citere{Lombardi:2020wju}.

The evaluation of these two-loop amplitudes turns out to be the
bottleneck of the calculation. In fact, it takes on average $\bar{t}_{\tt
  VVAMP}\approx 1.9$\,s to evaluate a single phase-space point, while
the evaluation of the tree- and one-loop amplitudes are orders of magnitude faster.
Therefore, even though we provide the option to run the code
using the exact two-loop amplitudes, all of the results of
this paper have been obtained using a four-dimensional cubic spline interpolation
procedure for the set of independent two-loop coefficient functions that
are required for the evaluation of the two-loop helicity
amplitudes. In the following, we present this procedure in detail.

\subsubsection[Coefficient functions of the $q\bar{q}\to\ell^+\nu_\ell \ell'^{-}\bar\nu_{\ell'}$ helicity amplitudes]{Coefficient functions of the \boldmath{$q\bar{q}\to\ell^+\nu_\ell \ell'^{-}\bar\nu_{\ell'}$} helicity amplitudes}
\label{sec:helicity}
We start by recalling some relevant formulae in \citere{Gehrmann:2015ora}
for the helicity amplitudes. Specifically, the physical process is denoted by:
\begin{equation}
q(p_1) + \bar{q}(p_2) \rightarrow W^+(p_3) + W^-(p_4)
\rightarrow \ell^+(p_5) + \nu_{\ell}(p_6) + \ell'^-(p_7) + \bar{\nu}_{\ell'}(p_8)\,,
\end{equation}
where $p_i$ are the momenta of the corresponding particles and each of the two off-shell $W$ bosons decays
into a neutrino--lepton pair, such that $p_3= p_5+p_6$ and $p_4=p_7+p_8$. We denote by $\M_{\lambda \lambda_1 \lambda_2}$ the bare helicity
amplitudes of a general vector-boson pair production process, where $\lambda$ represents the handedness of the partonic current, while $\lambda_1$
and $\lambda_2$ stand for the helicities of the two leptonic currents. There are just two
independent helicity amplitudes $\M_{LLL}$ and $\M_{RLL}$, since all the other helicity configurations can be recovered by permutations of
external legs~\cite{Gehrmann:2015ora}. The bare helicity amplitudes are the building blocks of the dressed helicity amplitudes
$\mathcal{M}_{\lambda LL}$, which are process specific and for \ww{} production read
\begin{align}
\mathcal{M}_{\lambda LL}^{W^+W^-}(p_1, p_2;p_5,p_6,p_7,p_8) &=
 \frac{(4 \pi \aew)^2}{2\, \sin^2\thW} \;
\frac{ \M_{\lambda LL}(p_1, p_2;p_5,p_6,p_7,p_8) }
{(p_3^2 - m^2_W + i\,\GW \mw)(p_4^2 - m^2_W + i\,\GW \mw)}\,,
\label{HelAmpl}
\end{align}
where $\lambda = L,R$. In the previous expression, $\aew$  refers to the EW coupling constant, $\thW$ to
the mixing angle, and $\mw$ and $\GW$ to the $W$-boson mass and decay width, respectively. Since a $W$ boson can just couple
to left-handed lepton currents, it is clear that $\mathcal{M}_{\lambda RL}=\mathcal{M}_{\lambda LR}=\mathcal{M}_{\lambda RR}=0$.
As shown in \citere{Gehrmann:2015ora}, for four-dimensional external states the expression of the bare
helicity amplitudes can be written in a compact form using the spinor-helicity formalism:
\begin{align}
\M_{\lambda LL}(p_1,p_2;p_5,p_6,p_7,p_8) &=
 ( [i5] \langle 5j \rangle  + [i6] \langle 6j \rangle )\, \Big\{
   E_1\, \langle 15 \rangle \langle 17 \rangle [16][18] \nonumber \\
&
+ E_2\, \langle 15 \rangle \langle 27 \rangle [16][28]
+ E_3\, \langle 25 \rangle \langle 17 \rangle [26][18] \nonumber \\
& +
 E_4\, \langle 25 \rangle \langle 27 \rangle [26][28] \,
 + E_5 \langle 5 7 \rangle [ 68 ] \Big\}\nonumber \\
&+ E_6\, \langle 15 \rangle \langle j7 \rangle [16][i8]
+   E_7\, \langle 25 \rangle \langle j7 \rangle [26][i8] \nonumber \\
&+ E_8\, \langle j5 \rangle \langle 17 \rangle [i6][18]
+ E_9\, \langle j5 \rangle \langle 27 \rangle [i6][28]\,,   \label{MLLL}
\end{align}
where the two indices $i$ and $j$ are determined by the handedness of the partonic current: $(i,j)=(1,2)$ for $\lambda=L$ and
$(i,j)=(2,1)$ for $\lambda=R$. \eqnc{MLLL} depends on
nine complex scalar coefficients $E_j$, which are functions of the invariant masses $p^2_3$ and $p^2_4$
of the two vector bosons and of the two Mandelstam invariants $\sm$ and $\tm$, defined as
\begin{align}
  \sm=(p_1+p_2)^2\,, \quad\quad\quad \tm=(p_1-p_3)^2\,.
\end{align}
Each coefficient $E_j$ receives a contribution from four different classes
of diagrams $\C$
\begin{align}\label{AdecompEW}
E_j &= \delta_{i_1 i_2} \sum_{\C} Q^{\lambda,W^+W^-,[\C]}_{q\,q} E_j^{[\C]}\,,
\qquad j = 1,\ldots,9\,,
\end{align}
\\[-1.15em]
where $i_1$, $i_2$ represent the colours of the incoming quark and anti-quark,
respectively, and $Q^{\lambda,W^+W^-,[\C]}_{q\,q}$ denotes a coupling factor, which is
the only process specific ingredient entering~\eqn{AdecompEW}.
Following the labeling introduced in \citere{Gehrmann:2015ora} for the diagram classes, we have
for \ww{} production:
\begin{itemize}
\item class $A$ and $B$, including all diagrams where the two vector bosons are
  attached to the fermion line, with the $W^+$ boson adjacent to the incoming quark or
  antiquark, respectively, whose coupling factors read
  \begin{align}
  Q^{L,W^+W^-,[A]}_{q\,q}=  Q^{L,W^+W^-,[B]}_{q\,q}=\frac{1}{2\sin^2\thW}\,,
  \end{align}
  which is identical to zero for $\lambda=R$;
\item class $C$, containing diagrams where both vector bosons are attached to a fermion loop,
  where
  \begin{align}
  Q^{\lambda,W^+W^-,[C]}_{q\,q}=\frac{n_g}{4\,\sin^2\thW}\,,\quad\quad\text{for $\lambda=L,R$}\,,
  \end{align}
  with $n_g$ being the number of massless quark generations;
\item class $F_V$, collecting form-factor diagrams where the production of the two $W$ bosons is mediated either by a virtual photon ($V=\gamma^*$) or a $Z$ boson
  ($V=Z$), as shown in~\fig{subfig:res2}.\footnote{Note that another class of form-factor diagrams exists, containing two-loop corrections to DY-type
production (see~\fig{subfig:res3}). This class is evaluated by \Matrix{} 
using also the corresponding form factor returned by {\tt VVAMP}. 
Since those form factors are constants, as discussed below, 
their contribution is handled without interpolation.}
In that case we have
\begin{align}
      Q^{L,W^\pm W^\mp,[F_{Z}]}_{q\,q}=&\frac{\mp1}{\sin^2\thW}\frac{(I^3_q-e_q\,\sin^2\thW)}{\sm-m^2_Z-i\GZ\mz}\,,\quad Q^{R,W^\pm W^\mp,[F_{Z}]}_{q\,q}=\frac{\pm e_q}{\sm-m^2_Z-i\GZ\mz}\,,\nonumber\\
      Q^{\lambda,W^\pm W^\mp,[F_{\gamma^*}]}_{q\,q}=&\frac{\mp e_q}{\sm}\,,\quad\quad\text{for $\lambda=L,R$}\,,
    \end{align}
    where $e_q$ and $I^3_q$ are the electric charge and isospin number of the incoming quark $q$, and  $\mz$ and $\GZ$ the $Z$-boson mass and decay width, respectively. 
  \end{itemize}
Since the functions $E_j$ admit a perturbative expansion as
\begin{align}
  \label{eq:Eexp}
  E_j=E_j^{(0)}+\biggl(\frac{\as}{2\pi}\biggr)E_j^{(1)}+\biggl(\frac{\as}{2\pi}\biggr)^2E_j^{(2)}+\mathcal{O}(\as^3)\,,
  \end{align}
the two-loop contribution to the helicity amplitude $\M_{\lambda LL}$ is fully determined once the $45$ complex coefficients $E_j^{[\C],(2)}$ are known. In contrast with the helicity amplitude itself, which is a complex-valued function of the full kinematics, the coefficients $E_j^{[\C],(2)}$ just depend on four Lorentz scalars. Therefore, an interpolation procedure that approximates the $E_j^{[\C],(2)}$ coefficients is clearly more feasible. This choice considerably reduces the complexity of the interpolation problem, since it decreases the dimensionality of the space on which the functions are interpolated, at the minor cost of increasing the number of functions to approximate. In essence, this turns our problem into a four-dimensional interpolation of $90$ real-valued functions.

However, one should bear in mind that $E_j^{[F_V], (2)}$ does not depend on the type of the vector boson $V$, so that in our case $E_j^{[F_{\gamma^*}], (2)}=E_j^{[F_Z], (2)}=E_j^{[F], (2)}$.
Moreover, any loop correction to the corresponding form-factor diagrams just amounts to a function $\FF(\sm)$ which multiplies the tree level structure, so that at two loops
\begin{align}
E_j^{[F],(2)} &= \FF^{(2)}(\sm) E_j^{[F],(0)}\,. \label{HelAmplFV}
\end{align}
The tree-level coefficients evaluate to constants:
\begin{align}
    E_j^{[F],(0)} &= 0\,, \quad j=1,...,4\,, \nonumber\\
E_6^{[F],(0)} &= E_7^{[F],(0)} = +4\,, &
  E_5^{[F],(0)}&=E_8^{[F],(0)} = E_{9}^{[F],(0)} = -4\,. & \label{HelAmplTree}
\end{align}
The dependence on $\sm$ in $\FF^{(2)}(\sm)$ just enters through the
ratio of $\sm$ with the squared of the renormalization scale $\muR$. 
By setting $\muR=\sqrt{\sm}$, the non-vanishing $E_j^{[F],(2)}$
coefficients also become constants, which are known
\cite{Gehrmann:2005pd,Gonsalves:1983nq,vanNeerven:1985xr,Kramer:1986sr}.
Note that the correct renormalization scale dependence will be recovered 
through the \minnlo{} formulae (cf.\ appendix~D of \citere{Monni:2019whf}).
Therefore, only the coefficient functions belonging to families
$\C=\{A,B,C\}$ need to be interpolated.
This reduces the number of
real-valued functions that need to be interpolated from $90$ to $54$.

\subsubsection{Generation of interpolation grids}
\label{sec:grids}
As a first step, we have generated rectilinear grids (i.e comprised of congruent parallelotopes) for each of the $54$ non-constant two-loop coefficient functions $E_j^{[\C],(2)}$ defined in eqs.\,\eqref{AdecompEW} and \eqref{eq:Eexp}, whose exact values
 have been computed through {\tt VVAMP} on a set of given phase-space points $(p^2_3,\,p^2_4,\,\sm,\,\tm)$
and stored. All results have been obtained by fixing the centre-of-mass energy to $\sqrt{s}=13\,$TeV.

As it turns out, a suitable parametrization of the four-dimensional phase-space points $(p^2_3,\,p^2_4,\,\sm,\,\tm)$ is crucial to obtain a good
interpolation performance. Moreover, a finer binning is required in those phase-space
regions that have a large contribution to the overall integral of the multi-differential cross section, such as
resonance-enhanced regions in $p^2_3$ and $p^2_4$ around the two $W$-boson masses.
To this end, our grids are defined on
a four-dimensional unit hypercube $[0,1]^4$ with fifty equally spaced bins, where each element $(x_1,\,x_2,\,x_3,\,x_4)\in [0,1]^4$
is uniquely mapped to a physical phase-space point. The first two axes $x_1$ and $x_2$ are related to the invariant masses $p^2_3$ and $p^2_4$
through the transformations
\begin{align}
  p^2_3=\mw\GW\,\tan(z_1(x_1))+m^2_W\,, & \quad\text{with}\quad z_{1,\,\text{\scalefont{0.77}min}}<z_1(x_1)<z_{1,\,\text{\scalefont{0.77}max}}\,, &\nonumber\\
  p^2_4=\mw\GW\,\tan(z_2(x_2))+m^2_W\,, & \quad\text{with}\quad z_{2,\,\text{\scalefont{0.77}min}}(p^2_3)<z_2(x_2)<z_{2,\,\text{\scalefont{0.77}max}}(p^2_3)\,, &\label{zdefinitions}
  \end{align}
where $z_1$ and $z_2$ are continuous functions of $x_1$ and $x_2$, respectively.
The lower and upper bounds on $z_1$ and $z_2$ have been chosen in such a way
 that the physical range of
invariant mass values is covered. Specifically, $z_{1,\,\text{\scalefont{0.77}min}}$ and $z_{1,\,\text{\scalefont{0.77}max}}$
are fixed by the choice $40\,\text{GeV}^2<p^2_3<s/100$. 
Through energy conservation $z_{2,\,\text{\scalefont{0.77}min}}$ and $z_{2,\,\text{\scalefont{0.77}max}}$ depend directly on $p^2_3$, as it has been made
explicit in~\eqn{zdefinitions}.
However, their exact expressions, which we omit here, 
have been tuned such that the physical mass range of $p^2_4$ is covered 
efficiently.
The two functions $z_1(x_1)$ and $z_2(x_2)$ are defined piecewise on three subranges of the two intervals in~\eqn{zdefinitions}. In the central subrange, $z_1$ and $z_2$ correspond to
a linear mapping, which guarantees that $p^2_3$ and $p^2_4$ follow a Breit-Wigner distribution.
For the other two subranges of $z_1$ and $z_2$ polynomial functions are used such that
the off-shell regions are covered by a sufficient number of grid points. 

The variables $x_3$ and $x_4$ also have a physical interpretation, since they are related to the relativistic velocity
$\beta_{\text{\scalefont{0.77}$W^+$}}$ and the cosine of the scattering angle $\cos\theta_{\text{\scalefont{0.77}$W^+$}}$ of one of the vector
bosons in the center of mass frame. In particular, we define
\begin{align}\label{betacoshyp}
  \beta_{\text{\scalefont{0.77}$W^+$}}=a_s+(b_s-a_s)\,x_3\,, \quad\quad\quad
  \cos\theta_{\text{\scalefont{0.77}$W^+$}}=1-2\,(a_t+(b_t-a_t)\,x_4)\,,
\end{align}
where $a_{s/t}$ and $b_{s/t}$ determine the range of values allowed for the two physical quantities. Instead of setting
$a_{s/t}=0$ and $b_{s/t}=1$, we use small cutoffs to avoid numerical instabilities at the kinematic edges.
$\beta_{\text{\scalefont{0.77}$W^+$}}$ and $\cos\theta_{\text{\scalefont{0.77}$W^+$}}$ can be expressed 
in terms of $\sm$ and $\tm$ as
\begin{align}\label{betacosW}
  \beta_{\text{\scalefont{0.77}$W^+$}}=\frac{\kappa\left(\sm,p^2_3,p^2_4\right)}{\sm+p^2_3-p^2_4}\,,\quad\quad \cos\theta_{\text{\scalefont{0.77}$W^+$}}=\frac{2\tm+\sm-p^2_3-p^2_4}{\kappa\left(\sm,p^2_3,p^2_4\right)}\,,
  \end{align}
with the K\"all\'en function
\begin{align}\label{kaellen}
\kappa\left(\sm,p_3^2,p_4^2\right) \equiv
  \sqrt{\sm^2 + p_3^4 + p_4^4 - 2 (\sm\,p_3^2 + p_3^2\,p_4^2 + p_4^2\,\sm)}\,.
\end{align}
By inverting~\eqn{betacosW} in the physical region of the process, which is defined by
\begin{align}
\sm \geq \Big(\sqrt{p_3^2} + \sqrt{p_4^2}\Big)^2\,,\qquad
\frac{1}{2}\big(p_3^2+p_4^2-\sm -\kappa\big) \leq \tm \leq  \frac{1}{2}\big(p_3^2+p_4^2-\sm +\kappa\big)\,,
\end{align}
we can express $\sm$ and $\tm$ in terms of the hypercube variables $x_3$ and $x_4$.

As illustrated in \citere{Gehrmann:2015ora}, the behaviour of the coefficients $E_j^{[\C], (2)}$
is not always smooth over the two-dimensional phase space  $\left(\beta_{\text{\scalefont{0.77}$W^+$}},\,\cos\theta_{\text{\scalefont{0.77}$W^+$}}\right)$ and it can even be divergent close to the highly relativistic ($\beta_{\text{\scalefont{0.77}$W^+$}}\to 1$) or highly collinear
($|\cos\theta_{\text{\scalefont{0.77}$W^+$}}|\to 1$) regions. One possibility to improve the description of this rapidly
changing functional behaviour is to combine different grids to cover the whole phase space $(p_3^2,\,p^2_3,\,\sm,\,\tm)$,
instead of simply increasing the number of bins for selected axes. For the case at hand, using four precomputed grids
for each of the $54$ real-valued functions has proven to significantly improve the performance of the interpolator in some phase-space regions.
Even though the definition of the grids is unchanged for $x_1$ and $x_2$, by adjusting the values of $a_{s/t}$ and $b_{s/t}$ in~\eqn{betacoshyp} we defined
four slightly overlapping grids in the $\left(\beta_{\text{\scalefont{0.77}$W^+$}},\,\cos \theta_{\text{\scalefont{0.77}$W^+$}}\right)$
phase space, to properly cover the above mentioned singular regions.

\subsubsection{Interpolation and validation}
\label{sec:validation}
At the beginning of each run the grids just need to be read and loaded into memory. Then, for each 
$E_j^{[\C], (2)}$ any value can be computed by properly interpolating between the values stored in the precomputed
grids. To perform this task, we make use of the $N$-dimensional interpolation library {\tt Btwxt}~\cite{btwxturl}, which just requires
the input grids to be rectilinear. The interpolation is achieved through
$N$-dimensional cubic splines~\cite{doi:10.1002/sapm1960391258}, which are multivariate piecewise polynomials of degree three. Specifically, {\tt Btwxt} employs
cubic Hermite splines, where each polynomial in a given $N$-dimensional interval is specified by its values and its
first derivatives at the corners of the interval itself. The values of the first derivatives are computed according to the Catmull-Rom
implementation~\cite{CATMULL1974317}. 

\begin{figure}[t]
\begin{center}\vspace{-0.2cm}
\begin{tabular}{cc}
\includegraphics[width=.31\textheight]{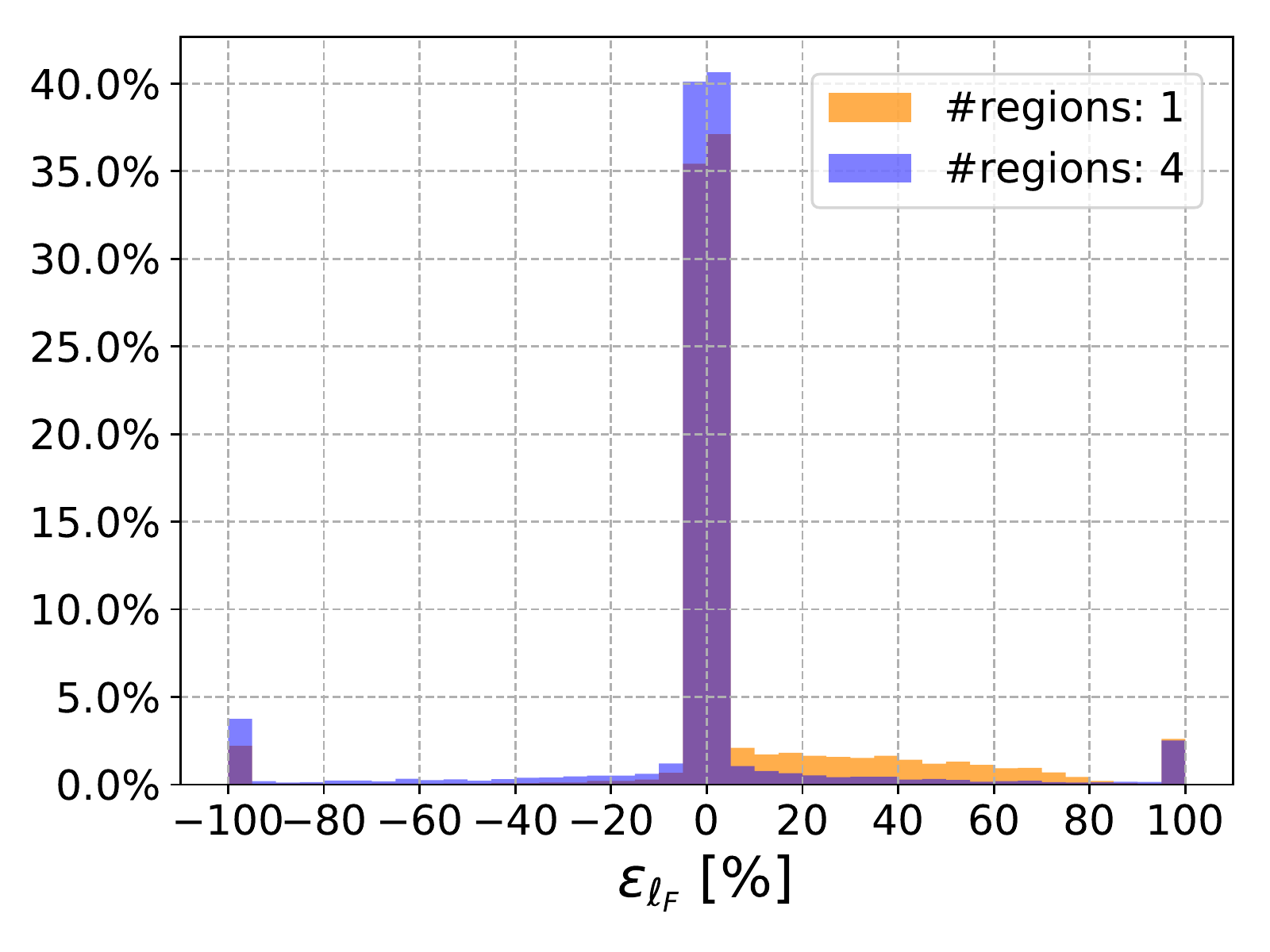}
&
\includegraphics[width=.31\textheight]{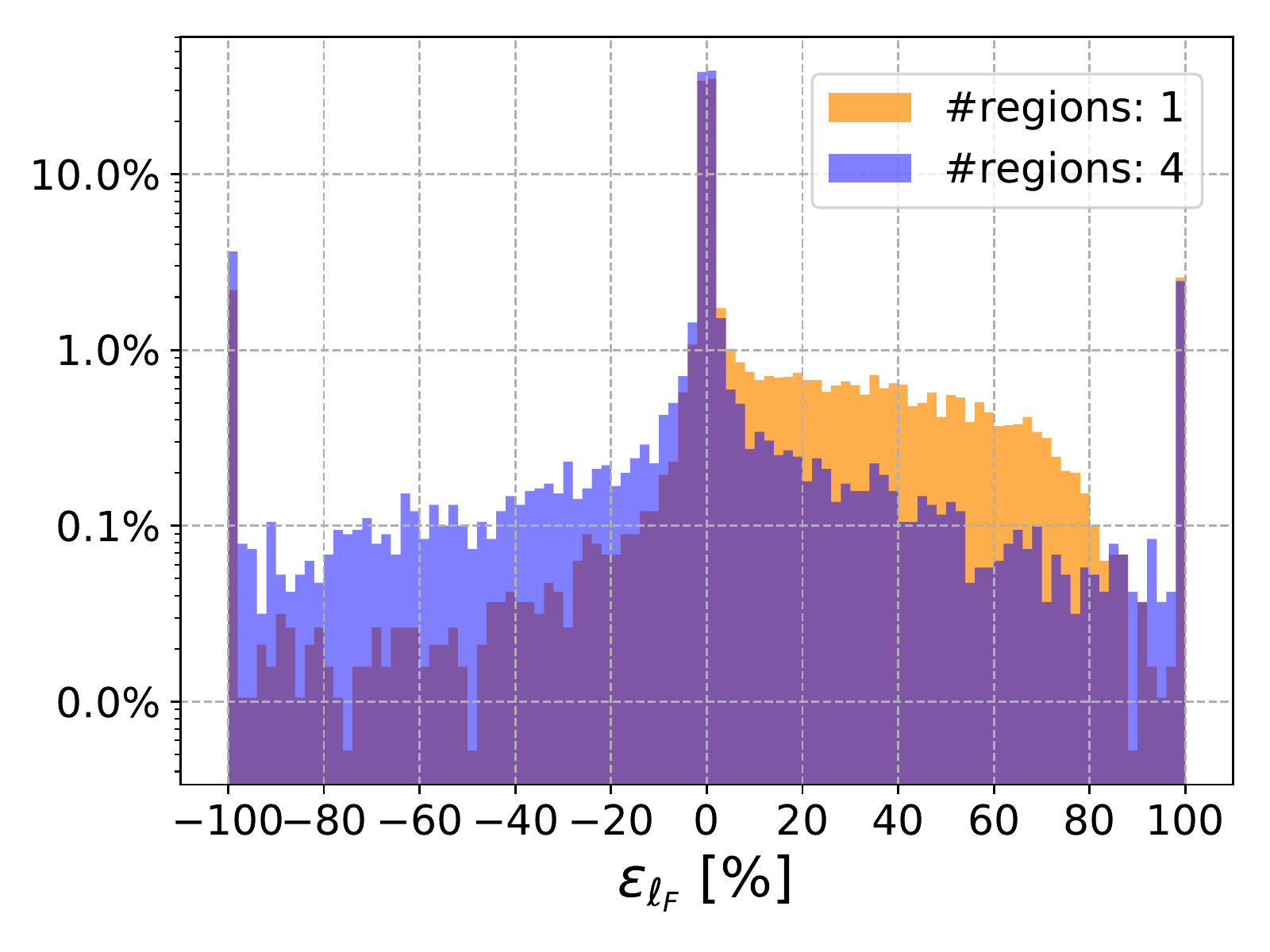}
\end{tabular}%\vspace{-0.10cm}
\caption{\label{fig:epsilonhistos} Results for the $\epsilon_{\flavB}$ accuracy parameter in one sample flavour channel
  (specifically the $\bar{u}u$ one) using either one (orange histogram)
  or four (blue histogram) precomputed grids. The same distribution is shown both in linear (left plot) and logarithmic (right plot)
  scale, %on a different range of $\epsilon_{\flavB}$ values
  where the bins at the edges account for overflow ($|\epsilon_{\flavB}|>100\%$).}
\end{center}
\end{figure}

In order to quantify the accuracy of our four-dimensional interpolation strategy we use an adimensional parameter $\epsilon$, which describes the deviation
of the interpolated result for the two-loop contribution from its exact expression on a given phase-space point and is defined as
\begin{align}\label{epsilonfact}
  \epsilon_{\flavB}[\%]=\frac{H_{(\text{int})\flavB}^{\qt(2)}-H_{(\text{ex})\flavB}^{\qt(2)}}{H_{(\text{ex})\flavB}^{\qt(2)}}\cdot 100\,,
\end{align}
where the four independent Born flavour configurations $\flavB=\{q\bar{q},\,\bar{q}q; \; \text{for}\: q\!=\!\text{u-type}\:{\rm or}\; q\!=\!\text{d-type}\}$ 
have been considered separately. In~\eqn{epsilonfact},
$H_{(\text{ex})\flavB}^{\qt(2)}$ refers to the hard function
in the $\qt$-scheme \cite{Catani:2013tia} returned by \Matrix{} using the evaluation
of the exact two-loop coefficients through {\tt VVAMP}, 
while $H_{(\text{int})\flavB}^{\qt(2)}$ stands for its value obtained by \Matrix{} 
using the interpolation of the two-loop coefficients from the precomputed grid results. 
Note that the conversion between the $\qt$-scheme
$H_{(\text{int})\flavB}^{\qt(2)}$ and the \minnlo{} $H^{(2)}_{\flavB}$
in \sct{sec:nnlo} has been given in eq.\,(3.22) of \citere{Lombardi:2020wju}.

In \fig{fig:epsilonhistos}, the distribution of the values of $\epsilon_{\flavB}$ is displayed for a selected flavour channel
(namely the $\bar{u}u$ one). All the other flavour channels have the same qualitative
behaviour.
The figure shows the impact of increasing the number of precomputed grids on the $\epsilon_{\flavB}$ parameter from one (orange histogram) to four (blue histogram).
Besides being essential for a simultaneously accurate description of physical observables over a wide phase-space region, our choice
of covering the phase space with four separate grids improves the accuracy of the predictions for the
two-loop contribution and yields a more symmetric $\epsilon_{\flavB}$ distribution.

We further notice that the bulk of the interpolator predictions (roughly $\gtrsim 80\%$) has an accuracy greater than $5\%$ (i.e
$|\epsilon_{\flavB}|\le 5\%$), while almost $95\%$ lie inside the interval $|\epsilon_{\flavB}|\le 100\%$. The remaining
fraction of $\epsilon_{\flavB}$ values
consists of phase-space points where the interpolator poorly reproduces the correct
two-loop result.
In \fig{fig:epsilonhistos}, where the $\epsilon_{\flavB}$ distribution is reported both in linear and in logarithmic scale, this fraction is clearly visible in the overflow bins at the edges of the
histograms. In most cases, these poorly predicted values are associated
to phase-space points falling outside the grid boundaries and thus requiring extrapolation of the two-loop coefficient functions $E_j^{[\C], (2)}$
outside the grid edges. However, this also means that most of these
points lie in kinematical regions where the cross section is strongly suppressed.

\begin{figure}[t]
\begin{center}\vspace{-0.2cm}
\begin{tabular}{cc}
\hspace{-0.5cm}
\includegraphics[width=.33\textheight]{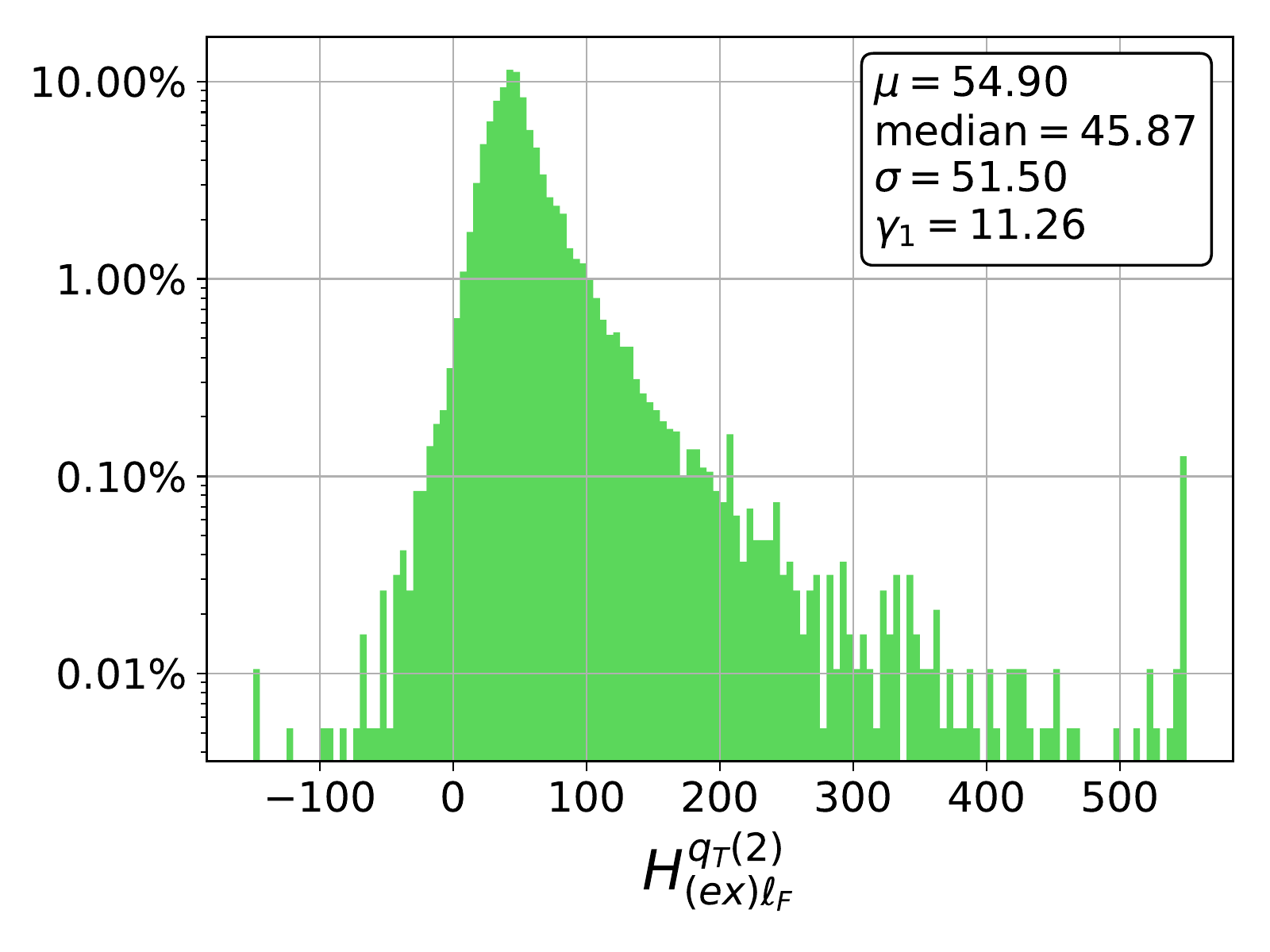}
\end{tabular}\vspace{-0.30cm}
\caption{\label{fig:H2histos} Distribution of the values of the two-loop hard function computed using the interpolator without
  rescue system for the $\bar{u}u$ channel. In the legend  the mean ($\mu$), the median,
  the standard deviation ($\sigma$) and the skewness ($\gamma_1$) are reported.
  The edge bins account for overflow.}
\end{center}
\end{figure}

\begin{figure}[p]
\begin{center}\vspace{-0.3cm}
\begin{tabular}{cc}
  \includegraphics[width=.31\textheight]{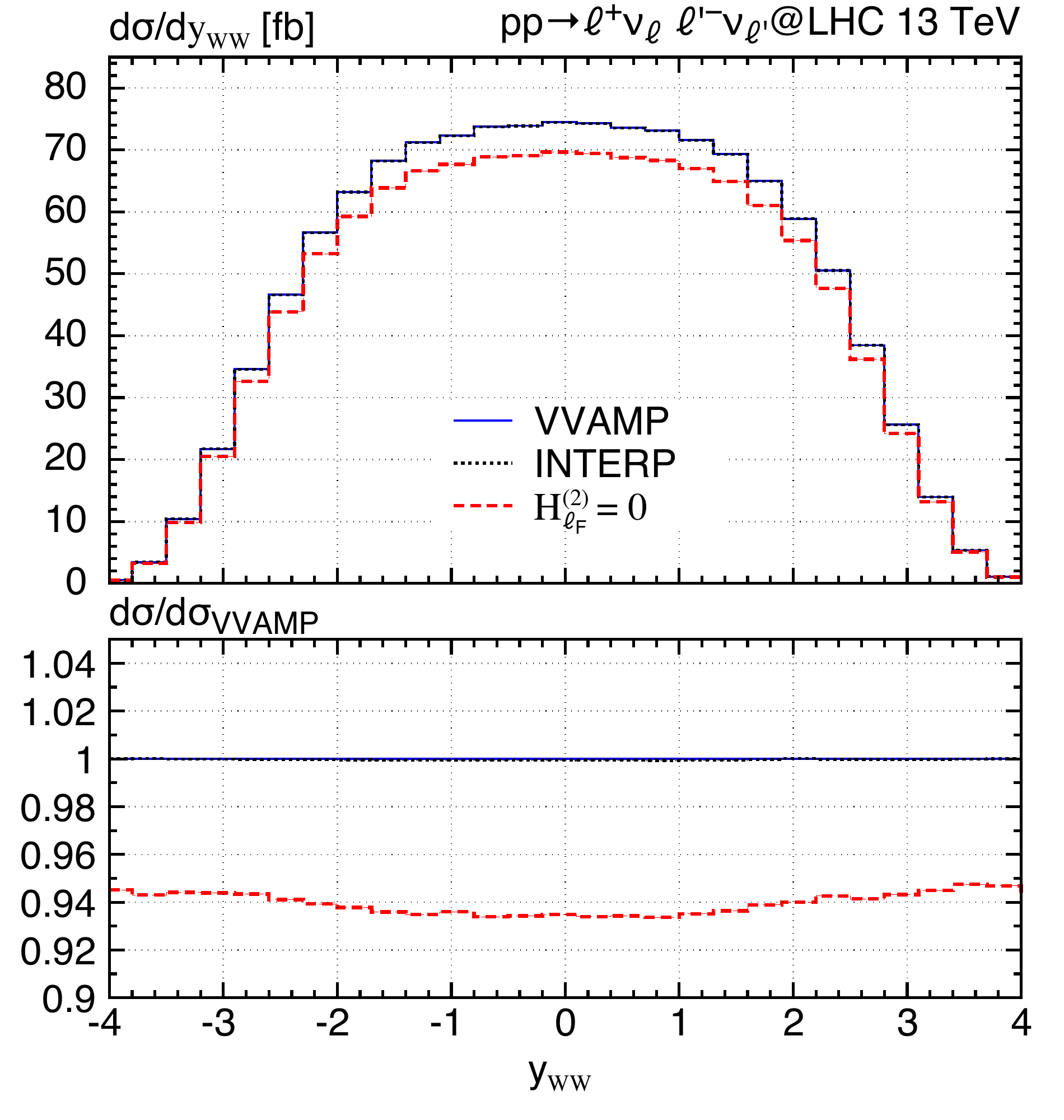}
&
  \includegraphics[width=.31\textheight]{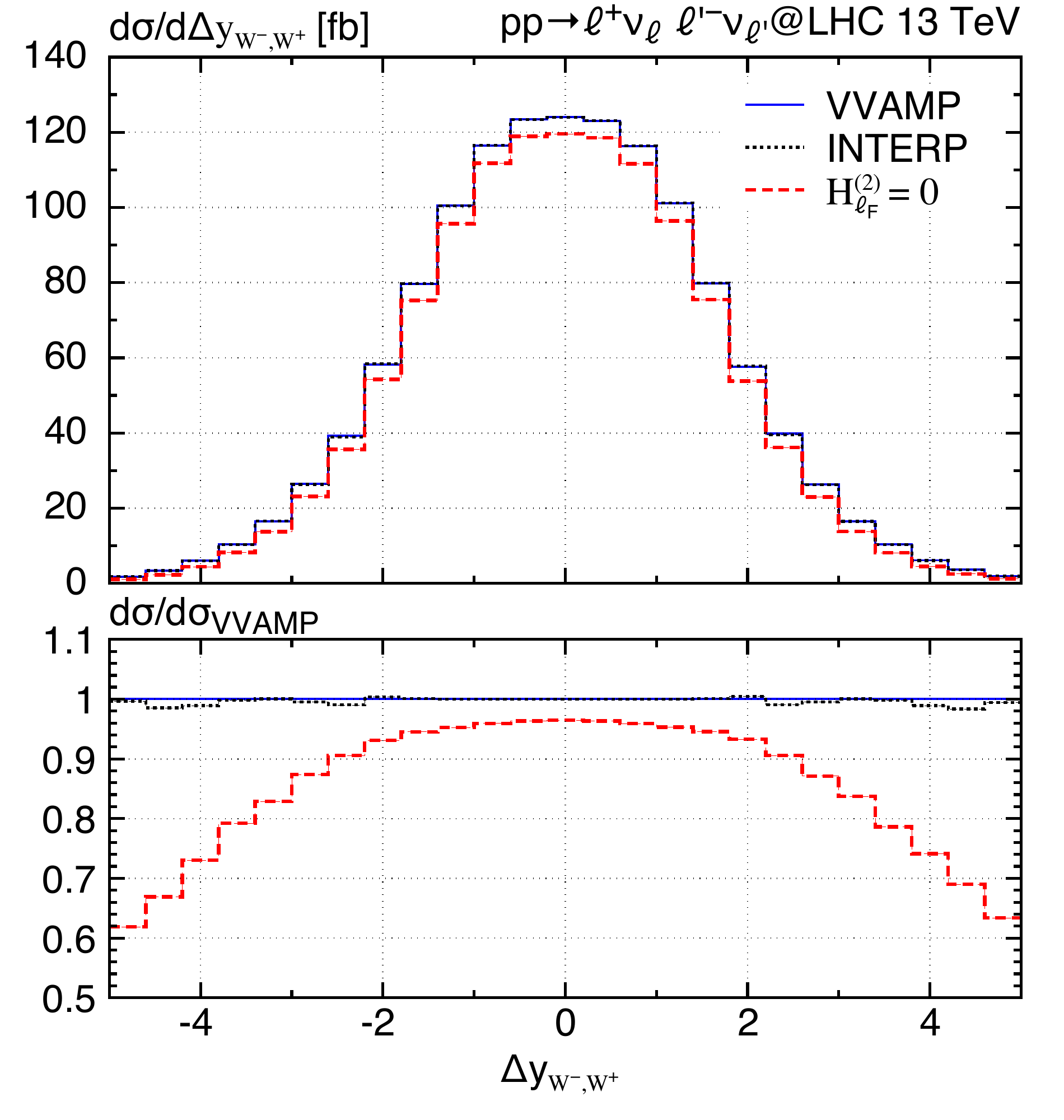}
\end{tabular}\vspace{-0.15cm}
\begin{tabular}{cc}
\includegraphics[width=.31\textheight]{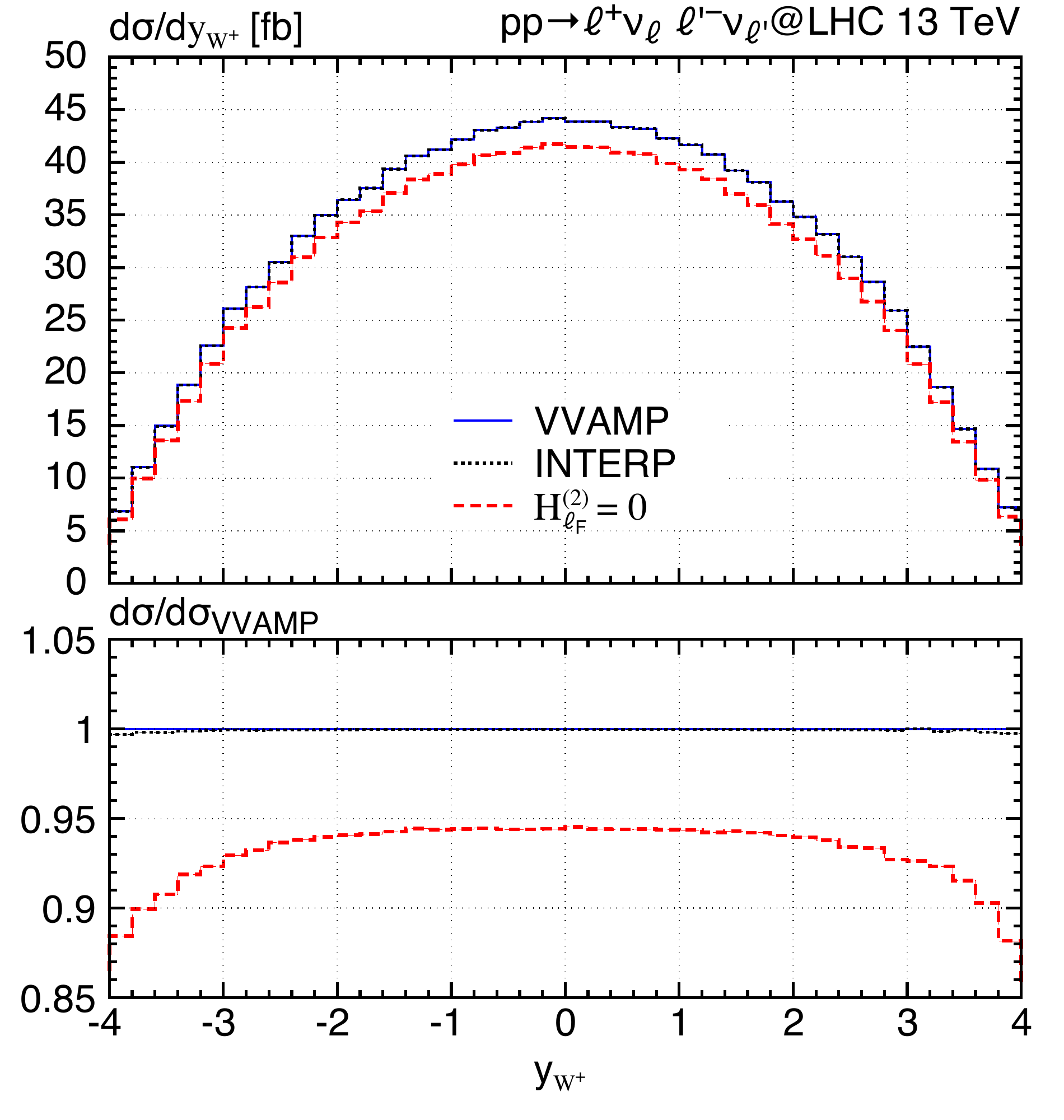}
&
\includegraphics[width=.31\textheight]{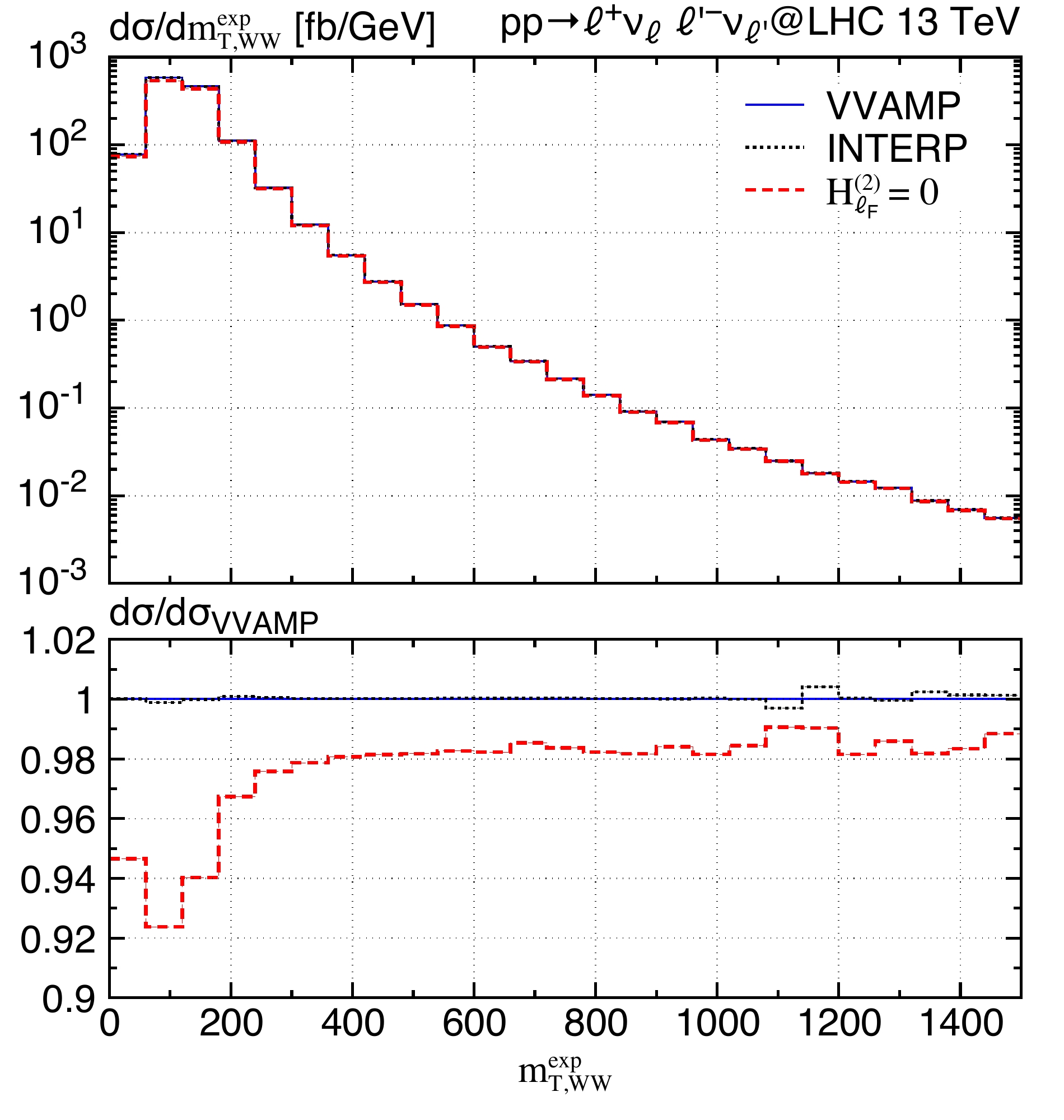}
\end{tabular}\vspace{-0.15cm}
\begin{tabular}{cc}
    \includegraphics[width=.31\textheight]{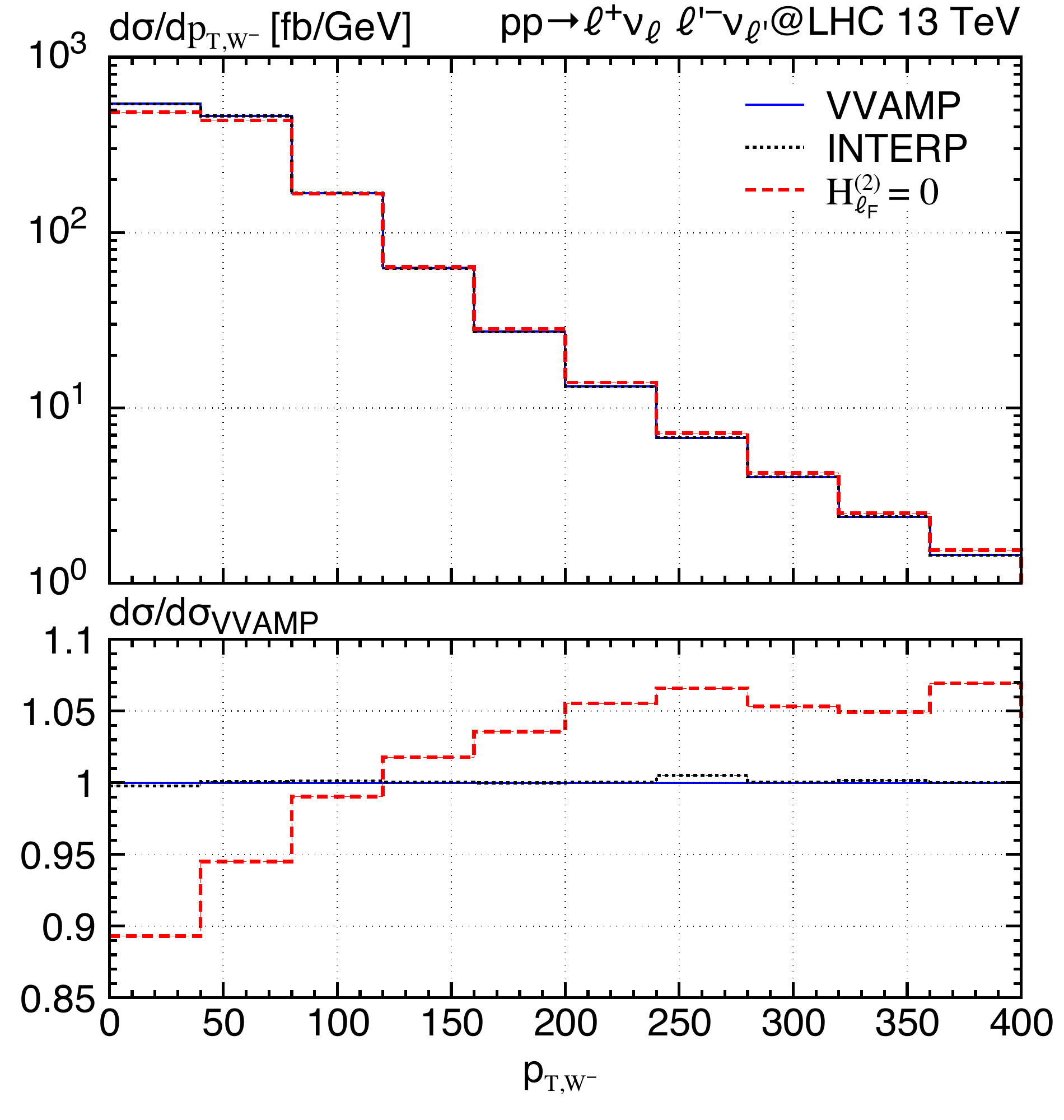}
  &
  \includegraphics[width=.31\textheight]{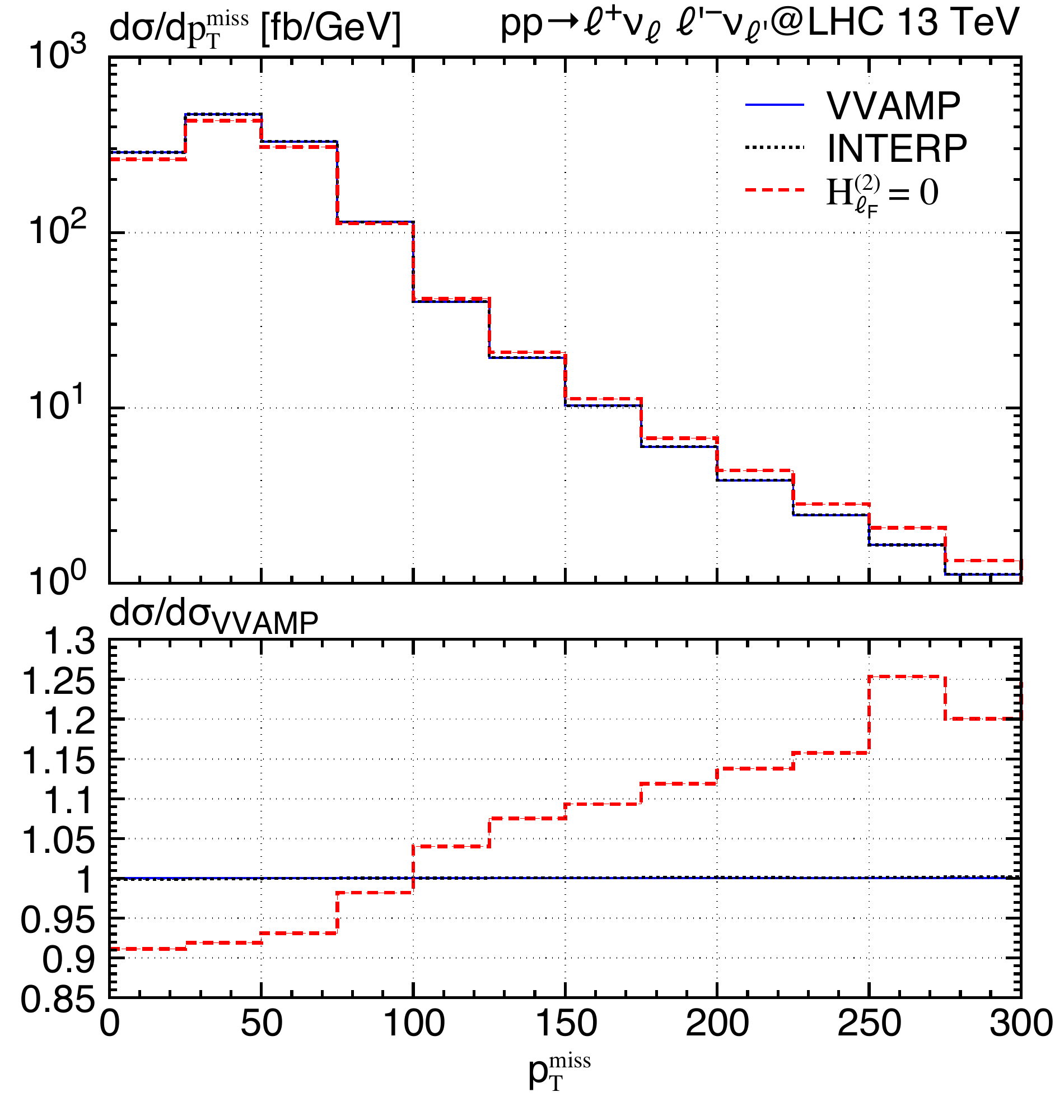}
\end{tabular}
\vspace{-0.2cm}
\caption{\label{fig:validation} \POWHEG{} stage 2 distributions (from left to right and top to bottom) in $\yww$,  $\dyww$, 
  $\ywp$,  $\mtwwexp$,  $\ptwm$ and  $\ptmiss$ for
  the {\tt VVAMP} (blue, solid line), the interpolator (black, dotted line) and the $H^{(2)}_{\flavB}=0$ (red, dotted line) results.
}
\end{center}
\end{figure}

To deal with instabilities, a basic rescue-system is introduced. This mechanism 
takes care of computing the exact $E_j^{[\C], (2)}$ coefficient functions with {\tt VVAMP} in all cases
where $H_{(\text{int})\flavB}^{\qt(2)}$ falls outside a process-specific range, where the bulk of
the $H_{(\text{ex})\flavB}^{\qt(2)}$ values lies. Specifically, we have required $-100<H_{(\text{int})\flavB}^{\qt(2)}<500$
as an acceptance interval, where roughly $99.8\%$ of the $H_{(\text{ex})\flavB}^{\qt(2)}$ distribution is concentrated.
In \fig{fig:H2histos} this distribution is shown together with
the median and the value of the first three moments of the distribution.  
As it can be
inferred from the positive skewness value (or from the fact that the mean and median do not coincide), the distribution is asymmetric, which is why
our acceptance interval for $H_{(\text{ex})\flavB}^{\qt(2)}$
is not centered around the mean, but rather it extends to higher values to partially account for the long distribution tail on the right of the peak.
This simple criterium suffices to catch the small fraction of $\epsilon_{\flavB}$ outliers that would compromise the stability of
the results. Some phase-space points remain that elude the rescue-system and where the two loop coefficients are not computed accurately, but we have checked that they have a negligible impact on the physical results, as it will be discussed below (see~\fig{fig:validation}).

\begin{figure}[t]
    \includegraphics[height=5cm]{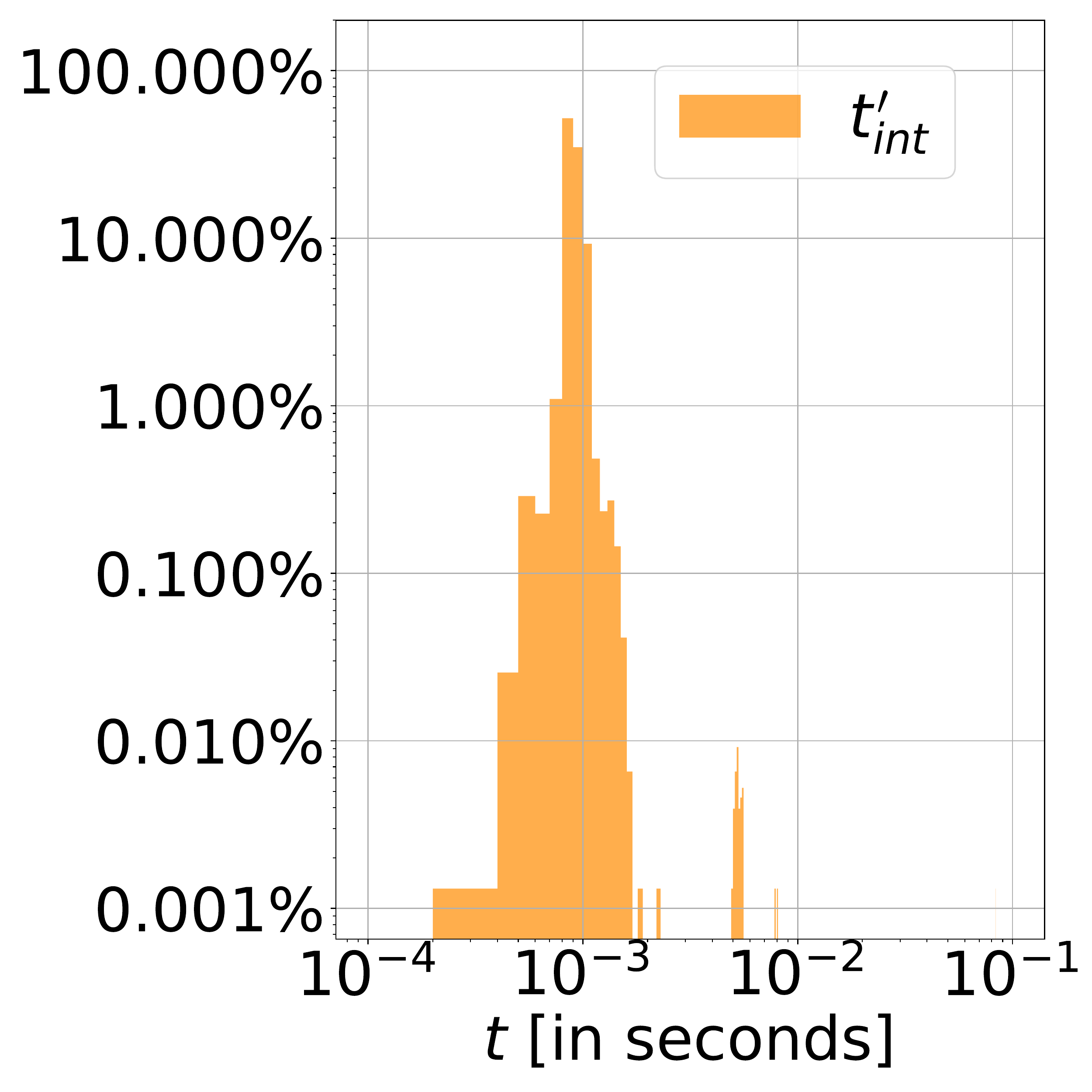}
    \includegraphics[height=5cm]{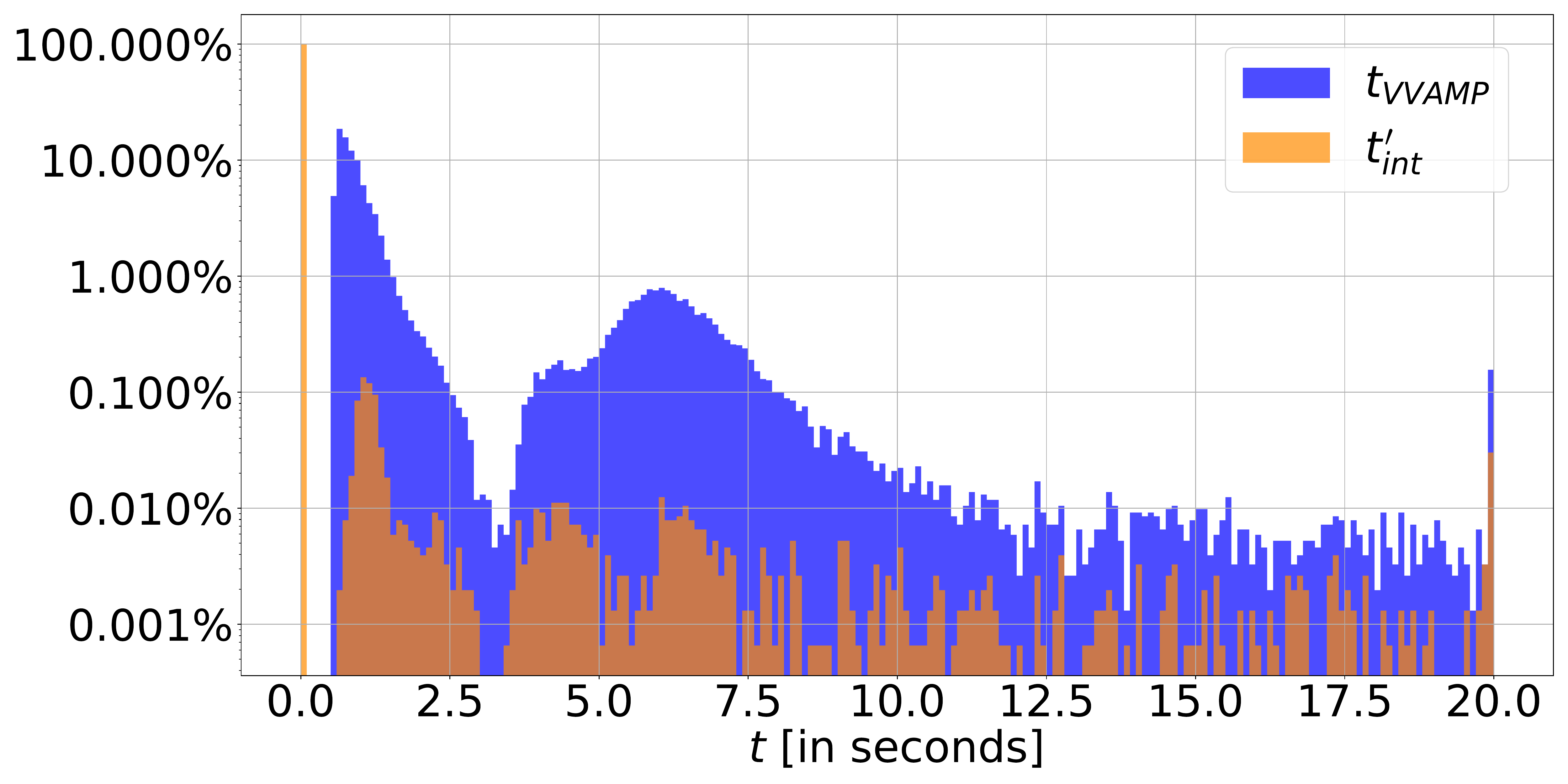}
  \caption{\label{fig:timing} Distributions in the evaluation time $t_{\text{\tt VVAMP}}$ (blue histogram) and
    $t^\prime_{\text{int}}$ (orange histogram). The last bin accounts for overflow ($t_{\text{\tt VVAMP}},t^\prime_{\text{int}}> 20$\,s).
    The left plot resolves the peak of the $t^\prime_{\text{int}}$ distribution in double logarithmic scale.
  }
\end{figure}

Clearly, the advantage of using the interpolation approach compared to
the full evaluation of the two-loop coefficients is the
time performance. Indeed, the average time
required by {\tt VVAMP} ($ \bar{t}_{\text{\tt VVAMP}}$) and the
interpolator ($ \bar{t}_{\text{int}}$) to evaluate the two-loop
contribution turns out to
differ by three orders of magnitude, while the improvement is still roughly a
factor forty if one uses the
rescue-system ($ \bar{t}^\prime_{\text{int}}$):

\begin{align}\label{meant}
  \bar{t}_{\text{\tt VVAMP}}\approx 1.9\,{\rm s}\,, \quad\quad    \bar{t}_{\text{int}}\approx 0.9\times10^{-3}\,{\rm s}\,, \quad\quad    \bar{t}^\prime_{\text{int}}\approx 4.5\times10^{-2}\,{\rm s}\,.
\end{align}
As complementary information, \fig{fig:timing} shows the time distributions of $t_{\text{\tt VVAMP}}$ (blue histogram) and
$t^\prime_{\text{int}}$ (orange histogram). The bulk of the $\tt VVAMP$ evaluation times
(roughly $80\%$) is located in the time interval $0.5s<t_{\text{\tt VVAMP}}<2.0$\,s, with a small, but not negligible fraction of
phase-space points requiring a CPU time between $5.0s<t_{\text{\tt VVAMP}}<7.5$\,s, and about $0.1\%$
exceeding $20$\,s (visible from the overflow bin). When using the interpolator, more than $99\%$ of the evaluations
just require some hundredths of a second. The small number of phase-space points with a CPU time $t^\prime_{\text{int}}>0.5$\,s
are associated to the values catched by the rescue-system.
Those timings have been obtained on machines with 
Intel Haswell Xeon E5-2698 processors with $2.3$\,GHz per core.

\afterpage{\clearpage}

Our implementation of a faster evaluation of the two-loop amplitudes through interpolation is tested by looking at
its impact on physical predictions, especially on some relevant differential distributions in
the inclusive phase space. All results have been obtained at the level of the Monte Carlo integration of the cross section (i.e
\POWHEG{} stage 2), so that no parton shower radiation or hadronization effects, which would not be relevant for the validation, have been included.

First, it is worth mentioning that the code with the interpolation of the two-loop amplitudes reproduces accurately the exact inclusive cross section,
with discrepancies of the order of about $0.4$ permille, which are well within statistical uncertainties.
Then, in \fig{fig:validation} we show representative plots that compare the exact {\tt VVAMP} predictions (blue, solid line) against the results with interpolation (black, dotted line) for 
the rapidity of the \ww{} pair ($\yww$), the rapidity difference between the two $W$-bosons ($\dyww$), the
rapidity of positively-charged $W$ boson ($\ywp$), the experimental definition of the
transverse mass of the \ww{} pair 
\begin{equation}\label{eq:mTWW}
  \mtwwexp{} = \sqrt{\left(E_{\text{\scalefont{0.77}T,$\ell_1$}}+E_{\text{\scalefont{0.77}T,$\ell_2$}}+\ptmiss\right)^2-\left({\bf p}_{\text{\scalefont{0.77}T,$\ell_1$}}+{\bf p}_{\text{\scalefont{0.77}T,$\ell_2$}}+
    {\mathbf\ptmiss}\right)^2}\,,
\end{equation}
the transverse momentum
of the negatively-charged $W$ boson ($\ptwm$) and the missing transverse momentum ($\ptmiss$).
We stress that many more distributions
than those included in this manuscript have been carefully examined and verified to show a very good
agreement between the analytic and interpolated results.
Moreover, in order to highlight the phase-space regions where the two-loop contribution gives
a large contribution to the cross section, a third curve (red, dotted line) has been included in all plots, obtained by
setting the \minnlo{} $H^{(2)}_{\flavB}=0$. The lower panel of the plots displays the bin-by-bin ratio using the {\tt VVAMP} curve as a reference.

From all plots, it is evident that the interpolator reproduces correctly the
differential distributions in all kinematical regimes, with only small fluctuations
at very high values
of $\dyww$ (at most of the order of $2\%$) or high values of $\mtww$ or
$\ptwm$ (where differences are well below one percent). These are the regions
where the two-loop contribution has the largest impact on the cross section. Indeed, from the
$\yww$ distribution it is evident that $H^{(2)}_{\flavB}$ has a $5$-$6\%$ effect on the cross section, and 
contributes uniformly to this observable, while for instance for $|\dyww|>4.5$
and $|\ywp|>4$ it induces a positive correction that reaches more than $30\%$ and $10\%$, respectively.
For transverse-momentum distributions, such as $\ptwm$ or $\ptmiss$, the two-loop contribution has a positive impact for relatively low transverse momenta (roughly for $\pt<100$\,GeV)
of at most $10\%$, while it yields an increasingly negative correction for very large transverse momenta.

\section{Phenomenological results}
\label{sec:phenomenology}

\subsection{Input parameters and settings}
\label{sec:setup}

We consider $\sqrt{s}=13$\,TeV proton--proton collisions at the LHC and present predictions 
for $p p  \to \ell^+\nu_\ell\,\ell^{\prime -}{\bar\nu}_{\ell^\prime} + X$ production
with $\ell=e$ and $\ell^\prime = \mu$. The EW parameters are determined in the 
$G_\mu$ scheme, therefore computing the EW coupling as $\alpha_{G_\mu}=\sqrt{2}\,G_\mu m_W^2(1-m_W^2/m_Z^2)/\pi$ and the mixing angle as $\cos\theta_W^2=m_W^2/m_Z^2$. We use the following PDG~\cite{Patrignani:2016xqp} values as inputs: 
$G_F = 1.16639\times 10^{-5}$\,GeV$^{-2}$, $m_W=80.385$\,GeV,
$\Gamma_W=2.0854$\,GeV, $m_Z = 91.1876$\,GeV, and 
$\Gamma_Z=2.4952$\,GeV.
We set the CKM matrix to unity, which, because of unitarity and the
fact that we consider only massless external quarks 
is a very good approximation, as explained
in~\citere{Re:2018vac}.
As described in \sct{sec:process}, the four-flavour scheme with $N_f=4$
massless quark flavours and massive bottom and top quarks is used to
define a top-free \ww{} cross section by removing all contributions
with final-state bottom quarks.  Accordingly, we use the $N_f=4$ NNLO set of
the NNPDF3.0~\cite{Ball:2014uwa} parton densities.
More precisely, in case of \minlo{} and \minnlo{} the PDF grids are
read from the \noun{lhapdf} interface~\cite{Buckley:2014ana}, copied
into \noun{hoppet} grids~\cite{Salam:2008qg} and evaluated by
\noun{hoppet} for scales below the internal PDF infrared cutoff
through DGLAP evolution with the number of active flavours kept fixed
to the one at the internal PDF infrared cutoff, as described in
\citere{Monni:2020nks}.  The central renormalization and factorization
scales are set following the usual setting for \minnlo{} and \minlo{}
discussed in \sct{sec:implementation}. 
Perturbative
uncertainties are estimated from customary 7-point variations, i.e.
by varying $\muR$ and $\muF$ around the central scale by a factor of
two while respecting the constraint $0.5\le \muR/\muF\le 2$.

We compare our \minnlo{} (and \minlo{}) predictions to the NNLOPS results presented in \citere{Re:2018vac}. Those are based on a \minlo{} calculation 
with $\muR=\muF=\ptww{}$, but use NNLO predictions for the reweighting with
\begin{align}
\label{eq:nnlopsscale}
\muR=\muF=\mu_{\text{\scalefont{0.77}0}}\equiv \frac12\,\left(m_{\text{\scalefont{0.77}T,$W^+$}}+m_{\text{\scalefont{0.77}T,$W^-$}}\right), \quad m_{\text{\scalefont{0.77}T,$W^\pm$}}=\sqrt{m_{\ell^{(\prime)}\nu_{\ell^{(\prime)}}}^2+p^2_{{\rm T},\ell^{(\prime)}\nu_{\ell^{(\prime)}}}}\,,
\end{align}
where $m_{\ell\nu_\ell}$ and $p_{{\rm T},\ell\nu_\ell}$ ($m_{\ell^\prime\nu_{\ell^\prime}}$ and $p_{{\rm T},\ell^\prime\nu_{\ell^\prime}}$) are
the invariant masses and the transverse momenta of the reconstructed $W$ bosons. The setting in \eqn{eq:nnlopsscale} is therefore 
the effective scale used in the NNLOPS calculation of \citere{Re:2018vac}, where the perturbative uncertainties are obtained from 7-point scale 
variations that are assumed correlated in the reweighting.
For the $\ptww{}$ spectrum and the jet-vetoed cross section we also compare against more accurate analytically resummed predictions obtained 
with {\sc Matrix+RadISH} \cite{Kallweit:2020gva,Wiesemann:2020gbm,MatrixRadishurl}, where we have chosen
\begin{align}
\label{eq:radishscale}
\muR=\muF=\mu_{\text{\scalefont{0.77}0}}\equiv m_{\text{\scalefont{0.77}T,$WW$}},\quad Q_{\text{\scalefont{0.77}res}}=m_{\text{\scalefont{0.77}$WW$}},\quad m_{\text{\scalefont{0.77}T,$WW$}}=\sqrt{m_{\text{\scalefont{0.77}$WW$}}^2+p_{\text{\scalefont{0.77}T,$WW$}}^2}\,,
\end{align}
with the invariant mass of the \ww{} pair $\mww$.
Here, scale uncertainties are obtained not just from customary $7$-point variations, but also by varying the resummation scale $Q_{\text{\scalefont{0.77}res}}$
by a factor of two around its central value, while keeping $\muR$ and $\muF$ fixed to $\mu_{\text{\scalefont{0.77}0}}$.
For some observables it is instructive to also compare to fixed-order NNLO predictions with both the scale settings 
in \eqn{eq:nnlopsscale} and the ones in \eqn{eq:radishscale}, which we have obtained with \Matrix{}~\cite{Grazzini:2017mhc,Matrixurl}.
In this case, perturbative uncertainties are again estimated from 7-point scale variations.

As pointed out before, we do not include the loop-induced gluon-fusion
contribution in all NNLO results throughout this paper and study the
genuine NNLO corrections to the $q\bar{q}$ initiated process. The
leading-order gluon-gluon initiated contribution enters at NNLO and
NLO QCD corrections to it are known and can be incoherently added to the
predictions presented here through a dedicated calculation, which is
beyond the scope of this paper.  Finally, for the matching to the
parton shower we employ \PYTHIA{8}~\cite{Sjostrand:2014zea} with a A14
tune~\cite{TheATLAScollaboration:2014rfk} (specifically {\tt py8tune
  21}).  Since in this study our focus is on the comparison with other
theory predictions, we do not include any effect from hadronization,
underlying event models, or a QED shower.  Such effects can, however,
be directly included and studied by any user of our program through the
\PYTHIA{8} interface of \POWHEGBOXRES{}.

\renewcommand\arraystretch{1.3}
\begin{table}[t]
\centering
\begin{tabular}{| c || c | c |}
\hline 
  & \setupone~\cite{Aaboud:2017qkn} & \setuptwo~\cite{Aaboud:2019nkz} \\
\hline
\hline 
\multirow{2}*{Lepton cuts} & $\ptl>25$\,GeV\quad$|\etal|<2.4$ & $\ptl>27$\,GeV\quad$|\etal|<2.5$\\
&$\mll>10$\,GeV & $\ptll>30$\,GeV\quad$\mll>55$\,GeV\\
\hline
Neutrino cuts & $\ptmiss>20$\,GeV\quad$\ptmissrel>15$\,GeV & $\ptmiss>20$\,GeV \\
\hline
& anti-$k_{\text{\scalefont{0.77}T}}$ algorithm with $R=0.4$ &  anti-$k_{\text{\scalefont{0.77}T}}$ algorithm with $R=0.4$\\
& $N_{\text{\scalefont{0.77}jet}} = 0$ for $\ptj>25$\,GeV & $N_{\text{\scalefont{0.77}jet}} = 0$ for $\ptj>35$\,GeV \\
Jet cuts &$|\etaj|<2.5$ \quad $\drej>0.3$ & \\
& $N_{\text{\scalefont{0.77}jet}} = 0$ for $\ptj>30$\,GeV &  \\
&$|\etaj|<4.5$ \quad $\drej>0.3$ & \\
\hline 
\end{tabular}
\caption{\label{tab:cuts} Fiducial cuts used in two different setups, see text for details.}
\end{table}
\renewcommand\arraystretch{1}

Since the \ww{} cross section is finite at the LO without any cuts, we present results 
in the fully inclusive \ww{} phase space, referred to as \setupinclusive{}. 
Additionally, we consider two sets of fiducial cuts, 
which are summarized in \tab{tab:cuts}.
The first one corresponds to an earlier ATLAS 13\,TeV analysis~\cite{Aaboud:2017qkn} and it is identical 
to that used in the NNLOPS calculation of \citere{Re:2018vac}, which allows us to compare directly
our \minnlo{} predictions with the fiducial NNLOPS results of \citere{Re:2018vac}. 
We refer to it as \setupone{} in the following.
We note that \setupone{} involves a two-fold jet-veto requirement,
vetoing all jets in the rapidity region
$|\etaj|<2.5$ and separated from the leptons by $\drej>0.3$  with $\ptj > 25$\,GeV
and all jets in the rapidity region
$|\etaj|<4.5$ and separated from the leptons by $\drej>0.3$  with $\ptj > 30$\,GeV.
The second setup instead corresponds to the most recent ATLAS 13\,TeV measurement of \citere{Aaboud:2019nkz}, and it was used to study high-accuracy resummed predictions for \ww{} 
production in \citere{Kallweit:2020gva}. This setup, referred to as \setuptwo{} in the following, is useful for two reasons.
First, at variance with \setupone{}, it includes a single jet-veto cut
for jets with $\ptj > 35$\,GeV. This allows us to 
directly compare against the NNLO+NNLL resummed predictions for the $\ptww{}$ spectrum
with a jet veto \cite{Kallweit:2020gva}. Note that to facilitate this comparison, we have removed 
the jet rapidity ($\etaj$) requirement from \setuptwo{} \cite{Aaboud:2019nkz}, which has a numerically tiny effect.
Second, \setuptwo{} is used to compare against data, since \citere{Aaboud:2019nkz} measured the fiducial cross section as a 
function of the jet-veto cut to validate theory predictions for the jet-vetoed \ww{} cross section.
Let us recall that jet-veto requirements are crucial for experimental \ww{} analyses in order 
to suppress the large top-quark backgrounds.
In addition, we introduce \setuponenoJV{} and \setuptwonoJV{} for the same fiducial
setups as given in \tab{tab:cuts}, but without any restriction on the jet activity. Those are 
relevant to study the $\ptww{}$ distribution inclusive over jet radiation as well as the cross 
section as a function of the jet-veto cut.
Besides jet-veto requirements, the two setups in \tab{tab:cuts} involve 
standard cuts on the transverse momentum ($\ptl$) and pseudorapidity ($\etal$) 
of the charged leptons as well as a lower cut on the invariant-mass of
the dilepton pair ($\mll$) and on the missing transverse momentum ($\ptmiss$). 
Setup \setuptwo{} includes also a lower cut on the transverse momentum of the dilepton pair ($\ptll$),
while setup \setupone{} cuts on the so-called relative missing transverse ($\ptmissrel$).
The latter denotes the component of the missing transverse momentum vector perpendicular 
to the direction of the closest lepton in the azimuthal plane, and is defined as
\begin{align}
\ptmissrel = \left\{\begin{array}{ll}\ptmiss\cdot \sin|\Delta\phi| & \; \;{\rm for}\; \Delta\phi<\pi/2\,, \\ \ptmiss & \; \;{\rm for}\; \Delta\phi>\pi/2\,,\end{array}\right.
\end{align}
where $\Delta\phi$ denotes the azimuthal angle between the missing transverse momentum vector
vector and the nearest lepton.

\subsection{Integrated cross sections}
\label{sec:crosssection}

\renewcommand\arraystretch{1.3}
\begin{table}[t]
\begin{center}
\resizebox{\columnwidth}{!}{%
\begin{tabular}{| l  ||   c | c | c |}
\hline 
 $\sigma(pp\to \ell^+ \nu_\ell\, \ell'^- \nu_{\ell'})$ [fb] & \setupinclusive{} & \setupone{} & \setuptwo{} \\
\hline
\hline
\minlo{}  & $1156.6(4)_{-5.7\%\phantom{0}}^{+5.4\%\phantom{0}}$ & $185.0(2)_{-6.5\%}^{+8.8\%}$ & $143.2(2)_{-8.1\%}^{+4.9\%}$  \\
\minnlo{}  & $1292.2(7)_{-0.7\%\phantom{0}}^{+0.6\%\phantom{0}}$ & $207.7(2)_{-1.7\%}^{+1.6\%}$ & $159.2(4)_{-1.4\%}^{+1.0\%}$  \\
NNLOPS \cite{Re:2018vac}  & $  1308.9(3)_{-1.6\%\phantom{0}}^{+1.7\%\phantom{0}}$ & $206.4(1)_{-2.3\%}^{+2.2\%}$ & $159.0(1)_{-1.8\%}^{+1.7\%}$  \\
%NLO  {{\scalefont{0.77}$\mu_{\text{\scalefont{0.77}0}}=(m_{\text{\scalefont{0.77}T,$W^+$}}^2+m_{\text{\scalefont{0.77}T,$W^-$}}^2)^{0.5}/2$}} & $1217.9(1)_{-2.8\%\phantom{0}}^{+3.5\%\phantom{0}}$ & $212.3(3)_{-2.1\%}^{+2.5\%}$ & ---  \\
NNLO  {{\scalefont{0.77}$\mu_{\text{\scalefont{0.77}0}}=(m_{\text{\scalefont{0.77}T,$W^+$}}^2+m_{\text{\scalefont{0.77}T,$W^-$}}^2)^{0.5}/2$}} & $1306.5(5)_{-1.6\%\phantom{0}}^{+1.6\%\phantom{0}}$ & $206.5(1)_{-0.7\%}^{+1.0\%}$ & $158.9(5)_{-0.6\%}^{+0.8\%}$  \\
%NNLO  {{\scalefont{0.77}$\mu_{\text{\scalefont{0.77}0}}=\mww{}$}}  & $1286.7(10)_{-1.3\%}^{+1.4\%}$ & --- & ---  \\
NNLO  {{\scalefont{0.77}$\mu_{\text{\scalefont{0.77}0}}=m_{\text{\scalefont{0.77}T,$WW$}}$}}& $1284.9(10)_{-1.3\%}^{+1.4\%}$ & --- & $160.8(3)_{-0.8\%}^{+1.0\%}$  \\
\hline
\multirow{2}*{ATLAS$-gg$ \cite{Aaboud:2017qkn}} & $1481 \pm 59 \text{\scalefont{0.55}(stat)}$ & $236.5 \pm 10 \text{\scalefont{0.55}(stat)}$ & \multirow{2}*{---} \\
&  $\pm 154 \text{\scalefont{0.55}(syst)} \pm 108\text{\scalefont{0.55}(lumi)}$ & $\pm 25 \text{\scalefont{0.55}(syst)} \pm 5.5\text{\scalefont{0.55}(lumi)}$ & \\
\multirow{2}*{ATLAS$-gg$ \cite{Aaboud:2019nkz}} &  \multirow{2}*{---} & \multirow{2}*{---} & $178.5 \pm 2.5 \text{\scalefont{0.55}(stat)}$ \\
&   & & $\pm 12.7 \text{\scalefont{0.55}(syst)} \pm 4\text{\scalefont{0.55}(lumi)}$ \\
\multirow{2}*{CMS$-gg$ \cite{CMS:2016vww}} & $1289 \pm 68 \text{\scalefont{0.55}(stat)}$ & \multirow{2}*{---} &  \multirow{2}*{---}\\
& $^{\pm 67 \text{\scalefont{0.55}(exp.\,syst)}}_{\pm 76 \text{\scalefont{0.55}(th.\,syst)}} \pm 42\text{\scalefont{0.55}(lumi)}$ & & \\
\multirow{2}*{CMS$-gg$ \cite{Sirunyan:2020jtq}} & $1316 \pm 65 \text{\scalefont{0.55}(stat)}$ & \multirow{2}*{---} &  \multirow{2}*{---}\\
& $\pm 23 \text{\scalefont{0.55}(syst)} \pm 38\text{\scalefont{0.55}(lumi)}$ & & \\
\hline
\end{tabular}}
\end{center}
\renewcommand{\baselinestretch}{1.0}
\caption{ \label{tab:crosssection}  \ww{} cross sections in the fully inclusive phase space and
in the fiducial phase spaces defined in \tab{tab:cuts}. We compare our \minlo{} and \minnlo{}
predictions to the NNLOPS results of \citere{Re:2018vac} and to the NNLO cross section obtained 
with different settings of $\muR{}$ and $\muF{}$. All NNLO corrections to $q\bar{q}$-induced \ww{} production are taken
  into account, while the loop-induced $gg$ contribution is excluded.
In the last rows, the comparison to CMS and ATLAS data is shown.
For the measured inclusive cross sections we have assumed a branching 
fraction of $\textrm{BR}(W^\pm \rightarrow \ell^\pm\nu_\ell) = 0.108987$, consistently
evaluated with our inputs, and applied one factor for each of the two $W$ bosons.
The measured fiducial cross sections have been divided by a factor two so that 
they correspond to $pp\to \ell^+ \nu_\ell\, \ell'^- \nu_{\ell'}$ production with $\ell=e$ and $\ell^\prime = \mu$.
In addition, we have subtracted the loop-induced gluon-fusion contribution from the central value of the data.
For the inclusive results and the \setupone{} result we used the prediction for $gg$ (non-resonant) cross section 
quoted in table 5 of the ATLAS analysis in \citere{Aaboud:2017qkn}.
For the \setuptwo{} result we used the $gg$LO result in table 2 of \citere{Grazzini:2020stb}.
The ATLAS measurement of \citere{Aaboud:2017qkn} includes resonant Higgs contributions,
which have been subtracted from that data as well, using the corresponding prediction quoted in table 5 of that paper.}
\end{table}
\renewcommand\arraystretch{1}

We start the presentation of phenomenological results by discussing 
integrated cross sections in \tab{tab:crosssection}. In particular, we report predictions in the fully inclusive and the two 
fiducial phase spaces introduced in \sct{sec:setup} for \minlo{}, \minnlo{}, 
NNLOPS \cite{Re:2018vac} as well as two
fixed-order NNLO predictions obtained with \Matrix{}~\cite{Grazzini:2017mhc,Matrixurl} 
using the scale settings of \eqn{eq:nnlopsscale} and \eqn{eq:radishscale}.
We summarize our main observations in the following:
\begin{itemize}
\item It is clear that NNLO accuracy is crucial for an accurate
  prediction of the \ww{} cross section, since the \minlo{} result is
  about 12\% lower than the \minnlo{} one not only for
  \setupinclusive{}, but also after including the \setupone{} and
  \setuptwo{} cuts. In fact, in all cases the \minnlo{} prediction is
  outside the upper uncertainty boundary of the \minlo{} one. This
  is not surprising since for \ww{} production also at fixed order
  the NLO uncertainty band does typically not include the central value
  of the NNLO
  prediction~\cite{Grazzini:2016ctr,Re:2018vac}. Additionally, the
  precision at NNLO is substantially improved, with scale uncertainties
  reduced by almost an order of magnitude.
\item The NNLO-accurate predictions compare well against one
  another. They are compatible within their respective scale
  uncertainties and the central predictions are all within less than
  2\%.  Indeed, small differences are expected from the fact that
  those predictions differ by terms beyond NNLO accuracy. Note that
  the NNLOPS and the NNLO calculations with
  $\mu_{\text{\scalefont{0.77}0}}=
  \frac12\,\left(m_{\text{\scalefont{0.77}T,$W^+$}}+m_{\text{\scalefont{0.77}T,$W^-$}}\right)$
  are very close, both in terms of central values and uncertainties,
  since the former is actually reweighted to the latter prediction in
  the inclusive phase space.  The fact that the inclusive \minnlo{}
  result is about $1.2\%$ below the NNLOPS one is due to the
  different scale choice and treatment of terms beyond
  accuracy. Indeed, the second NNLO prediction with a scale choice of
  $\mu_{\text{\scalefont{0.77}0}}=\mtww$ is even slightly lower than
  the \minnlo{} one. 
\item The agreement among predictions with NNLO accuracy gets even better
as far as fiducial cross sections are concerned. This could be an indication that the jet vetos that are
applied in the fiducial phase spaces reduce the impact of terms beyond NNLO accuracy.
\item Some caution is advised regarding the quoted scale uncertainties. First of all,  
the quoted uncertainties generally appear to be quite small, and potentially at the edge 
of providing a realistic estimate of the true uncertainty. Clearly, the \minlo{} uncertainty 
does not cover the inclusion of NNLO corrections through \minnlo{}.
As far as the NNLO-accurate results are concerned,
the \minnlo{} uncertainties in the inclusive phase space are even a factor of two 
smaller than the ones of the other predictions. Moreover, the fixed-order NNLO 
uncertainties further decrease when the jet-veto requirements are imposed.
Such behaviour is not new \cite{Stewart:2011cf}, but at least the showered results show an increased 
uncertainty when imposing fiducial cuts. Still, especially for the jet-vetoed predictions
one may consider more conservative approaches to estimate the perturbative uncertaintes,
see for instance \citeres{Stewart:2011cf,Banfi:2012jm}.
\item Finally, there is a good agreement of \minnlo{} results 
with data from ATLAS and CMS in both inclusive and fiducial 
phase-space regions. The measured cross sections agree mostly within 
one and at most within two standard deviations.
\end{itemize}

\subsection{Differential distributions}
\label{sec:distributions}
We now turn to discussing differential distributions. We start in \sct{sec:incl} and \sct{sec:fid}
with comparing our \minnlo{} to \minlo{} and NNLOPS predictions in the inclusive and the fiducial 
phase space, respectively. This allows us, on the one hand, to study the effect of NNLO corrections through \minnlo{}
with respect to \minlo{} and, on the other hand, to assess the compatibility of the
\minnlo{} predictions with the known NNLOPS results. Then in \sct{sec:IRsensitive} we move 
to distributions sensitive to soft-gluon radiation that require the inclusion of large logarithmic 
corrections to all orders in QCD perturbation theory either through a parton shower or 
through analytic resummation.

Unless indicated otherwise, the plots are organized as follows: 
there is a main frame, which shows differential distributions for the
\minnlo{} (blue, solid), \minlo{} (black, dotted), and NNLOPS (magenta, dash-dotted) predictions.
In a lower frame we show bin-by-bin ratios of all curves to the central \minnlo{} result.
In some cases, where it is instructive to compare to the fixed-order results, we show
fixed-order NNLO distributions for $\mu_{\text{\scalefont{0.77}0}}= \frac12\,\left(m_{\text{\scalefont{0.77}T,$W^+$}}+m_{\text{\scalefont{0.77}T,$W^-$}}\right)$ (green, long-dashed) and/or for $\mu_{\text{\scalefont{0.77}0}}=\mtww$ (red, dashed).
We note that we refrain from showing fixed-order NNLO predictions for 
most observables as the NNLOPS results correspond to a scale setting of
$\mu_{\text{\scalefont{0.77}0}}= \frac12\,\left(m_{\text{\scalefont{0.77}T,$W^+$}}+m_{\text{\scalefont{0.77}T,$W^-$}}\right)$ and are, in general, numerically very close to the respective 
fixed-order NNLO cross section.
Additionally, for the $\ptww{}$ distribution 
we show NNLO+N$^3$LL (green, double-dash dotted) and NNLO+NNLL
(brown, dash-double dotted) predictions, and for the $\ptww{}$ spectrum with a jet veto as well as  the jet-vetoed cross 
section we show NNLO+NNLL (green, double-dash dotted) and NLO+NLL (brown, dash-double dotted) results.

\subsubsection{Inclusive phase space}
\label{sec:incl}

\begin{figure}[t]
\begin{center}\vspace{-0.2cm}
\begin{tabular}{cc}
\includegraphics[width=.31\textheight]{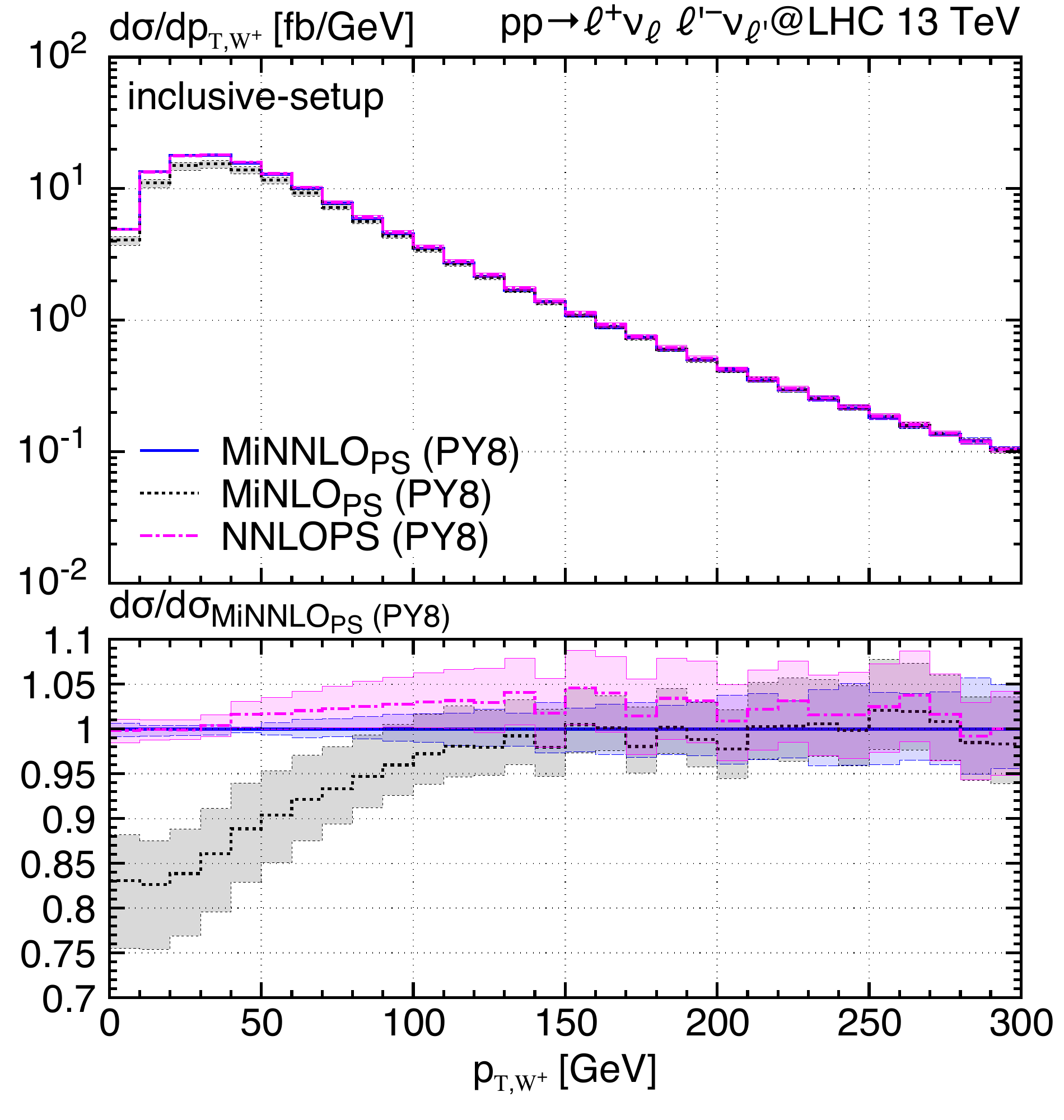} 
&
\includegraphics[width=.31\textheight]{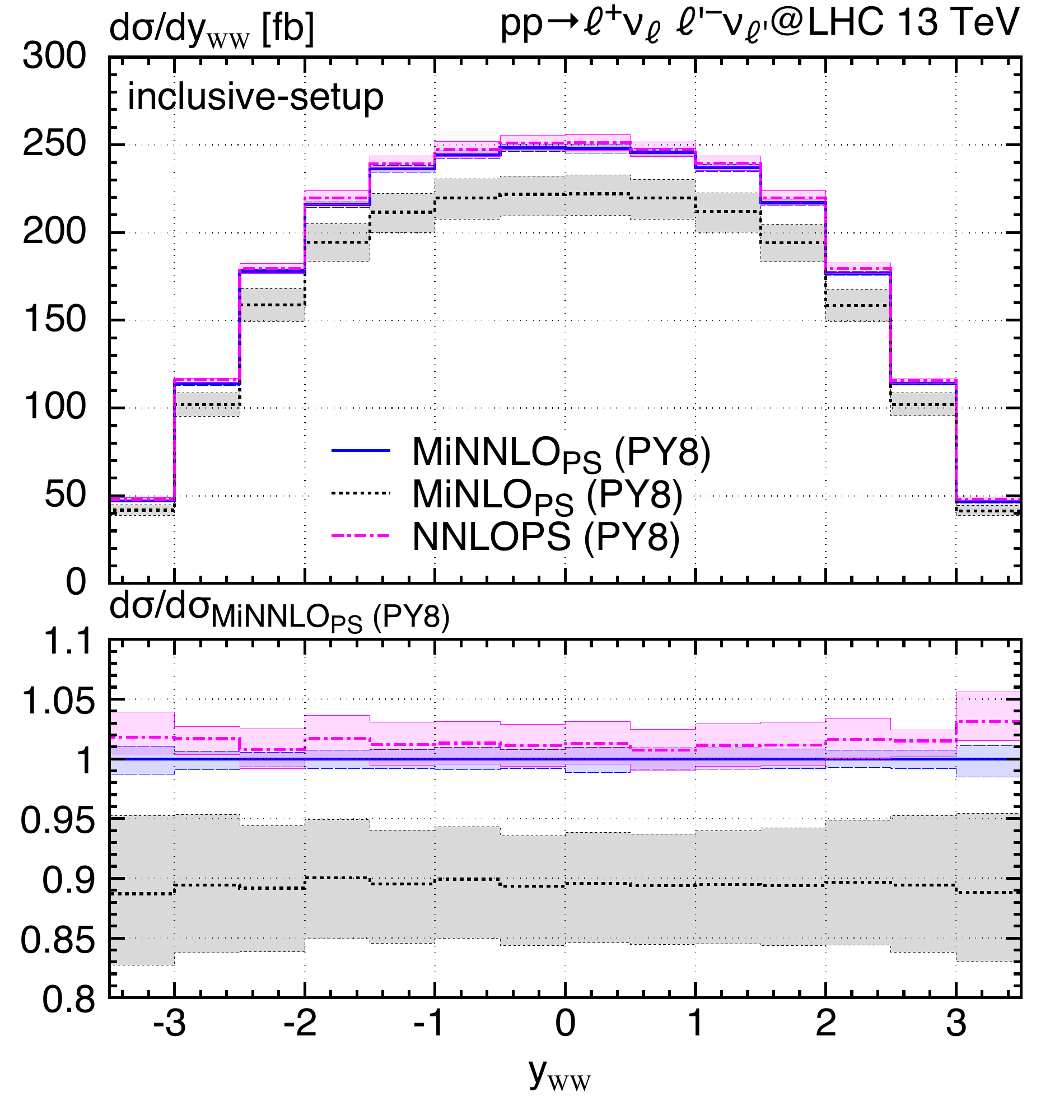}
\end{tabular}\vspace{-0.15cm}
\begin{tabular}{cc}
\includegraphics[width=.31\textheight]{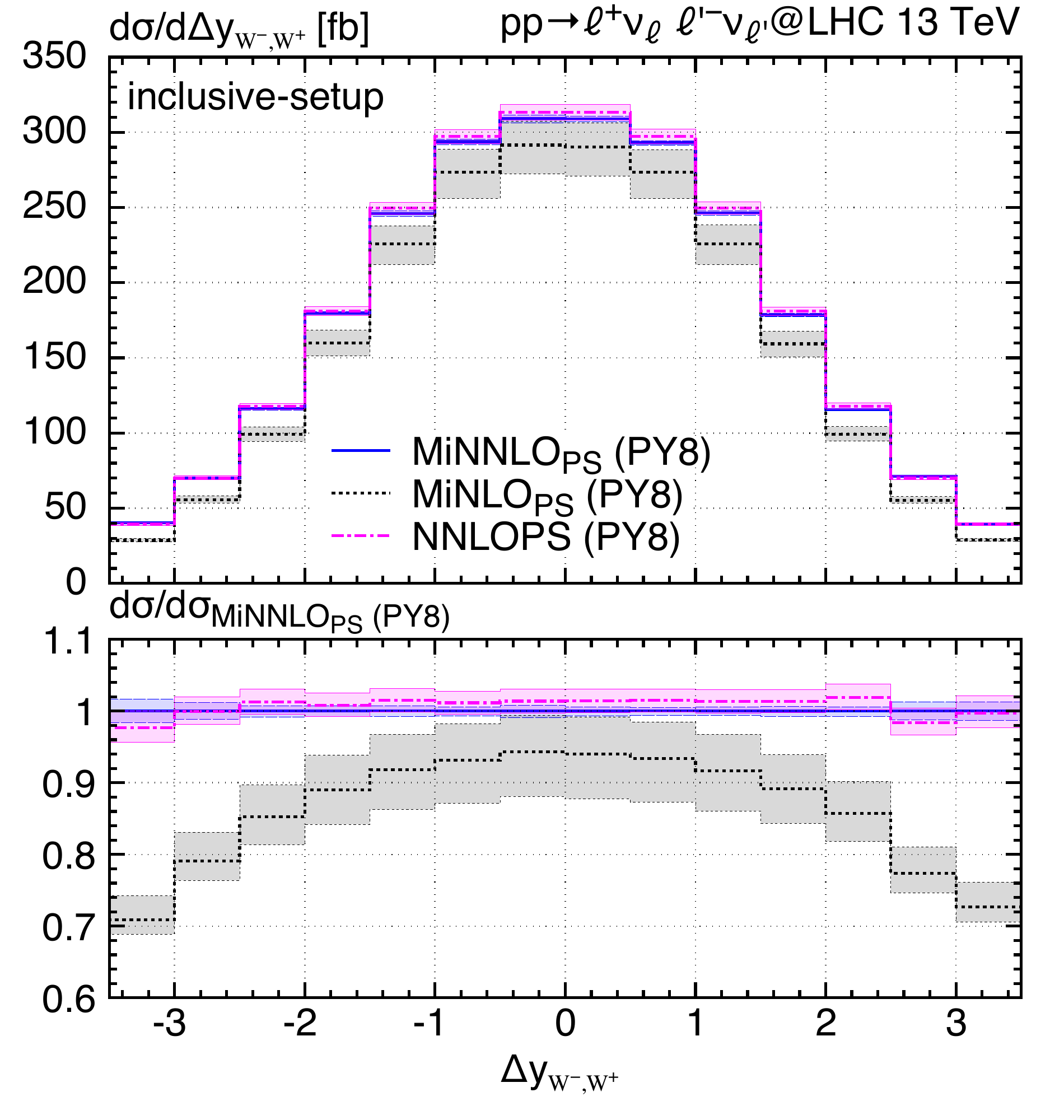}
&
\includegraphics[width=.31\textheight]{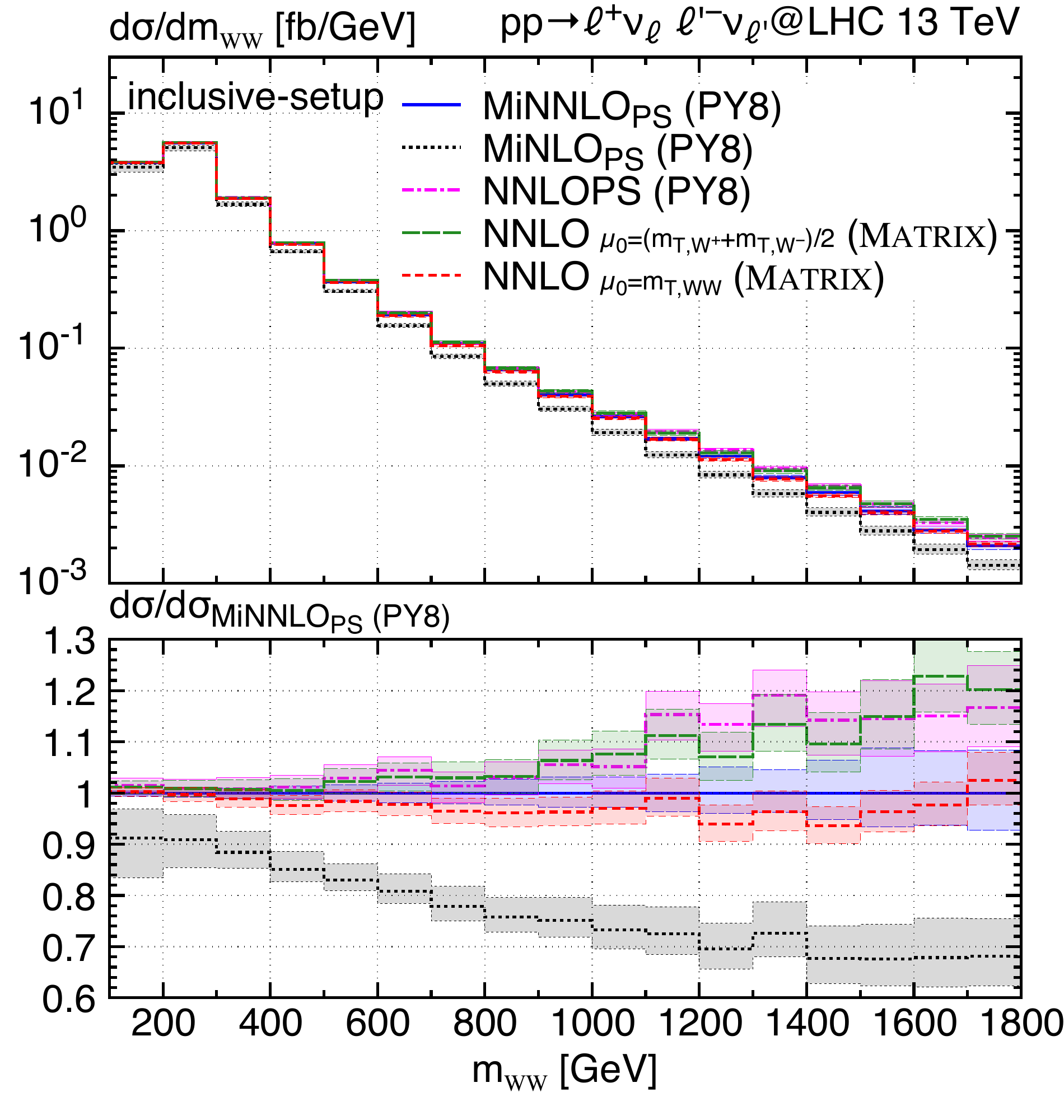}
\end{tabular}\vspace{-0.15cm}
\caption{\label{fig:inclusive} Differential distributions in the \setupinclusive{}.
}
\end{center}
\end{figure}

We start by discussing distributions in the inclusive phase space. We have considered a 
large number of relevant distributions of both the leptonic final states and of the reconstructed 
$W$-bosons. A selection of those, which reflect some general 
features, is presented in \fig{fig:inclusive}. Since experimentally the $W$ bosons can not be directly reconstructed and the 
fully inclusive phase space can not be covered by the detectors in any case, we follow here
a more theoretical motivation and study observables related to the reconstructed 
$W$ bosons rather than their decay products. In particular,
\fig{fig:inclusive} shows the transverse-momentum spectrum of the
$W^+$ boson ($\ptwp$), the rapidity distribution of the $W$-boson pair ($\yww$),
the rapidity difference between the two $W$ bosons ($\dyww$),
and the invariant-mass distribution of the $W$-boson 
pair ($\mww$).

For the $\ptwp$ spectrum, the \minnlo{} prediction is in full agreement with 
the NNLOPS result, which is particularly striking in the low-$\ptwp$ region since 
scale uncertainties are only at the level of $\pm 1\%$. At larger values of $\ptwp$,
the uncertainty bands of the NNLO+PS accurate predictions widen
and reach about $\pm 5\%$. This indicates that this region is predominantly 
filled by higher-order (real) radiative corrections with at least one jet, and that 
the formal accuracy is somewhat decreased by one order.
Indeed, in the region $\ptwp\gtrsim 100$\,GeV the NNLO+PS predictions become fully compatible 
with the \minlo{} result, also in terms of the size of the uncertainty bands. By contrast, for 
smaller $\ptwp$ we observe large 
NNLO corrections with respect to \minlo{} that reach almost $20\%$ and 
substantially reduced scale uncertainties.

Also for the $\yww{}$ and $\dyww{}$ distributions we find fully compatible 
results with overlapping uncertainty bands when comparing
\minnlo{} and NNLOPS predictions. While the NNLO corrections compared 
to \minlo{} are relatively flat for $\yww{}$, we find that the corrections
increase substantially at larger values of $\dyww{}$, reaching $\sim +30\%$ 
for $\yww{}\gtrsim 3$. This behaviour was observed already in \citere{Re:2018vac} 
and it is reassuring to see that this large effect is not a remnant of the scale setting 
in the NNLOPS calculation, but a genuine NNLO correction.

Similarly sizeable NNLO corrections are observed also at large values of $\mww$.
This is also one of the few regions of phase space that we found 
where \minnlo{} and NNLOPS predictions do not agree at the level of a few percent.
While up to $\mww{}\lesssim 500$\,GeV the \minnlo{} and NNLOPS results are fully compatible,
they start deviating at larger invariant masses, reaching differences of about $20\%$ at
$\mww{}=1.8$\,TeV. Those differences can be traced back to the different scale settings
in the \minnlo{} and NNLOPS calculations. Indeed, comparing the additional NNLO 
results shown for the $\mww{}$ distribution, we notice a relatively large spread between 
the green long-dashed curve with scale setting $\mu_{\text{\scalefont{0.77}0}}= \frac12\,\left(m_{\text{\scalefont{0.77}T,$W^+$}}+m_{\text{\scalefont{0.77}T,$W^-$}}\right)$  and 
the red dashed curve with $\mu_{\text{\scalefont{0.77}0}}=\mtww$, which is of the same 
size as (or even slightly larger than) 
the observed differences between \minnlo{} and NNLOPS. As expected, the NNLOPS
result is close to the NNLO one with $\mu_{\text{\scalefont{0.77}0}}= \frac12\,\left(m_{\text{\scalefont{0.77}T,$W^+$}}+m_{\text{\scalefont{0.77}T,$W^-$}}\right)$, while
the \minnlo{} prediction is somewhat in-between the two NNLO results, but slightly closer
to the one with $\mu_{\text{\scalefont{0.77}0}}=\mtww$. Thus, the origin of the observed
differences are terms beyond NNLO accuracy. Although the uncertainty bands increase to
about $10\%$ towards large $\mww$, the two NNLO+PS accurate predictions do not (or 
only barely) overlap for $\mww\gtrsim 1$\,TeV, indicating that plain $7$-point scale variations
do not represent a realistic estimate of the actual size of uncertainties in that region of phase space.
One may ask the question whether one of the two scale choices can be preferred. 
Although one might assume that $\mww$ would be the natural scale of the 
$\mww$ distribution, the situation is actually not that clear. This was discussed 
in some detail in \citere{Re:2018vac}, and it boils down to the fact that for $s$-channel 
topologies $\mww$ would be the more natural scale, while for $t$-channel topologies
the transverse masses of the W bosons reflect better the natural scale of the process.
Since both topologies appear in \ww{} production already at the LO and they interfere,
it is hard to argue in favour of any of the two scale choices. As a result, and since now 
there are two NNLO+PS accurate predictions available, the difference between the 
two should be regarded as an uncertainty induced by terms beyond NNLO accuracy.
Moreover, one could introduce a different setting of the hard scales 
at high transverse momenta in the \minnlo{} calculation (i.e.\ in the \ww{}+jet part) 
as a further probe of missing higher-order terms.

In summary, we find that \minnlo{} and NNLOPS predictions are in excellent agreement
for essentially all observables we considered in the inclusive phase space 
that are genuinely NNLO accurate. This indicates the robustness
of NNLO+PS predictions for such observables.
For the few exceptions, like large $\mww$, we could trace back the origin 
of the differences to terms beyond accuracy that are induced by the different scale settings. 
Moreover, in all cases the NNLO corrections substantially reduce the scale uncertainties with 
respect to the \minlo{} prediction. Notice, however, that in the bulk region of the inclusive
phase space the \minnlo{} uncertainty bands
are about a factor of two smaller than the NNLOPS ones, as already observed for the fully 
inclusive cross section.

\subsubsection{Fiducial phase space}
\label{sec:fid}

\begin{figure}[p]
\begin{center}\vspace{-0.2cm}
\begin{tabular}{cc}
\includegraphics[width=.31\textheight]{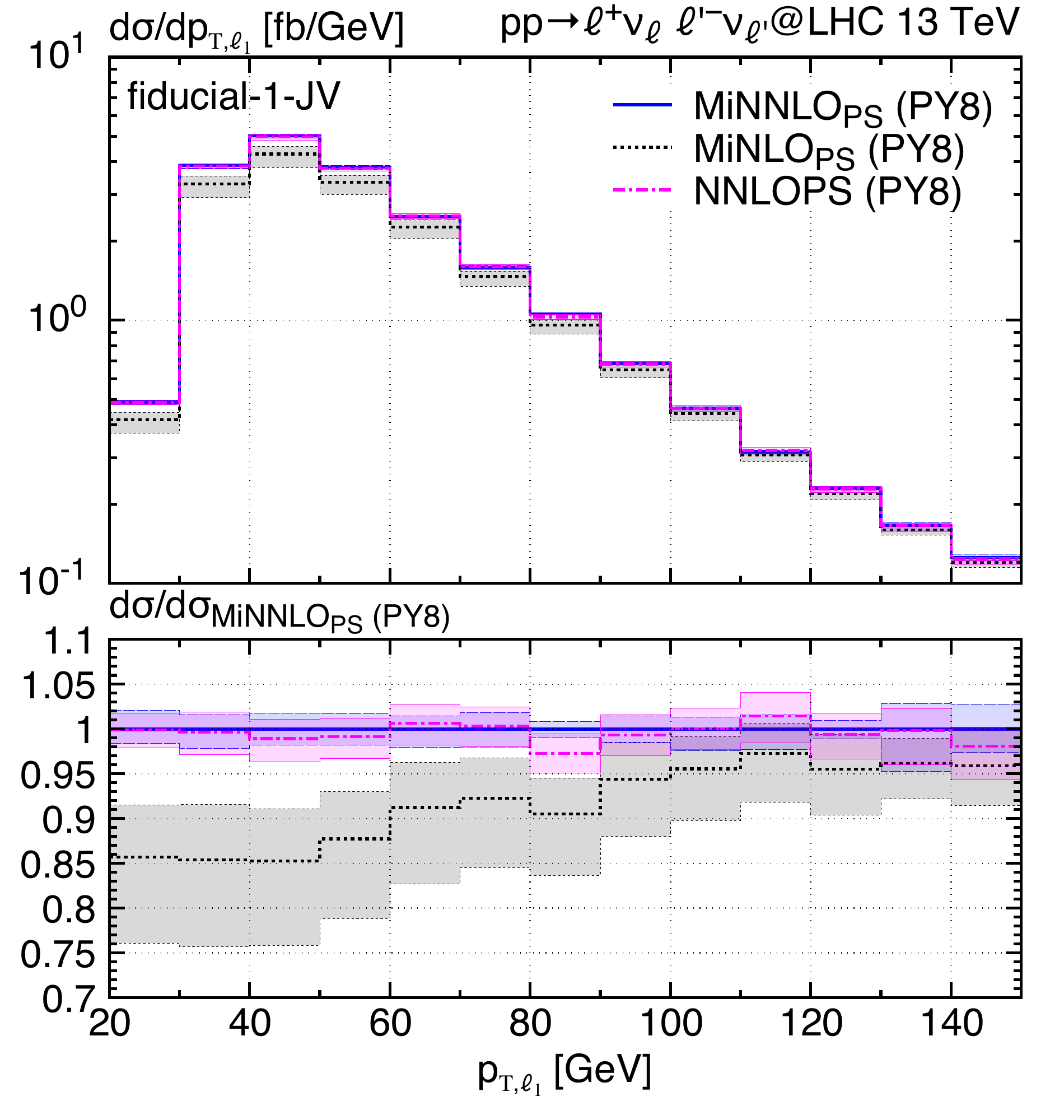} 
&
\includegraphics[width=.31\textheight]{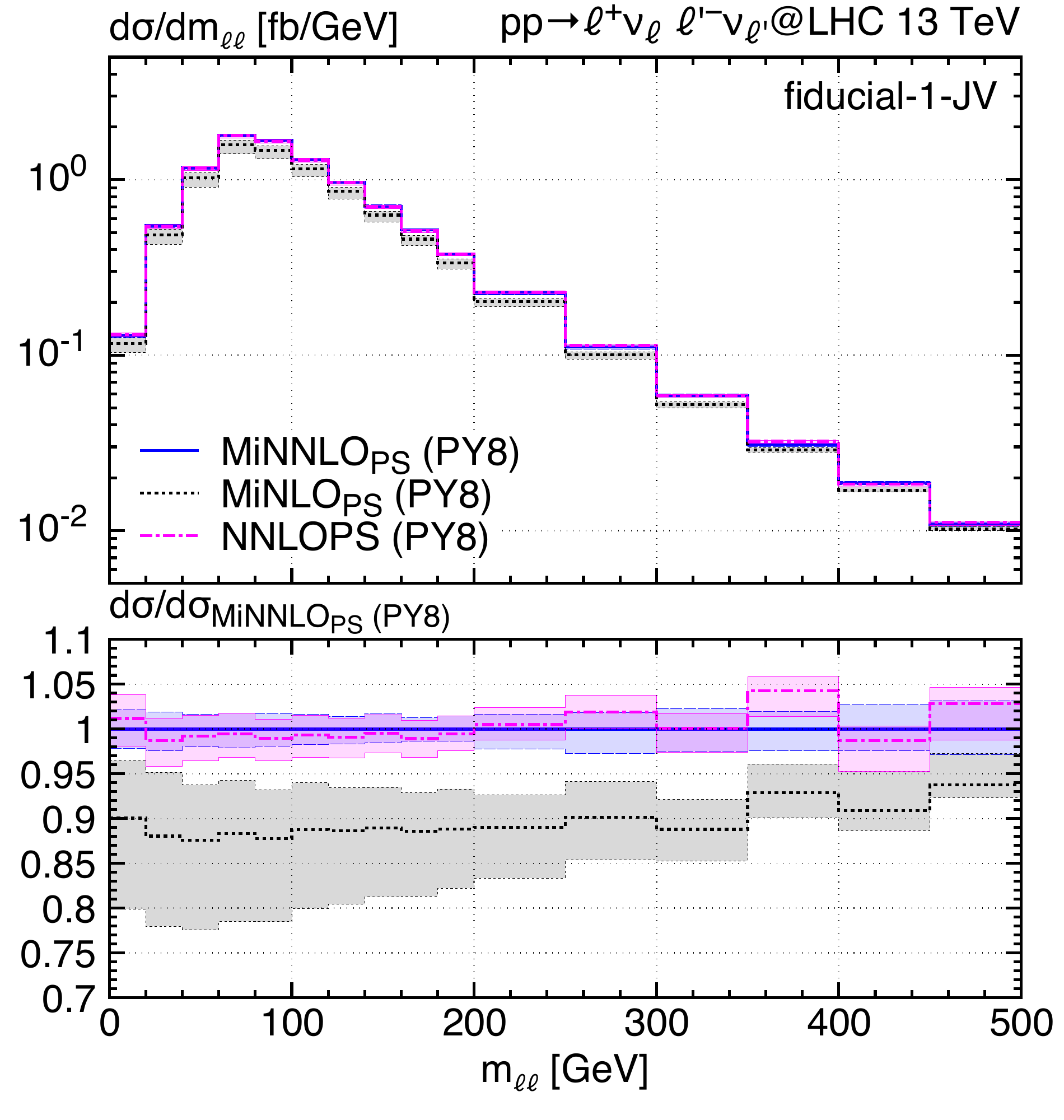}
\end{tabular}\vspace{-0.15cm}
\begin{tabular}{cc}
\includegraphics[width=.31\textheight]{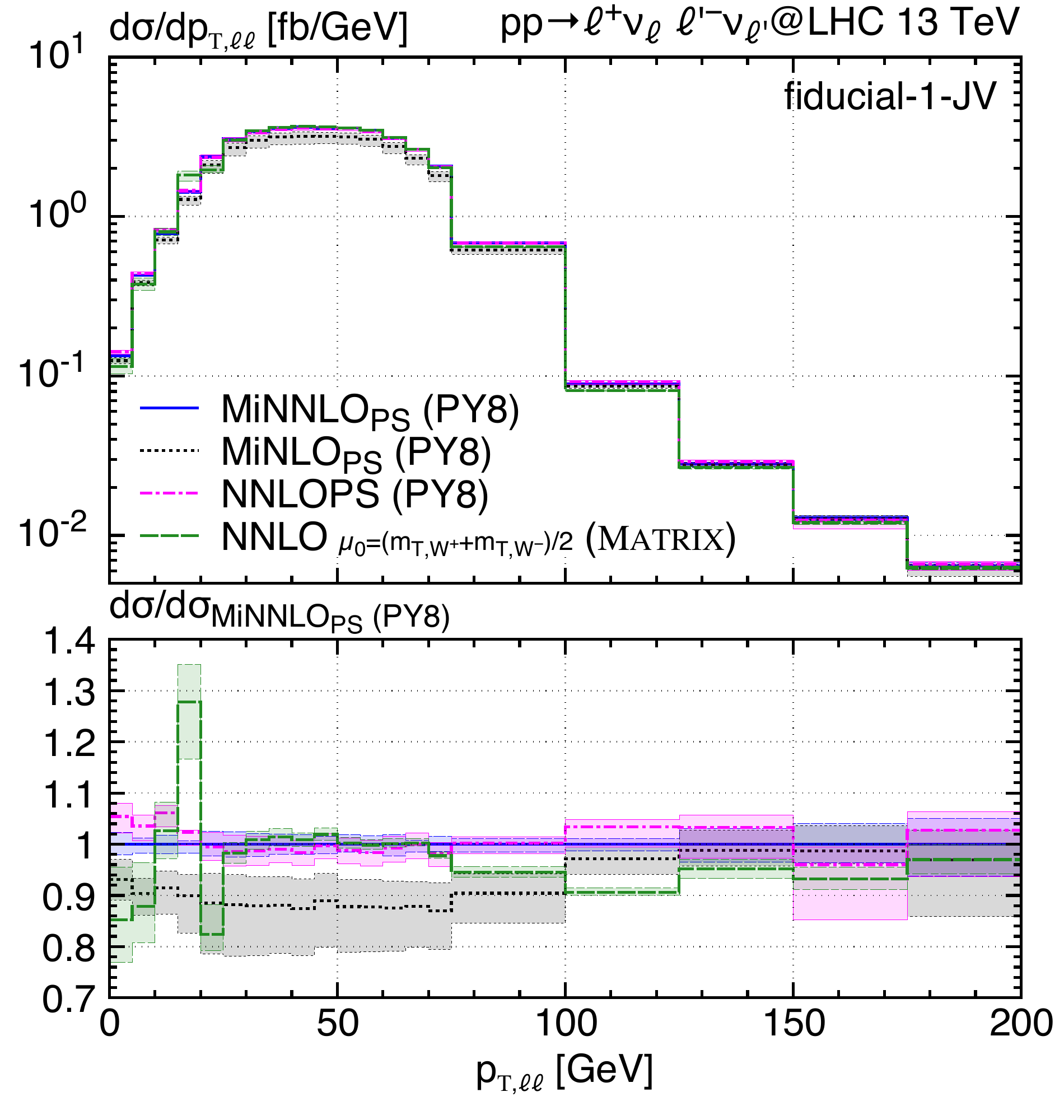}
&
\includegraphics[width=.31\textheight]{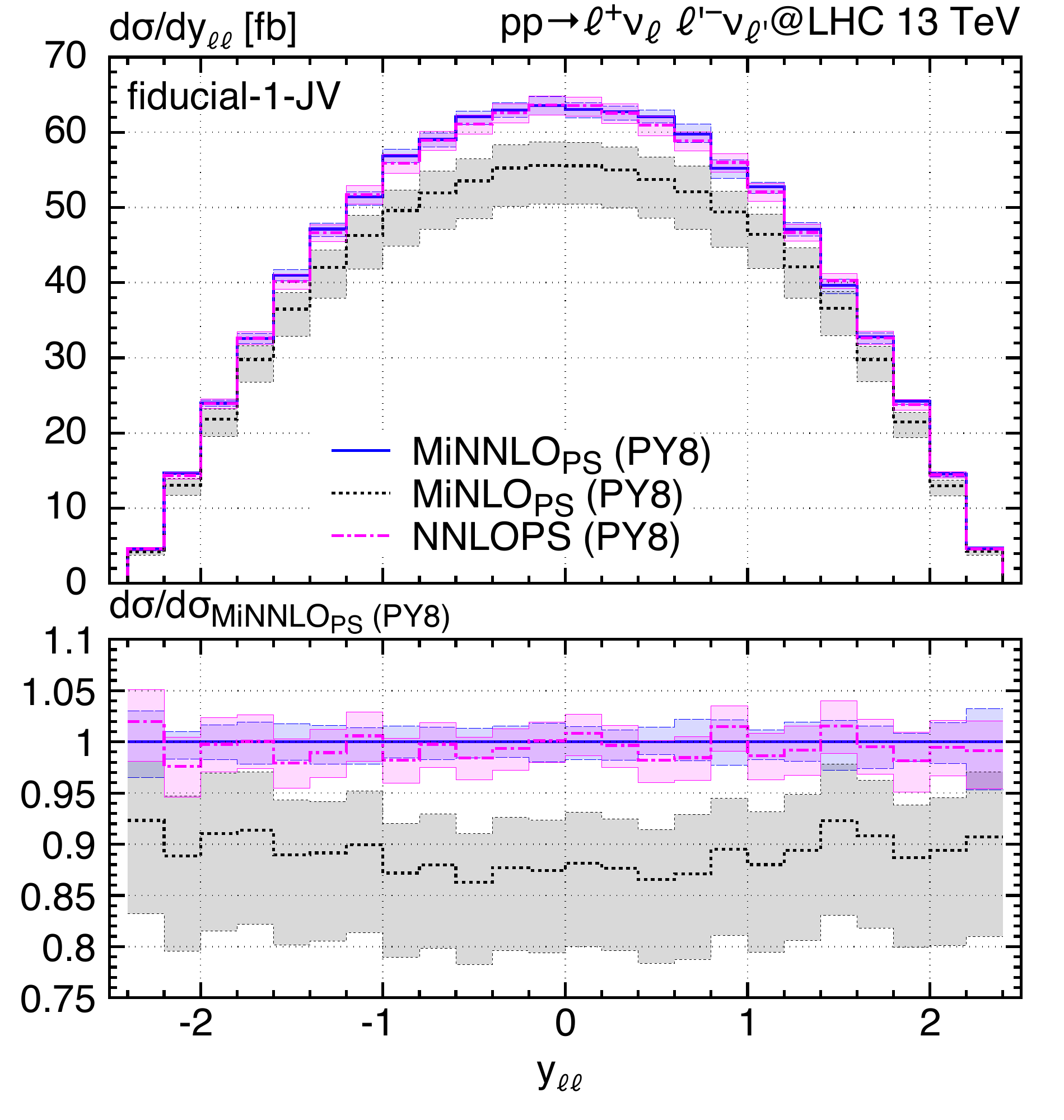}
\end{tabular}\vspace{-0.15cm}
\begin{tabular}{cc}
\includegraphics[width=.31\textheight]{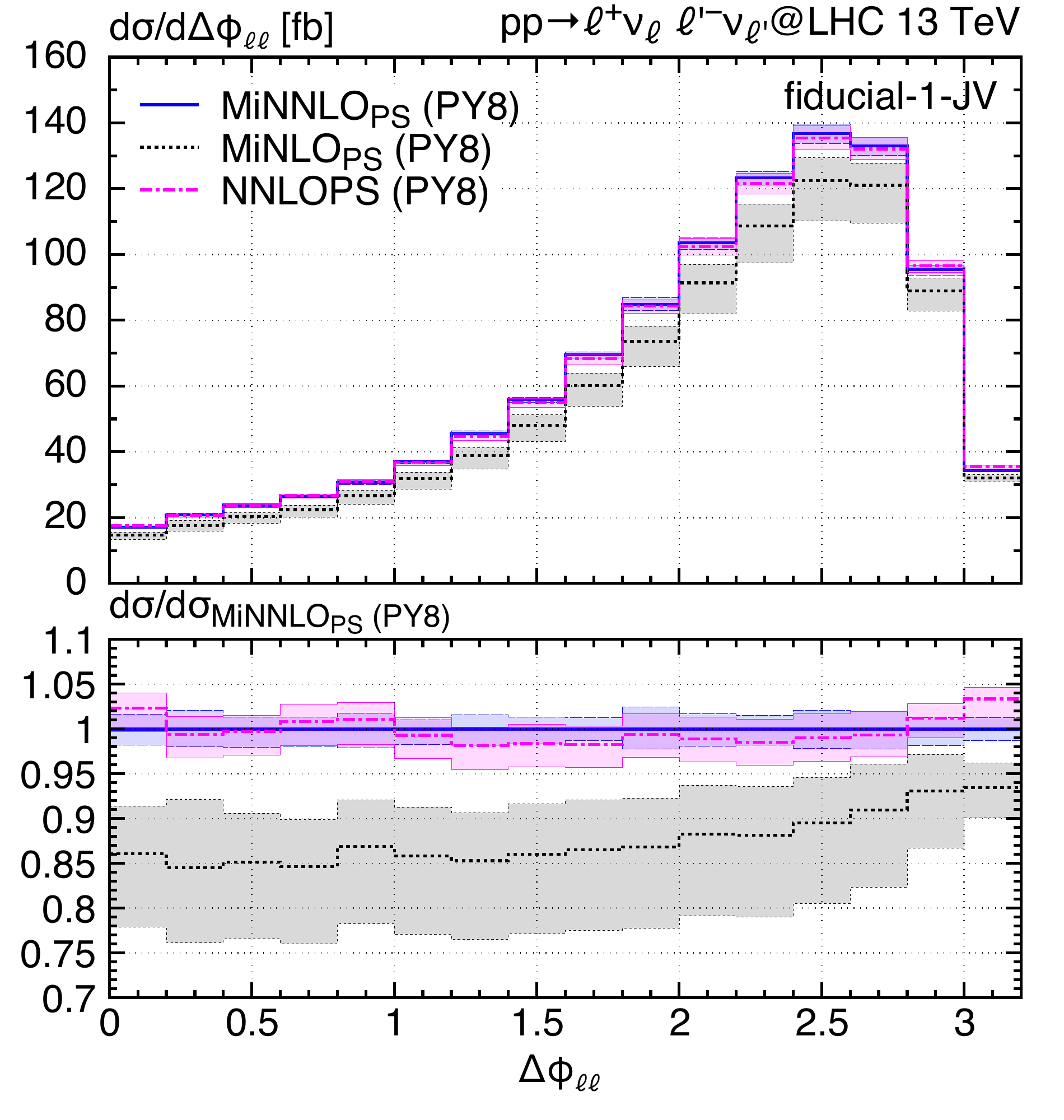}
&
\includegraphics[width=.31\textheight]{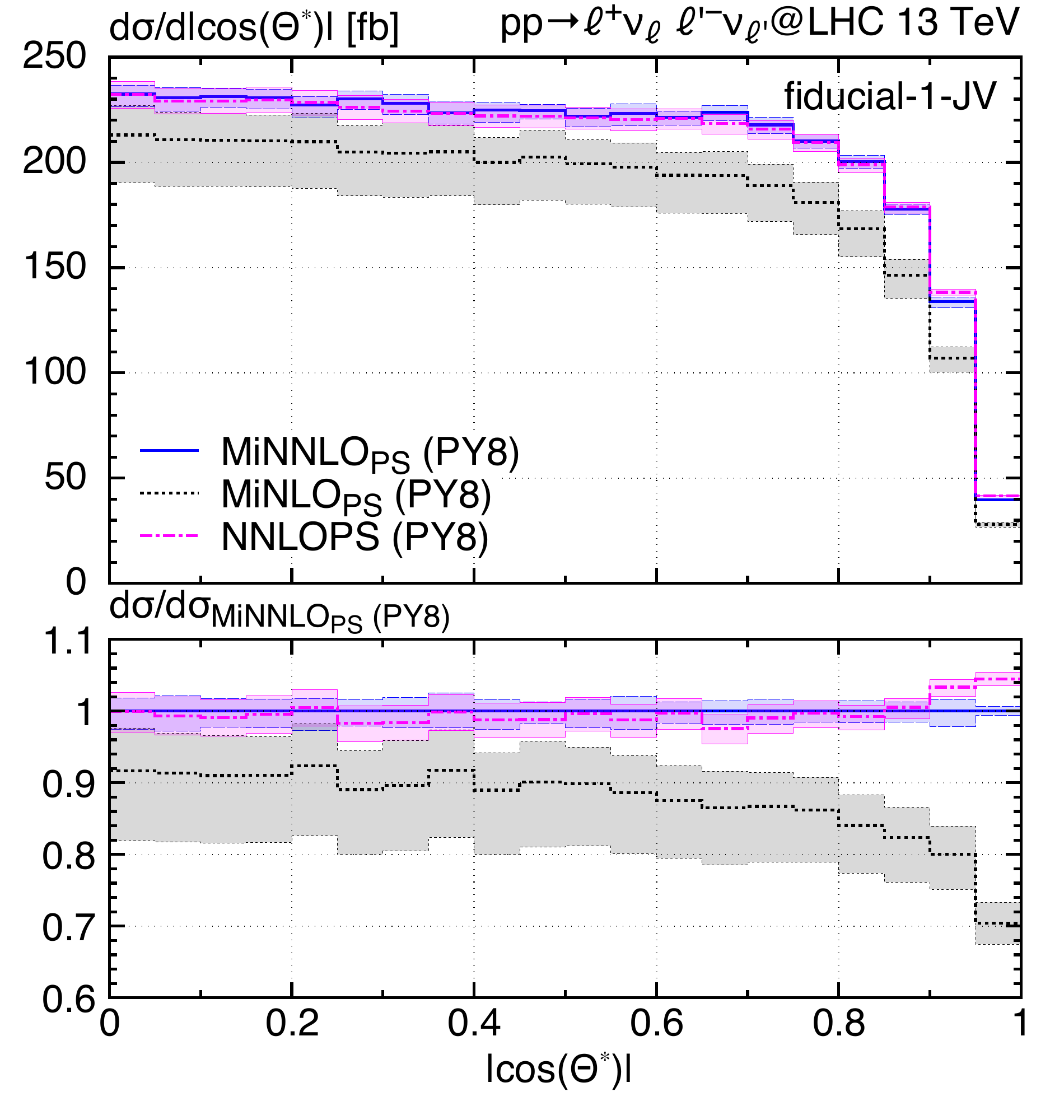} 
\end{tabular}
\caption{\label{fig:fiducial1} Differential distributions in the \setupone{} phase space.}
\end{center}
\end{figure}
\afterpage{\clearpage}

We continue our comparison by considering differential distributions in the \setupone{} phase space.
Here, we have selected a set of leptonic observables 
that are directly measured by the experimental analyses, cf.\ \citeres{Aaboud:2017qkn,Aaboud:2019nkz}, and which represent well the general features of all observables we considered.
To this end, \fig{fig:fiducial1} shows the distributions in the transverse momentum of the leading 
lepton ($\ptlone$), in the invariant mass ($\mll$), transverse momentum ($\ptll$), 
rapidity ($\yll$) and azimuthal difference ($\dphill$) of the dilepton pair, and in 
an observable particularly sensitive to new physics effects 
defined through the separation in $\eta{}$ of the two leptons:
\begin{align}\label{eq:costhetastart}
\left|\cos(\theta^\star)\right|=\left| \tanh\left(\Delta\eta_{\ell\ell}/2\right) \right|\,.
\end{align}

As for the \setupinclusive{} in the previous section, we find full compatibility
between \minnlo{} and NNLOPS predictions. With fiducial cuts, even the 
differences induced by terms beyond accuracy are reduced and the 
scale-uncertainty bands of the two calculations are of similar size.
Also in this case, an important observation is that
the inclusion of NNLO corrections on top of the \minlo{}
is crucial not only for the correct normalization, but for many observables
also to capture relevant shape effects. Moreover, the NNLO-accurate predictions
are substantially more precise due to their strongly reduced uncertainty bands with 
respect to \minlo{}. We further notice that the impact of parton-shower emissions
on observables with NNLO accuracy is quite moderate.
Nevertheless, at phase-space boundaries where the fixed-order accuracy is reduced
and the perturbative expansion breaks down due to effects from soft QCD radiation,
the parton shower is absolutely crucial for a physical description. For instance, 
this can observed in the $\ptll$ distribution, where we have added the fixed-order NNLO 
prediction for comparison.  Since the $\ptmiss>20$\,GeV requirement in \setupone{} setup translates 
directly into a $\ptll>20$\,GeV
cut at LO, where the two leptons are back-to-back with the two neutrinos, the region 
$\ptll\le20$\,GeV is filled only upon inclusion of higher-order corrections and is effectively 
only NLO accurate. As a result, the boundary region $\ptll\sim 20$\,GeV becomes sensitive
to soft-gluon effects that induce large logarithmic corrections  and a
perturbative instability \cite{Catani:1997xc} 
at $\ptll=20$\,GeV in the fixed-order NNLO prediction. This unphysical behaviour is
cured through the matching to the parton shower in the \minnlo{} and NNLOPS calculations.

It is clear that our new \minnlo{} predictions compare very well with the previous NNLOPS results,
and that the two tools can be used equivalently to produce \ww{} cross sections and distributions
at NNLO accuracy matched to parton showers. This is also an indication of the robustness
of NNLO+PS predictions for observables that are genuinely NNLO accurate.
Given the limitation of the NNLOPS calculation
regarding the necessity of multi-dimensional reweighting, the advantage of the \minnlo{}
generator is that those results can be obtained directly at the level of the event generation.
However, in the few phase-space regions where differences between the two calculations can 
be observed, those differences indicate relevant corrections beyond NNLO accuracy. Since 
plain $7$-point scale variations do not always cover those discrepancies, they should be 
regarded as a perturbative uncertainty.

In the next section we will move to observables that are subject to large logarithmic corrections
and where differences between the \minnlo{} and NNLOPS generator are larger.
Thus, their assessment as an uncertainty becomes particularly important.

\subsubsection{Observables sensitive to soft-gluon effects}
\label{sec:IRsensitive}

\begin{figure}[t]
\begin{center}
\begin{tabular}{cc}\hspace{-0.5cm}
\includegraphics[width=.31\textheight]{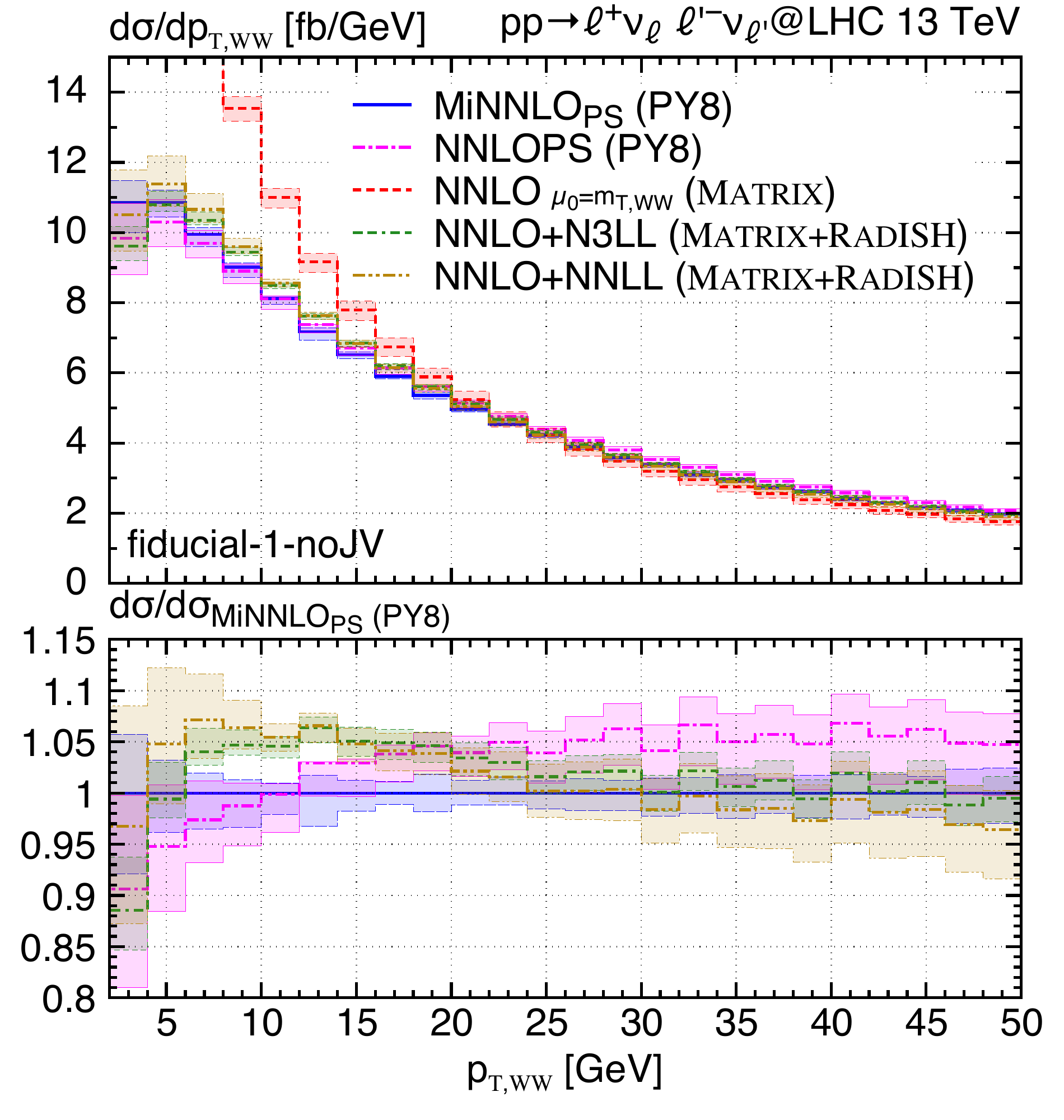}
&
\includegraphics[width=.31\textheight]{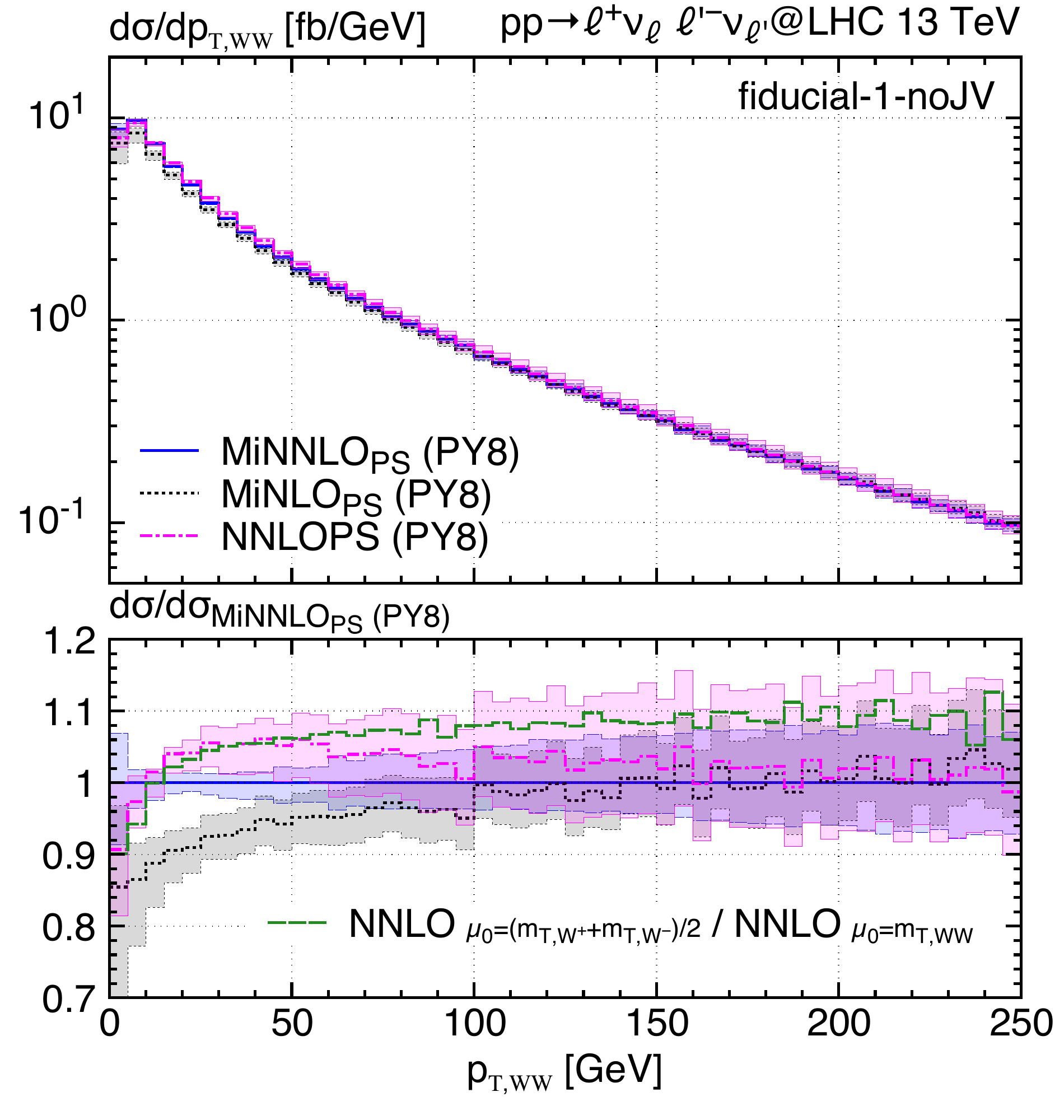}
\end{tabular}
\caption{\label{fig:ptww} Distribution in the transverse momentum of the
  \ww{} pair in the \setuponenoJV{} phase space, showing a smaller (left plot) and a wider range (right plot).}
\end{center}
\end{figure}

In \fig{fig:ptww}, we study the transverse-momentum spectrum of the \ww{} pair ($\ptww$)
in the \setuponenoJV{} phase space.
We refrain from showing the corresponding distribution in \setupinclusive{} and within the \setuptwonoJV{}
phase space, as we found them to be almost identical concerning the 
relative behaviour of the various predictions.
At small values of $\ptww$, large logarithmic contributions break the validity of the 
expansion in the strong coupling constant at a given fixed order, which requires their 
inclusion all orders in perturbation theory either through a parton 
shower or through an analytic resummation. The left figure displays the region 
$0\le\ptww\le 50$\,GeV and, indeed, the NNLO prediction, which is shown  
in the main frame only, becomes unphysical for small values of $\ptww$.
If we compare \minnlo{} and NNLOPS results in that region, we observe differences of 
about $-10$\% to $+5$\%. By and large, those are covered by the respective uncertainty bands.
However, it is clear (and expected) that for such an observable, which is sensitive to infrared physics, 
the differences between the two calculations become larger. In particular, both predictions are 
only NLO accurate in the tail of the $\ptww$ distribution and at small transverse momenta
the parton shower limits the accuracy of the calculation effectively to leading-logarithmic (LL) 
or partial (i.e.\ at leading colour) 
next-to-LL (NLL) accuracy. Therefore, differences of the order of those that we observe 
between \minnlo{} and NNLOPS are understood. Also the comparison against the 
high-accuracy analytic resummation results at NNLO+N$^3$LL and NNLO+NNLL is quite good, which 
also agree within $-10$\% to $+5$\% with the \minnlo{} prediction for $\ptww<20$\,GeV and 
are even fully compatible in the intermediate region up to $50$\,GeV. The resummed predictions
do not favour either \minnlo{} or NNLOPS results, but rather show similar differences 
to the two. On the other hand, the agreement is actually quite remarkable considering the fact
that the region $\ptww<20$\,GeV is entirely described by the substantially 
less accurate parton shower. Given the fact that for some bins the NNLO+N$^3$LL and NNLO+NNLL
predictions are outside the uncertainty bands of the NNLO+PS accurate predictions though, the 
estimated uncertainties from $\muR$ and $\muF$ variations appear insufficient to reflect the 
actual size of uncertainties and one should consider additional handles to better 
assess the uncertainties of the parton shower at small $\ptww$. Indeed, the NNLL  
prediction has a much larger uncertainty band in this region (induced by the variation of
$Q_{\text{\scalefont{0.77}res}}$) even though it is more accurate.

In the right plot of \fig{fig:ptww}, we show the range $0\le\ptww\le 250$\,GeV. 
In the tail of the distribution, \minnlo{} and NNLOPS (as well as
\minlo{}) predictions are in perfect agreement with fully overlapping
uncertainty bands.
In the lower frame we show an additional curve that is ratio of the central fixed-order
NNLO prediction 
with $\mu_{\text{\scalefont{0.77}0}}= \frac12\,\left(m_{\text{\scalefont{0.77}T,$W^+$}}+m_{\text{\scalefont{0.77}T,$W^-$}}\right)$ to the 
one with  $\mu_{\text{\scalefont{0.77}0}}=\mtww$.
It is very interesting to observe that the ratio corresponds almost exactly
to the NNLOPS/\minnlo{} ratio at smaller $\ptww{}$. We recall that  $\mu_{\text{\scalefont{0.77}0}}= \frac12\,\left(m_{\text{\scalefont{0.77}T,$W^+$}}+m_{\text{\scalefont{0.77}T,$W^-$}}\right)$ is the scale used in the reweighting of the NNLOPS prediction, while $\mu_{\text{\scalefont{0.77}0}}=\mtww$ is somewhat 
more similar to the one within the \minnlo{} approach. This suggests that the differences 
originating from terms beyond accuracy at small $\ptww{}$
between the \minnlo{} and NNLOPS are predominantly 
induced by the different scale settings in the two calculations.
In fact, for any distribution (of the various ones we considered) 
where the NNLOPS/\minnlo{} ratio becomes larger than a couple of 
percent, we observe that the corresponding ratio of fixed-order NNLO
predictions is either very similar or even larger.

\begin{figure}[t]
\begin{center}
\begin{tabular}{cc}\hspace{-0.5cm}
\includegraphics[width=.33\textheight]{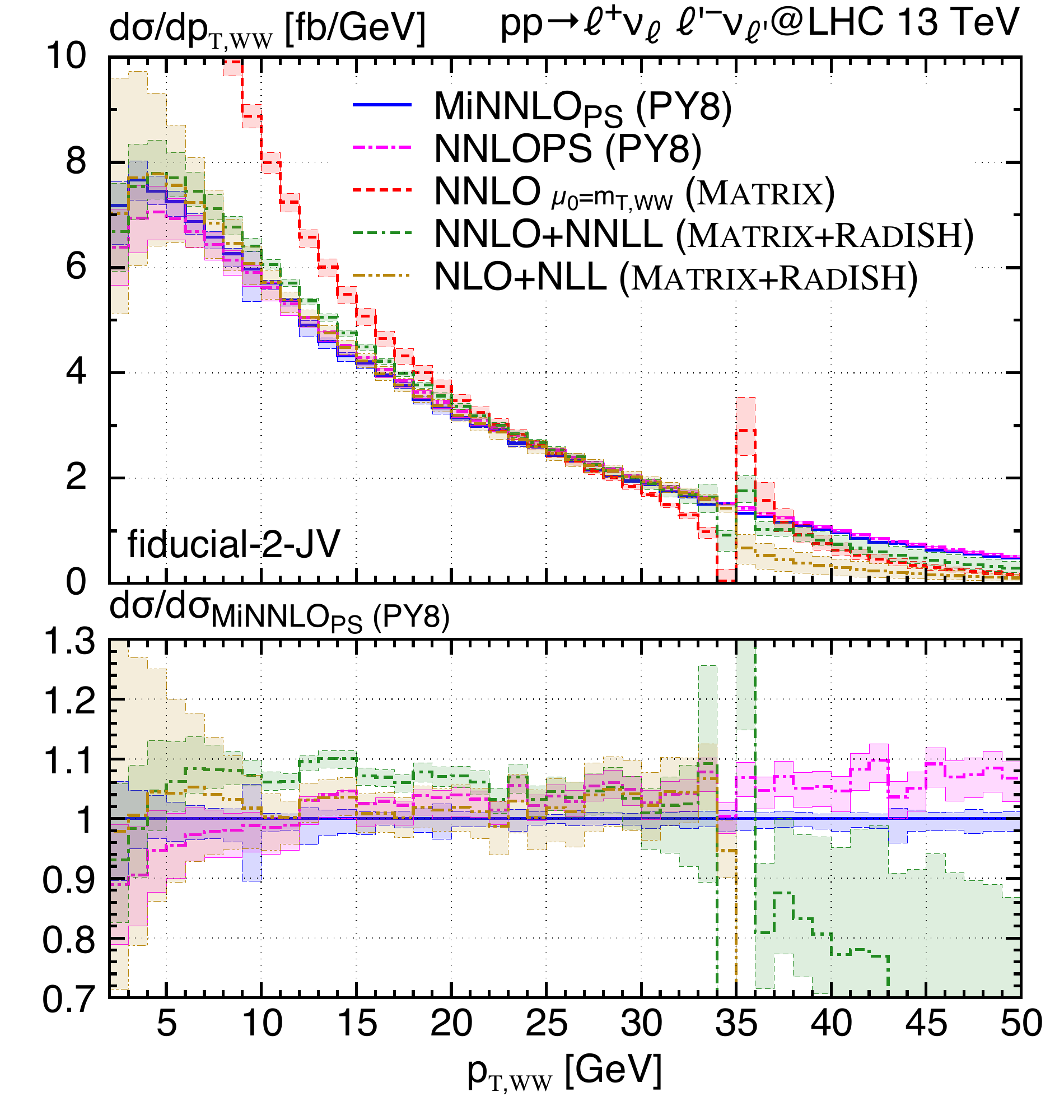}
&
\includegraphics[width=.33\textheight]{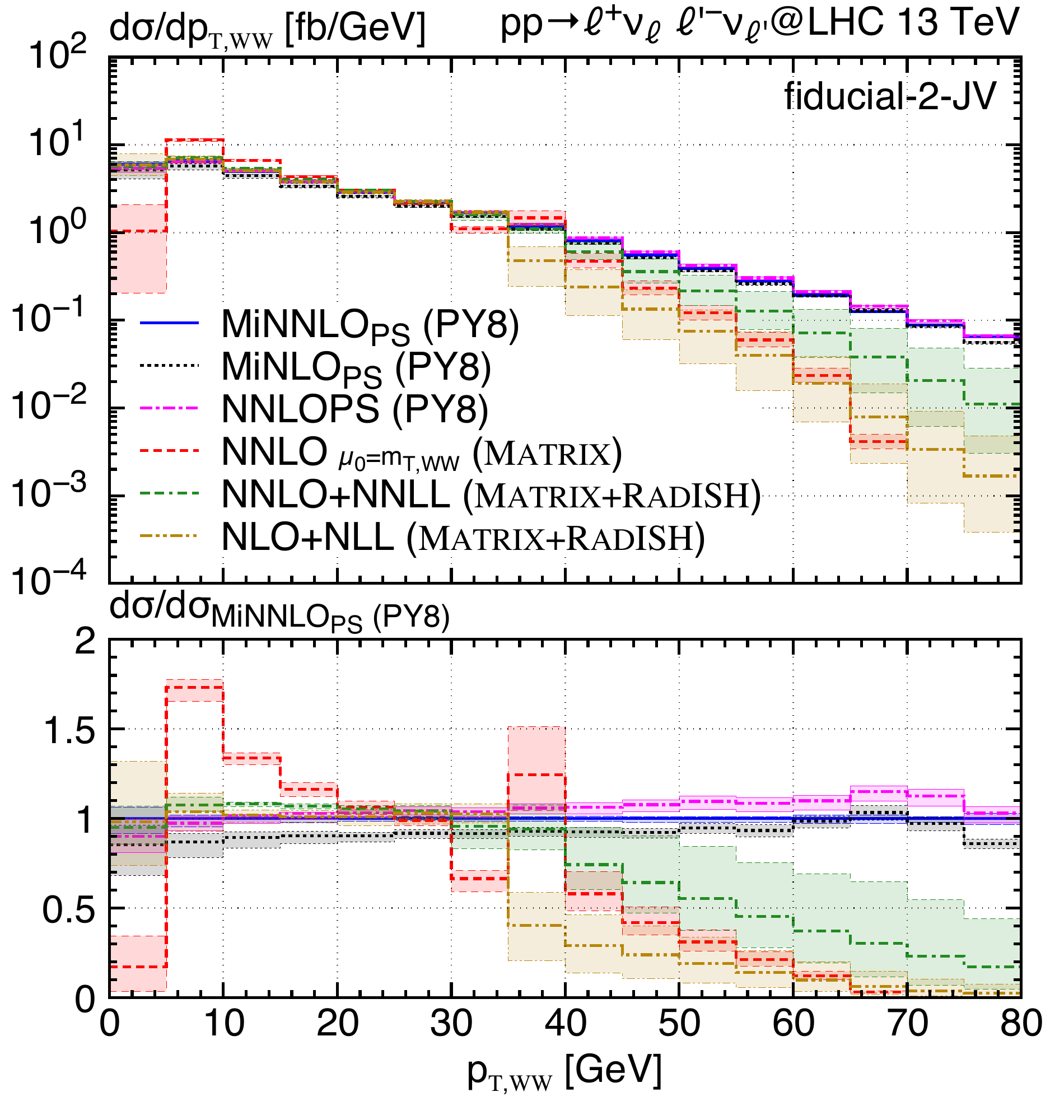}
\end{tabular}
\caption{\label{fig:ptwwJV} Distribution in the transverse momentum of the
  \ww{} pair in the presence of a jet veto (\setuptwo{} phase space), showing a smaller (left plot) and a wider range (right plot).}
\end{center}
\end{figure}

In \fig{fig:ptwwJV} we consider the \ww{} transverse momentum spectrum in the presence of 
a jet veto of $\ptjoneveto=35$\,GeV
using the \setuptwo{} setup. The relative behaviour between the \minnlo{}, NNLO+PS,
NNLO+NNLL and NLO+NLL results at small transverse momenta is relatively similar to the one observed 
for the $\ptww{}$ distribution without jet veto in setup \setuponenoJV{}. 
One main difference is that for this observable, which 
requires double differential resummation in $\ptww$ and $\ptjone$, the analytically resummed 
results are less accurate and therefore feature larger uncertainty bands, rendering them
more compatible with the showered results. 
Indeed, the NLL uncertainty band is strongly increased
at small $\ptww$ and much larger than the NNLO+PS one, which, as argued before, 
also points to the fact that the scale uncertainties of the latter are somewhat underestimated, 
given that the parton shower is less accurate than the NLL calculation in that region.
Another interesting region for this observable is around $\ptww$ values 
of $35$\,GeV, i.e. of the order of the jet-veto cut. The region $\ptww\ge \ptjoneveto$ is filled for 
the first time at NNLO, which is effectively only LO accurate, since at LO it is $\ptww=0$ 
and at NLO $\ptww=\ptjone$. Therefore, large logarithmic contributions challenge 
the perturbative expansion around $\ptww=\ptjoneveto$ and the fixed-order NNLO prediction 
develops a perturbative instability, as visible in the main frame of the left 
plot in \fig{fig:ptwwJV}. This instability is partially cured by the analytic resummation approach, 
which resums Sudakov logarithms in the limit where $\ptww$ and $\ptjone$ are much 
smaller than the hard scale, but not all logarithmic contributions of the 
form $\log(\ptww-\ptjoneveto)$,
which would require additional resummation when one or more hard jets
are present. 
By contrast, the NNLO+PS calculations cure this instability entirely 
as they resum all relevant classes of logarithms (although with limited accuracy).
Therefore, the \minnlo{} and NNLOPS calculations
provide a more physical prediction at and above threshold, while below 
the threshold they are in good agreement with the analytically resummed predictions.
If we look at region above threshold in the right plot of \fig{fig:ptwwJV}, we notice that 
the NNLO result drops substantially for $\ptww$ values above $\ptjoneveto$, and also
the NNLO+NNLL prediction is only slightly larger. Hence, this region of phase space is  
almost exclusively filled by the parton shower. Consequently, the transverse-momentum 
spectrum of a colour singlet in presence of  a jet veto could be a good observable
 to tune the parton shower in experimental analyses.

\begin{figure}[t]
\begin{center}
\begin{tabular}{cc}\hspace{-0.5cm}
\includegraphics[width=.31\textheight]{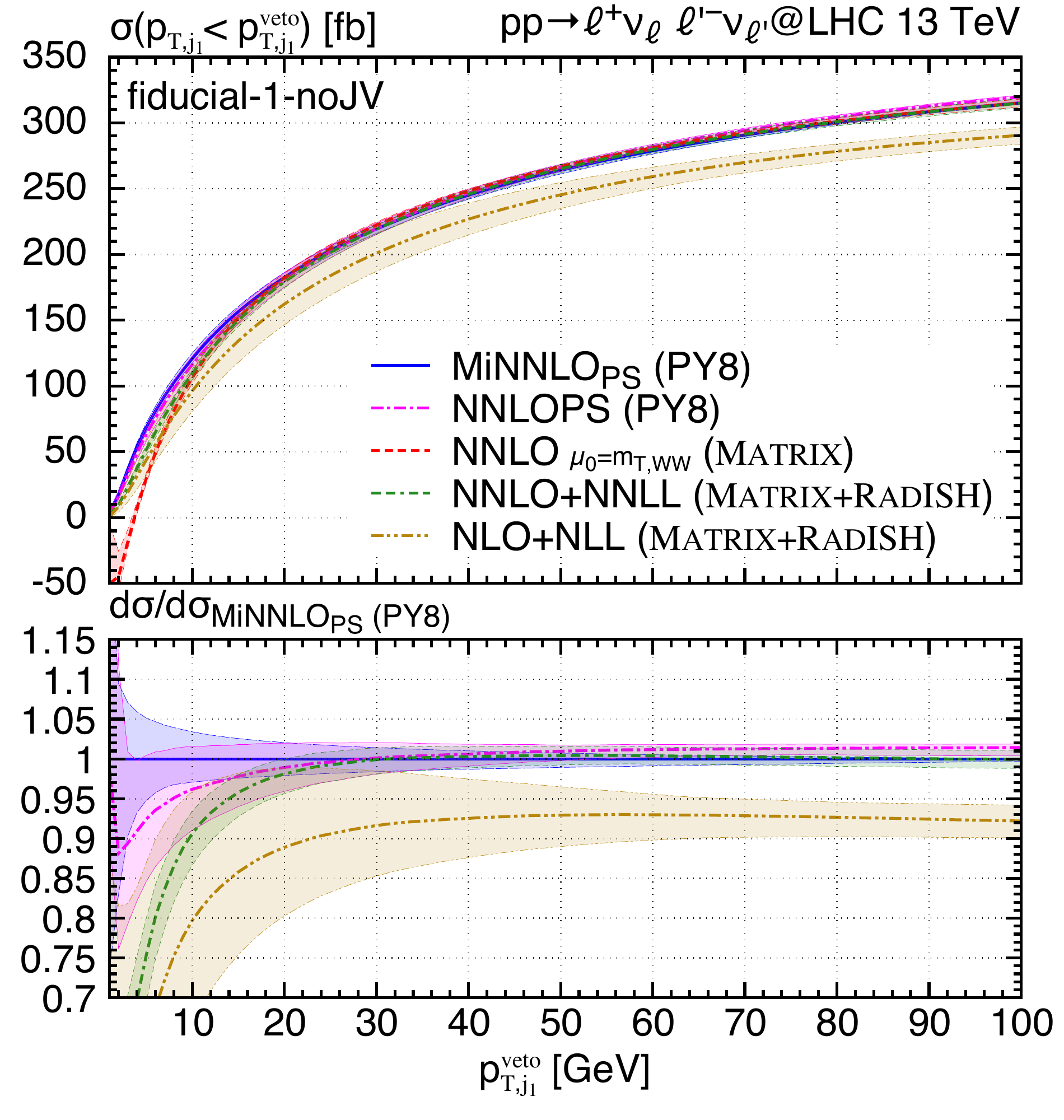}
&
\includegraphics[width=.31\textheight]{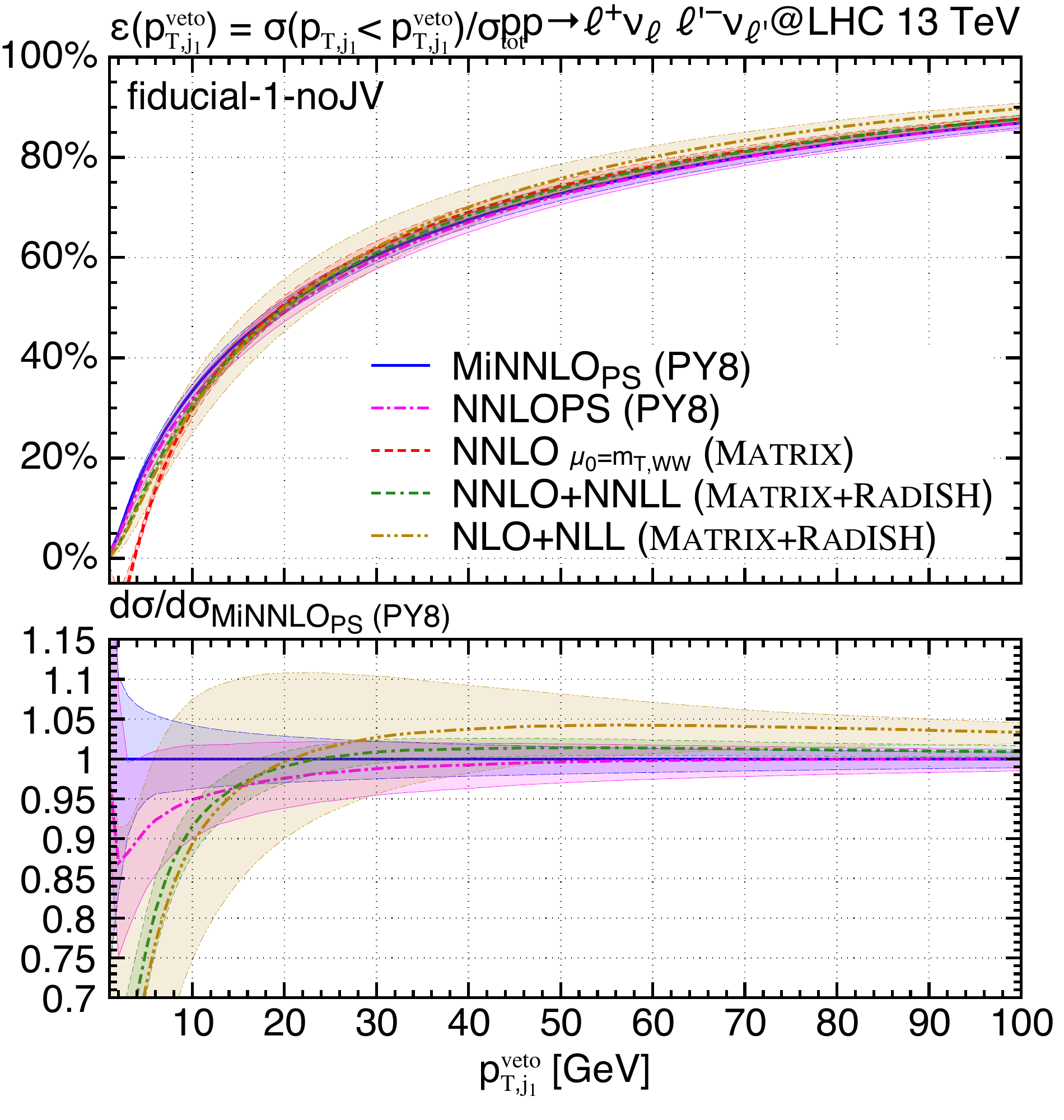}
\end{tabular}
\caption{\label{fig:jetveto} Jet-vetoed cross section (left plot) and jet-veto efficiency (right plot) in the \setuponenoJV{} phase space.}
\end{center}
\end{figure}

In \fig{fig:jetveto} we study the jet-vetoed cross section as a function of the jet-veto 
cut $\ptjoneveto$, which is defined as
\begin{align}
\sigma(\ptjone < \ptjoneveto) = \int_0^{\ptjoneveto}
\mathd\ptjone\,\frac{\mathd\sigma}{\mathd\ptjone}\,,
\end{align}
and the jet-veto efficiency given by
\begin{align}
\varepsilon(\ptjoneveto)=\sigma(\ptjone < \ptjoneveto)/\sigma_{\rm
  int}\,,
\end{align}
where $\sigma_{\rm int}$ is the integrated cross section in the \setuponenoJV{} phase space.
Again the results for \setupinclusive{} and \setuptwonoJV{} are very similar and are not shown.
The interesting region is at small jet-veto cuts, where the validity 
of the perturbative expansion is broken by large logarithmic contributions in $\ptjoneveto$,
while for larger values the results tend towards their respective integrated cross sections. 
As it can be seen from the main frame, in the low $\ptjoneveto$ region
the pure fixed-order result at NNLO becomes indeed unphysical and turns actually negative.
When comparing \minnlo{} and NNLOPS predictions, we find them to be in reasonable 
agreement within their respective uncertainties,
with the NNLOPS one tending a bit faster towards zero for $\ptjoneveto\lesssim 20$\,GeV.
In that region, the resummed NNLO+NNLL and NLO+NLL results tend even 
faster towards zero, with the NNLO+NNLL curve being about $20\%$ below the 
\minnlo{} one at $\ptjoneveto=5$\,GeV. This region is dominated by the parton shower, which
resums only the LL (partial NLL) contributions. Clearly, the actual uncertainties in the NNLO+PS
calculations
are not covered by plain $\muR$ and $\muF$ variations. As argued for the $\ptww{}$ 
spectrum, additional handles would need to be considered to better assess the parton-shower 
uncertainties for very small $\ptjoneveto$ cuts. Indeed, the NLL result features 
much wider uncertainties, despite being more (similarly) accurate in that region of phase space.
However, we stress that such small $\ptjoneveto$ cuts are usually not relevant for experimental 
\ww{} analyses. Moreover, as pointed out before, there have been suggestions 
to include more conservative uncertainty estimates for jet-vetoed predictions \cite{Stewart:2011cf,Banfi:2012jm}.
We leave their proper assessment to future work, as those effects are currently 
not accessed by any \ww{} measurement.
For instance, looking at the fiducial phase-space definitions of
 \citeres{Aaboud:2017qkn,Aaboud:2019nkz} that are considered 
in this paper, jet-veto cuts of $\ptjoneveto=25$\,GeV, $30$\,GeV and $35$\,GeV are used.
For those values, \minnlo{} predictions are in perfect agreement with the NNLO+NNLL 
resummation, and even down to $\ptjoneveto\sim15$\,GeV they differ by less than $5\%$ with 
overlapping uncertainties.

\begin{figure}[t]
\begin{center}
\includegraphics[width=.4\textheight]{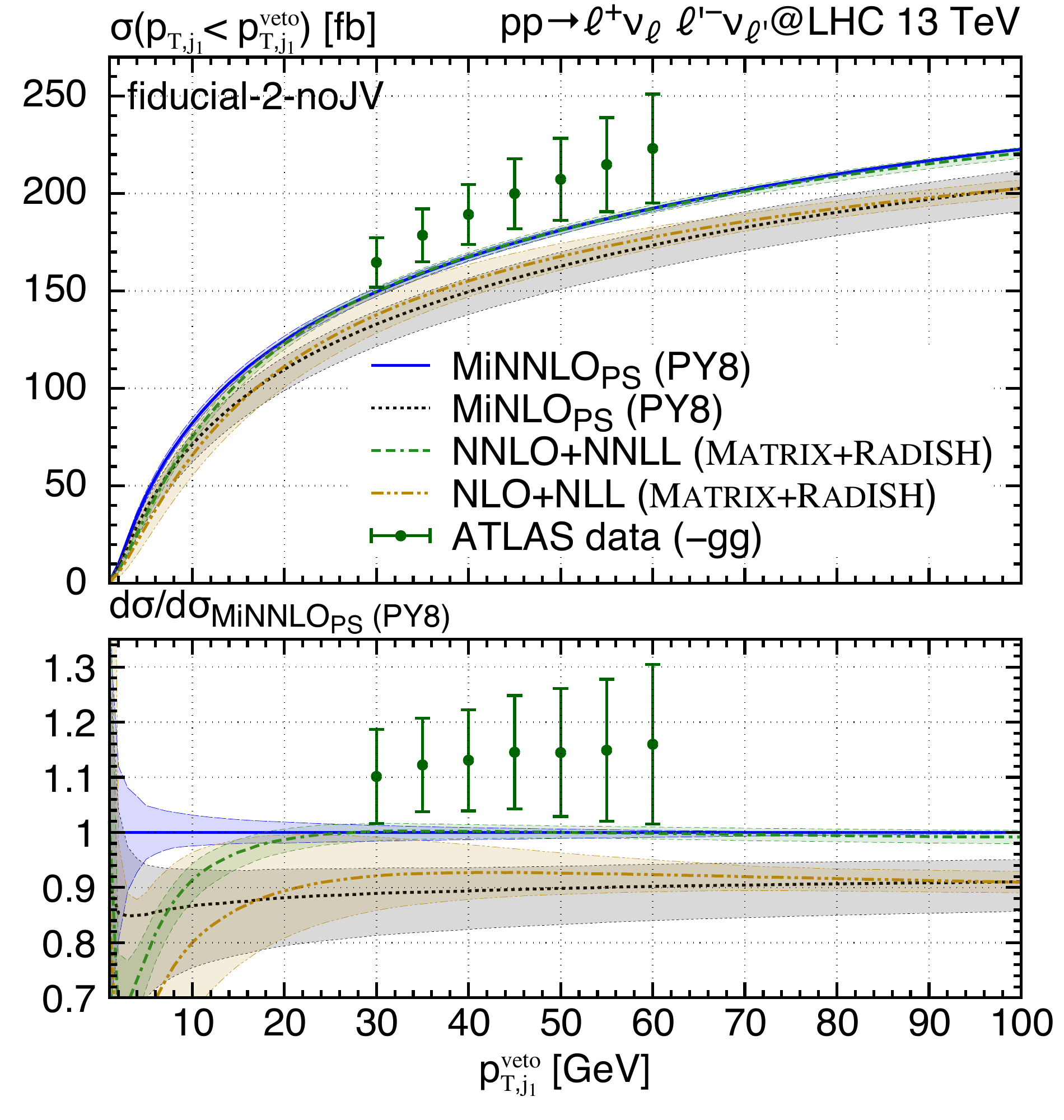}
\caption{\label{fig:jetvetodata} Jet-vetoed cross section in the \setuptwonoJV{} phase space compared to data. As described in the caption of \tab{tab:crosssection} the data has been adjusted by subtracting the $gg$LO contribution quoted in table 2 of \citere{Grazzini:2020stb} and by dividing out a factor of two.}
\end{center}
\end{figure}

When comparing the predicted jet-vetoed cross section as a function of $\ptjoneveto{}$
in the \setuptwonoJV{} setup
against data in \fig{fig:jetvetodata}, it is clear that the \minnlo{} and the NNLO+NNLL prediction are
fully compatible in the relevant region. The agreement with data is good in either case, 
with the data points either marginally overlapping within one standard deviation or being just 
outside this range. One should bear in mind however that the normalization of the theory prediction can be 
increased by $\sim 5$\% just by using a different PDF set, which yields even better agreement 
with data, as shown in \citere{Kallweit:2020gva}. Apart from that,
it is clear that the inclusion 
of NNLO corrections brings the theory predictions closer to data.

\section{Conclusions}
\label{sec:summary}
In this paper we have presented the matching of NNLO-accurate predictions
with parton showers for \ww{} production at the LHC using the \minnlo{} method. 
We have performed the calculation consistently in the four-flavour scheme
with massive bottom quarks. By dropping contributions with final-state 
bottom quarks, which are regulated by the finite bottom mass, we generate
top-free \ww{} events. 
We have presented an extensive comparison of our \minnlo{} predictions
against the NNLOPS results of~\citere{Re:2018vac}. 
We find excellent agreement with the latter
results, with only minor differences in phase-space regions where
they are expected from the different treatment of terms beyond accuracy,
most importantly the ones related to different scale settings.
Especially for genuine NNLO observables \minnlo{} and NNLOPS are compatible within less than a few percent. Larger differences (but still within uncertainties)
can be observed for observables that are sensitive to the limited accuracy 
of the parton shower, in particular for small values of the
\ww{} transverse momentum or for very small
jet-veto cuts. For those observables we also compared 
to high-accurate analytically resummed 
predictions. 
The agreement is very reasonable considering the
limited (leading logarithmic) accuracy of the parton shower.
In particular for phase-space regions relevant for experimental \ww{}
analyses, i.e\ for jet-veto cuts of $25$\,GeV to $35$\,GeV (and higher), 
the \minnlo{} prediction essentially
coincides with the NNLO+NNLL result. Indeed, we find good agreement
both for inclusive and fiducial cross sections with experimental data,
as well as for the cross section as a function of the jet-veto cut.

We found that the major
bottleneck in the computation is the evaluation of the two-loop amplitude.
To improve the speed, we have constructed four-dimensional grids of 
the coefficients that encode all the information required to 
reproduce the full two-loop contribution. We then use those grids
to obtain the coefficients at any given phase-space configuration 
through four-dimensional cubic spline interpolation and reconstruct 
the two-loop amplitude on-the-fly. As a result, the evaluation time of the two-loop 
contribution has been reduced by a factor of forty, becoming 
subleading with respect to the other parts of the calculation.
We have performed a thorough and extensive validation of the results with 
interpolation against the ones without.
The \minnlo{} code can be used either with or without interpolator, the 
latter option being about five times slower.

The \minnlo{} approach has various positive features. First, NNLO corrections
are calculated on-the-fly during the generation of the events, with no
need for further re-processing or reweighting of the events. Second, no
unphysical merging scale needs to be introduced to separate event
samples of different multiplicities, a concept already introduced and
discussed in detail in \citere{Hamilton:2012rf}. Third, when combined
with transverse-momentum ordered parton showers, the matching
guarantees that the logarithmic accuracy of the parton shower
simulation is preserved. We note that, because the logarithmic
accuracy of parton showers is only leading logarithmic, it is often
taken for granted that this accuracy is preserved. On the contrary,
this is a subtle, but crucial point in any approach that combines NNLO
and parton showers. In the \minnlo{} case this requirement is immediately
fulfilled for transverse-momentum ordered showers,
since zero, one, and two emissions are included in the NNLO weight,
where the second-hardest is generated following the \POWHEG{}
matching procedure, while additional emissions are generated
 by the parton shower.

We expect that the \minnlo{} code associated to the work presented in
this paper, which enables an accurate fully-exclusive
hadron-level generation of \ww{} events, will be highly valuable for 
experimental measurements, which require an accurate simulation of
\ww{} production either as signal or as background to other
processes. The code
 will be publicly released within \POWHEGBOXRES{}
and will supersede the previous NNLOPS approach based on 
multi-differential reweighting in \POWHEGBOXVTWO{}.
Nevertheless, we stress that it may be useful to take differences between 
\minnlo{} and NNLOPS predictions
(or fixed-order NNLO with different scale settings) to assess residual uncertainties in 
certain phase-space regimes. Alternatively, different scale settings in the \minnlo{} calculation
and handles in the parton shower could be used to probe the size of terms beyond accuracy.

Finally, any meaningful comparison to data of differential distributions
in charge-neutral vector-boson pair production processes 
should include not only NNLO QCD accuracy for the $q\bar{q}$ channel,
but also NLO QCD corrections to the loop-induced $gg$ process, and 
possibly NLO EW corrections in the high-energy tails. We  leave 
such studies to future work.

\section*{Acknowledgements}

We are indebted to Pier Monni, Paolo Nason and Emanuele Re for several 
fruitful discussions and comments. We are particularly thankful to Pier Monni
for a very careful reading of the manuscript and several useful comments.
We also thank Emanuele Re for sharing the NNLOPS results of \citere{Re:2018vac}.
Moreover, we are grateful to Stefan Kallweit for helpful discussions and  
to Luca Rottoli for providing insights on the resummation as well as 
the predictions of \citere{Kallweit:2020gva}.
Most of the simulations have been performed
at the Max Planck Computing and Data Facility (MPCDF) in Garching.

\bibliography{MiNNLO}

\providecommand{\href}[2]{#2}\begingroup\raggedright\begin{thebibliography}{100}

\bibitem{Aaltonen:2009aa}
{\scshape CDF} collaboration, T.~Aaltonen et~al., \emph{{Measurement of the
  $W^+ W^-$ Production Cross Section and Search for Anomalous $WW \gamma$ and
  $WWZ$ Couplings in $p \bar{p}$ Collisions at $\sqrt{s}=1.96$~TeV}},
  \href{http://dx.doi.org/10.1103/PhysRevLett.104.201801}{\emph{Phys. Rev.
  Lett.} {\bfseries 104} (2010) 201801},
  [\href{https://arxiv.org/abs/0912.4500}{{\ttfamily 0912.4500}}].

\bibitem{Abazov:2009ys}
{\scshape D0} collaboration, V.~M. Abazov et~al., \emph{{Measurement of the WW
  production cross section with dilepton final states in p anti-p collisions at
  s**(1/2) = 1.96-TeV and limits on anomalous trilinear gauge couplings}},
  \href{http://dx.doi.org/10.1103/PhysRevLett.103.191801}{\emph{Phys. Rev.
  Lett.} {\bfseries 103} (2009) 191801},
  [\href{https://arxiv.org/abs/0904.0673}{{\ttfamily 0904.0673}}].

\bibitem{Abazov:2011cb}
{\scshape D0} collaboration, V.~M. Abazov et~al., \emph{{Measurements of $WW$
  and $WZ$ production in $W$ + jets final states in $p\bar{p}$ collisions}},
  \href{http://dx.doi.org/10.1103/PhysRevLett.108.181803}{\emph{Phys. Rev.
  Lett.} {\bfseries 108} (2012) 181803},
  [\href{https://arxiv.org/abs/1112.0536}{{\ttfamily 1112.0536}}].

\bibitem{Aad:2012oea}
{\scshape ATLAS} collaboration, G.~Aad et~al., \emph{{Measurement of the $W W$
  cross section in $\sqrt{s}=7$ TeV $pp$ collisions with the ATLAS detector and
  limits on anomalous gauge couplings}},
  \href{http://dx.doi.org/10.1016/j.physletb.2012.05.003}{\emph{Phys. Lett. B}
  {\bfseries 712} (2012) 289--308},
  [\href{https://arxiv.org/abs/1203.6232}{{\ttfamily 1203.6232}}].

\bibitem{ATLAS:2012mec}
{\scshape ATLAS} collaboration, G.~Aad et~al., \emph{{Measurement of $W^+W^-$
  production in $pp$ collisions at $\sqrt{s}$=7\,TeV with the ATLAS detector
  and limits on anomalous $WWZ$ and $WW\gamma$ couplings}},
  \href{http://dx.doi.org/10.1103/PhysRevD.87.112001,
  10.1103/PhysRevD.88.079906}{\emph{Phys. Rev.} {\bfseries D87} (2013) 112001},
  [\href{https://arxiv.org/abs/1210.2979}{{\ttfamily 1210.2979}}].

\bibitem{Chatrchyan:2013yaa}
{\scshape CMS} collaboration, S.~Chatrchyan et~al., \emph{{Measurement of the
  $W^+W^-$ Cross section in $pp$ Collisions at $\sqrt{s} = 7$ TeV and Limits on
  Anomalous $WW\gamma$ and $WWZ$ couplings}},
  \href{http://dx.doi.org/10.1140/epjc/s10052-013-2610-8}{\emph{Eur. Phys. J.}
  {\bfseries C73} (2013) 2610},
  [\href{https://arxiv.org/abs/1306.1126}{{\ttfamily 1306.1126}}].

\bibitem{Aad:2014mda}
{\scshape ATLAS} collaboration, G.~Aad et~al., \emph{{Measurement of the
  $WW+WZ$ cross section and limits on anomalous triple gauge couplings using
  final states with one lepton, missing transverse momentum, and two jets with
  the ATLAS detector at $\sqrt{\rm{s}} = 7$ TeV}},
  \href{http://dx.doi.org/10.1007/JHEP01(2015)049}{\emph{JHEP} {\bfseries 01}
  (2015) 049}, [\href{https://arxiv.org/abs/1410.7238}{{\ttfamily 1410.7238}}].

\bibitem{Aad:2016wpd}
{\scshape ATLAS} collaboration, G.~Aad et~al., \emph{{Measurement of total and
  differential $W^+W^-$ production cross sections in proton-proton collisions
  at $\sqrt{s}=$ 8 TeV with the ATLAS detector and limits on anomalous
  triple-gauge-boson couplings}},
  \href{http://dx.doi.org/10.1007/JHEP09(2016)029}{\emph{JHEP} {\bfseries 09}
  (2016) 029}, [\href{https://arxiv.org/abs/1603.01702}{{\ttfamily
  1603.01702}}].

\bibitem{Chatrchyan:2013oev}
{\scshape CMS} collaboration, S.~Chatrchyan et~al., \emph{{Measurement of W+W-
  and ZZ production cross sections in pp collisions at sqrt(s) = 8 TeV}},
  \href{http://dx.doi.org/10.1016/j.physletb.2013.03.027}{\emph{Phys. Lett.}
  {\bfseries B721} (2013) 190--211},
  [\href{https://arxiv.org/abs/1301.4698}{{\ttfamily 1301.4698}}].

\bibitem{Khachatryan:2015sga}
{\scshape CMS} collaboration, V.~Khachatryan et~al., \emph{{Measurement of the
  ${{\mathrm{W} }^{+} }\mathrm{W}^{-} $ cross section in pp collisions at
  $\sqrt{s} =$ 8 TeV and limits on anomalous gauge couplings}},
  \href{http://dx.doi.org/10.1140/epjc/s10052-016-4219-1}{\emph{Eur. Phys. J.}
  {\bfseries C76} (2016) 401},
  [\href{https://arxiv.org/abs/1507.03268}{{\ttfamily 1507.03268}}].

\bibitem{Aaboud:2017cgf}
{\scshape ATLAS} collaboration, M.~Aaboud et~al., \emph{{Measurement of $WW/WZ
  \to \ell \nu q q^{\prime}$ production with the hadronically decaying boson
  reconstructed as one or two jets in $pp$ collisions at $\sqrt{s}=8$ TeV with
  ATLAS, and constraints on anomalous gauge couplings}},
  \href{http://dx.doi.org/10.1140/epjc/s10052-017-5084-2}{\emph{Eur. Phys. J.
  C} {\bfseries 77} (2017) 563},
  [\href{https://arxiv.org/abs/1706.01702}{{\ttfamily 1706.01702}}].

\bibitem{Aaboud:2019nkz}
{\scshape ATLAS} collaboration, M.~Aaboud et~al., \emph{{Measurement of
  fiducial and differential $W^+W^-$ production cross-sections at $\sqrt{s}=13$
  TeV with the ATLAS detector}},
  \href{http://dx.doi.org/10.1140/epjc/s10052-019-7371-6}{\emph{Eur. Phys. J.}
  {\bfseries C79} (2019) 884},
  [\href{https://arxiv.org/abs/1905.04242}{{\ttfamily 1905.04242}}].

\bibitem{Aaboud:2017qkn}
{\scshape ATLAS} collaboration, M.~Aaboud et~al., \emph{{Measurement of the
  $W^+W^-$ production cross section in $pp$ collisions at a centre-of-mass
  energy of $\sqrt{s}$ = 13 TeV with the ATLAS experiment}},
  \href{http://dx.doi.org/10.1016/j.physletb.2017.08.047}{\emph{Phys. Lett.}
  {\bfseries B773} (2017) 354--374},
  [\href{https://arxiv.org/abs/1702.04519}{{\ttfamily 1702.04519}}].

\bibitem{CMS:2016vww}
{\scshape CMS} collaboration, \emph{{Measurement of the WW cross section pp
  collisions at sqrt(s)=13 TeV}},
  \href{https://arxiv.org/abs/CMS-PAS-SMP-16-006}{{\ttfamily
  CMS-PAS-SMP-16-006}}.

\bibitem{Sirunyan:2020jtq}
{\scshape CMS} collaboration, A.~M. Sirunyan et~al., \emph{{W$^+$W$^-$ boson
  pair production in proton-proton collisions at $\sqrt{s} =$ 13 TeV}},
  \href{http://dx.doi.org/10.1103/PhysRevD.102.092001}{\emph{Phys. Rev. D}
  {\bfseries 102} (2020) 092001},
  [\href{https://arxiv.org/abs/2009.00119}{{\ttfamily 2009.00119}}].

\bibitem{Chatrchyan:2013fya}
{\scshape CMS} collaboration, S.~Chatrchyan et~al., \emph{{Measurement of the
  $W\gamma$ and $Z\gamma$ Inclusive Cross Sections in $pp$ Collisions at $\sqrt
  s=7$ TeV and Limits on Anomalous Triple Gauge Boson Couplings}},
  \href{http://dx.doi.org/10.1103/PhysRevD.89.092005}{\emph{Phys. Rev. D}
  {\bfseries 89} (2014) 092005},
  [\href{https://arxiv.org/abs/1308.6832}{{\ttfamily 1308.6832}}].

\bibitem{CMS:2015uda}
{\scshape CMS} collaboration, \emph{{Measurement of the $W^+W^-$ cross section
  in pp collisions at sqrt(s) = 8 TeV and limits on anomalous gauge
  couplings}},  \href{https://arxiv.org/abs/CMS-PAS-SMP-14-016}{{\ttfamily
  CMS-PAS-SMP-14-016}}.

\bibitem{Sirunyan:2017bey}
{\scshape CMS} collaboration, A.~M. Sirunyan et~al., \emph{{Search for
  anomalous couplings in boosted $\mathrm{ WW/WZ }\to\ell\nu\mathrm{ q \bar{q}
  }$ production in proton-proton collisions at $\sqrt{s} =$ 8 TeV}},
  \href{http://dx.doi.org/10.1016/j.physletb.2017.06.009}{\emph{Phys. Lett. B}
  {\bfseries 772} (2017) 21--42},
  [\href{https://arxiv.org/abs/1703.06095}{{\ttfamily 1703.06095}}].

\bibitem{Aad:2021dse}
{\scshape ATLAS} collaboration, G.~Aad et~al., \emph{{Measurements of
  $W^+W^-+\ge 1~$jet production cross-sections in $pp$ collisions at
  $\sqrt{s}=13~$TeV with the ATLAS detector}},
  \href{https://arxiv.org/abs/2103.10319}{{\ttfamily 2103.10319}}.

\bibitem{Binoth:2005ua}
T.~Binoth, M.~Ciccolini, N.~Kauer and M.~Kr\"amer, \emph{{Gluon-induced $WW$
  background to Higgs boson searches at the LHC}},
  \href{http://dx.doi.org/10.1088/1126-6708/2005/03/065}{\emph{JHEP} {\bfseries
  0503} (2005) 065}, [\href{https://arxiv.org/abs/hep-ph/0503094}{{\ttfamily
  hep-ph/0503094}}].

\bibitem{Campbell:2011cu}
J.~M. Campbell, R.~K. Ellis and C.~Williams, \emph{{Gluon-Gluon Contributions
  to $W^+ W^-$ Production and Higgs Interference Effects}},
  \href{http://dx.doi.org/10.1007/JHEP10(2011)005}{\emph{JHEP} {\bfseries 1110}
  (2011) 005}, [\href{https://arxiv.org/abs/1107.5569}{{\ttfamily 1107.5569}}].

\bibitem{Aad:2012me}
{\scshape ATLAS} collaboration, G.~Aad et~al., \emph{{Search for the Higgs
  boson in the $H \to W W \to$ lnujj decay channel at $\sqrt{s}=7$ TeV with the
  ATLAS detector}},
  \href{http://dx.doi.org/10.1016/j.physletb.2012.10.066}{\emph{Phys. Lett.}
  {\bfseries B718} (2012) 391--410},
  [\href{https://arxiv.org/abs/1206.6074}{{\ttfamily 1206.6074}}].

\bibitem{Aad:2013wqa}
{\scshape ATLAS} collaboration, G.~Aad et~al., \emph{{Measurements of Higgs
  boson production and couplings in diboson final states with the ATLAS
  detector at the LHC}},
  \href{http://dx.doi.org/10.1016/j.physletb.2014.05.011,
  10.1016/j.physletb.2013.08.010}{\emph{Phys. Lett.} {\bfseries B726} (2013)
  88--119}, [\href{https://arxiv.org/abs/1307.1427}{{\ttfamily 1307.1427}}].

\bibitem{Chatrchyan:2013iaa}
{\scshape CMS} collaboration, S.~Chatrchyan et~al., \emph{{Measurement of Higgs
  boson production and properties in the WW decay channel with leptonic final
  states}}, \href{http://dx.doi.org/10.1007/JHEP01(2014)096}{\emph{JHEP}
  {\bfseries 01} (2014) 096},
  [\href{https://arxiv.org/abs/1312.1129}{{\ttfamily 1312.1129}}].

\bibitem{Campbell:2013wga}
J.~M. Campbell, R.~K. Ellis and C.~Williams, \emph{{Bounding the Higgs width at
  the LHC: Complementary results from $H \to WW$}},
  \href{http://dx.doi.org/10.1103/PhysRevD.89.053011}{\emph{Phys. Rev.}
  {\bfseries D89} (2014) 053011},
  [\href{https://arxiv.org/abs/1312.1628}{{\ttfamily 1312.1628}}].

\bibitem{ATLAS:2014aga}
{\scshape ATLAS} collaboration, G.~Aad et~al., \emph{{Observation and
  measurement of Higgs boson decays to WW$^*$ with the ATLAS detector}},
  \href{http://dx.doi.org/10.1103/PhysRevD.92.012006}{\emph{Phys. Rev.}
  {\bfseries D92} (2015) 012006},
  [\href{https://arxiv.org/abs/1412.2641}{{\ttfamily 1412.2641}}].

\bibitem{Khachatryan:2014kca}
{\scshape CMS} collaboration, V.~Khachatryan et~al., \emph{{Constraints on the
  spin-parity and anomalous HVV couplings of the Higgs boson in proton
  collisions at 7 and 8 TeV}},
  \href{http://dx.doi.org/10.1103/PhysRevD.92.012004}{\emph{Phys. Rev.}
  {\bfseries D92} (2015) 012004},
  [\href{https://arxiv.org/abs/1411.3441}{{\ttfamily 1411.3441}}].

\bibitem{Aad:2015xua}
{\scshape ATLAS} collaboration, G.~Aad et~al., \emph{{Constraints on the
  off-shell Higgs boson signal strength in the high-mass $ZZ$ and $WW$ final
  states with the ATLAS detector}},
  \href{http://dx.doi.org/10.1140/epjc/s10052-015-3542-2}{\emph{Eur. Phys. J.}
  {\bfseries C75} (2015) 335},
  [\href{https://arxiv.org/abs/1503.01060}{{\ttfamily 1503.01060}}].

\bibitem{Aad:2015rwa}
{\scshape ATLAS} collaboration, G.~Aad et~al., \emph{{Determination of spin and
  parity of the Higgs boson in the $WW^*\rightarrow e \nu \mu \nu $ decay
  channel with the ATLAS detector}},
  \href{http://dx.doi.org/10.1140/epjc/s10052-015-3436-3}{\emph{Eur. Phys. J.}
  {\bfseries C75} (2015) 231},
  [\href{https://arxiv.org/abs/1503.03643}{{\ttfamily 1503.03643}}].

\bibitem{Aad:2015ona}
{\scshape ATLAS} collaboration, G.~Aad et~al., \emph{{Study of (W/Z)H
  production and Higgs boson couplings using $H \rightarrow WW^{\ast}$ decays
  with the ATLAS detector}},
  \href{http://dx.doi.org/10.1007/JHEP08(2015)137}{\emph{JHEP} {\bfseries 08}
  (2015) 137}, [\href{https://arxiv.org/abs/1506.06641}{{\ttfamily
  1506.06641}}].

\bibitem{Aad:2016lvc}
{\scshape ATLAS} collaboration, G.~Aad et~al., \emph{{Measurement of fiducial
  differential cross sections of gluon-fusion production of Higgs bosons
  decaying to $WW^{\ast}{\rightarrow\,}e\nu\mu\nu$ with the ATLAS detector at
  $\sqrt{s}=8$ TeV}},
  \href{http://dx.doi.org/10.1007/JHEP08(2016)104}{\emph{JHEP} {\bfseries 08}
  (2016) 104}, [\href{https://arxiv.org/abs/1604.02997}{{\ttfamily
  1604.02997}}].

\bibitem{Caola:2016trd}
F.~Caola, M.~Dowling, K.~Melnikov, R.~Röntsch and L.~Tancredi, \emph{{QCD
  corrections to vector boson pair production in gluon fusion including
  interference effects with off-shell Higgs at the LHC}},
  \href{http://dx.doi.org/10.1007/JHEP07(2016)087}{\emph{JHEP} {\bfseries 07}
  (2016) 087}, [\href{https://arxiv.org/abs/1605.04610}{{\ttfamily
  1605.04610}}].

\bibitem{Ferrera:2011bk}
G.~Ferrera, M.~Grazzini and F.~Tramontano, \emph{{Associated $WH$ production at
  hadron colliders: a fully exclusive QCD calculation at NNLO}},
  \href{http://dx.doi.org/10.1103/PhysRevLett.107.152003}{\emph{Phys. Rev.
  Lett.} {\bfseries 107} (2011) 152003},
  [\href{https://arxiv.org/abs/1107.1164}{{\ttfamily 1107.1164}}].

\bibitem{Ferrera:2014lca}
G.~Ferrera, M.~Grazzini and F.~Tramontano, \emph{{Associated $ZH$ production at
  hadron colliders: the fully differential NNLO QCD calculation}},
  \href{http://dx.doi.org/10.1016/j.physletb.2014.11.040}{\emph{Phys. Lett.}
  {\bfseries B740} (2015) 51--55},
  [\href{https://arxiv.org/abs/1407.4747}{{\ttfamily 1407.4747}}].

\bibitem{Ferrera:2017zex}
G.~Ferrera, G.~Somogyi and F.~Tramontano, \emph{{Associated production of a
  Higgs boson decaying into bottom quarks at the LHC in full NNLO QCD}},
  \href{http://dx.doi.org/10.1016/j.physletb.2018.03.021}{\emph{Phys. Lett.}
  {\bfseries B780} (2018) 346--351},
  [\href{https://arxiv.org/abs/1705.10304}{{\ttfamily 1705.10304}}].

\bibitem{Campbell:2016jau}
J.~M. Campbell, R.~K. Ellis and C.~Williams, \emph{{Associated production of a
  Higgs boson at NNLO}},
  \href{http://dx.doi.org/10.1007/JHEP06(2016)179}{\emph{JHEP} {\bfseries 06}
  (2016) 179}, [\href{https://arxiv.org/abs/1601.00658}{{\ttfamily
  1601.00658}}].

\bibitem{Harlander:2003ai}
R.~V. Harlander and W.~B. Kilgore, \emph{{Higgs boson production in bottom
  quark fusion at next-to-next-to leading order}},
  \href{http://dx.doi.org/10.1103/PhysRevD.68.013001}{\emph{Phys. Rev.}
  {\bfseries D68} (2003) 013001},
  [\href{https://arxiv.org/abs/hep-ph/0304035}{{\ttfamily hep-ph/0304035}}].

\bibitem{Harlander:2010cz}
R.~V. Harlander, K.~J. Ozeren and M.~Wiesemann, \emph{{Higgs plus jet
  production in bottom quark annihilation at next-to-leading order}},
  \href{http://dx.doi.org/10.1016/j.physletb.2010.08.038}{\emph{Phys. Lett.}
  {\bfseries B693} (2010) 269--273},
  [\href{https://arxiv.org/abs/1007.5411}{{\ttfamily 1007.5411}}].

\bibitem{Harlander:2011fx}
R.~Harlander and M.~Wiesemann, \emph{{Jet-veto in bottom-quark induced Higgs
  production at next-to-next-to-leading order}},
  \href{http://dx.doi.org/10.1007/JHEP04(2012)066}{\emph{JHEP} {\bfseries 04}
  (2012) 066}, [\href{https://arxiv.org/abs/1111.2182}{{\ttfamily 1111.2182}}].

\bibitem{Buehler:2012cu}
S.~B{\"u}hler, F.~Herzog, A.~Lazopoulos and R.~M{\"u}ller, \emph{{The fully
  differential hadronic production of a Higgs boson via bottom quark fusion at
  NNLO}}, \href{http://dx.doi.org/10.1007/JHEP07(2012)115}{\emph{JHEP}
  {\bfseries 07} (2012) 115},
  [\href{https://arxiv.org/abs/1204.4415}{{\ttfamily 1204.4415}}].

\bibitem{Marzani:2008az}
S.~Marzani, R.~D. Ball, V.~Del~Duca, S.~Forte and A.~Vicini, \emph{{Higgs
  production via gluon-gluon fusion with finite top mass beyond next-to-leading
  order}}, \href{http://dx.doi.org/10.1016/j.nuclphysb.2008.03.016}{\emph{Nucl.
  Phys.} {\bfseries B800} (2008) 127--145},
  [\href{https://arxiv.org/abs/0801.2544}{{\ttfamily 0801.2544}}].

\bibitem{Harlander:2009mq}
R.~V. Harlander and K.~J. Ozeren, \emph{{Finite top mass effects for hadronic
  Higgs production at next-to-next-to-leading order}},
  \href{http://dx.doi.org/10.1088/1126-6708/2009/11/088}{\emph{JHEP} {\bfseries
  11} (2009) 088}, [\href{https://arxiv.org/abs/0909.3420}{{\ttfamily
  0909.3420}}].

\bibitem{Harlander:2009my}
R.~V. Harlander, H.~Mantler, S.~Marzani and K.~J. Ozeren, \emph{{Higgs
  production in gluon fusion at next-to-next-to-leading order QCD for finite
  top mass}},
  \href{http://dx.doi.org/10.1140/epjc/s10052-010-1258-x}{\emph{Eur. Phys. J.}
  {\bfseries C66} (2010) 359--372},
  [\href{https://arxiv.org/abs/0912.2104}{{\ttfamily 0912.2104}}].

\bibitem{Pak:2009dg}
A.~Pak, M.~Rogal and M.~Steinhauser, \emph{{Finite top quark mass effects in
  NNLO Higgs boson production at LHC}},
  \href{http://dx.doi.org/10.1007/JHEP02(2010)025}{\emph{JHEP} {\bfseries 02}
  (2010) 025}, [\href{https://arxiv.org/abs/0911.4662}{{\ttfamily 0911.4662}}].

\bibitem{Neumann:2014nha}
T.~Neumann and M.~Wiesemann, \emph{{Finite top-mass effects in gluon-induced
  Higgs production with a jet-veto at NNLO}},
  \href{http://dx.doi.org/10.1007/JHEP11(2014)150}{\emph{JHEP} {\bfseries 11}
  (2014) 150}, [\href{https://arxiv.org/abs/1408.6836}{{\ttfamily 1408.6836}}].

\bibitem{deFlorian:2013jea}
D.~de~Florian and J.~Mazzitelli, \emph{{Higgs Boson Pair Production at
  Next-to-Next-to-Leading Order in QCD}},
  \href{http://dx.doi.org/10.1103/PhysRevLett.111.201801}{\emph{Phys. Rev.
  Lett.} {\bfseries 111} (2013) 201801},
  [\href{https://arxiv.org/abs/1309.6594}{{\ttfamily 1309.6594}}].

\bibitem{deFlorian:2016uhr}
D.~de~Florian, M.~Grazzini, C.~Hanga, S.~Kallweit, J.~M. Lindert,
  P.~Maierhöfer, J.~Mazzitelli and D.~Rathlev, \emph{{Differential Higgs Boson
  Pair Production at Next-to-Next-to-Leading Order in QCD}},
  \href{http://dx.doi.org/10.1007/JHEP09(2016)151}{\emph{JHEP} {\bfseries 09}
  (2016) 151}, [\href{https://arxiv.org/abs/1606.09519}{{\ttfamily
  1606.09519}}].

\bibitem{Grazzini:2018bsd}
M.~Grazzini, G.~Heinrich, S.~Jones, S.~Kallweit, M.~Kerner, J.~M. Lindert and
  J.~Mazzitelli, \emph{{Higgs boson pair production at NNLO with top quark mass
  effects}}, \href{http://dx.doi.org/10.1007/JHEP05(2018)059}{\emph{JHEP}
  {\bfseries 05} (2018) 059},
  [\href{https://arxiv.org/abs/1803.02463}{{\ttfamily 1803.02463}}].

\bibitem{Catani:2011qz}
S.~Catani, L.~Cieri, D.~de~Florian, G.~Ferrera and M.~Grazzini, \emph{{Diphoton
  production at hadron colliders: a fully-differential QCD calculation at
  NNLO}}, \href{http://dx.doi.org/10.1103/PhysRevLett.108.072001}{\emph{Phys.
  Rev. Lett.} {\bfseries 108} (2012) 072001},
  [\href{https://arxiv.org/abs/1110.2375}{{\ttfamily 1110.2375}}].

\bibitem{Campbell:2016yrh}
J.~M. Campbell, R.~K. Ellis, Y.~Li and C.~Williams, \emph{{Predictions for
  diphoton production at the LHC through NNLO in QCD}},
  \href{http://dx.doi.org/10.1007/JHEP07(2016)148}{\emph{JHEP} {\bfseries 07}
  (2016) 148}, [\href{https://arxiv.org/abs/1603.02663}{{\ttfamily
  1603.02663}}].

\bibitem{Grazzini:2013bna}
M.~Grazzini, S.~Kallweit, D.~Rathlev and A.~Torre, \emph{{$Z\gamma$ production
  at hadron colliders in NNLO QCD}},
  \href{http://dx.doi.org/10.1016/j.physletb.2014.02.037}{\emph{Phys. Lett.}
  {\bfseries B731} (2014) 204--207},
  [\href{https://arxiv.org/abs/1309.7000}{{\ttfamily 1309.7000}}].

\bibitem{Grazzini:2015nwa}
M.~Grazzini, S.~Kallweit and D.~Rathlev, \emph{{$W\gamma$ and $Z\gamma$
  production at the LHC in NNLO QCD}},
  \href{http://dx.doi.org/10.1007/JHEP07(2015)085}{\emph{JHEP} {\bfseries 07}
  (2015) 085}, [\href{https://arxiv.org/abs/1504.01330}{{\ttfamily
  1504.01330}}].

\bibitem{Campbell:2017aul}
J.~M. Campbell, T.~Neumann and C.~Williams, \emph{{$Z\gamma$ Production at NNLO
  Including Anomalous Couplings}},
  \href{http://dx.doi.org/10.1007/JHEP11(2017)150}{\emph{JHEP} {\bfseries 11}
  (2017) 150}, [\href{https://arxiv.org/abs/1708.02925}{{\ttfamily
  1708.02925}}].

\bibitem{Gehrmann:2020oec}
T.~Gehrmann, N.~Glover, A.~Huss and J.~Whitehead, \emph{{Scale and isolation
  sensitivity of diphoton distributions at the LHC}},
  \href{http://dx.doi.org/10.1007/JHEP01(2021)108}{\emph{JHEP} {\bfseries 01}
  (2021) 108}, [\href{https://arxiv.org/abs/2009.11310}{{\ttfamily
  2009.11310}}].

\bibitem{Cascioli:2014yka}
F.~Cascioli, T.~Gehrmann, M.~Grazzini, S.~Kallweit, P.~Maierh\"ofer, A.~von
  Manteuffel, S.~Pozzorini, D.~Rathlev, L.~Tancredi and E.~Weihs, \emph{{$ZZ$
  production at hadron colliders in NNLO QCD}},
  \href{http://dx.doi.org/10.1016/j.physletb.2014.06.056}{\emph{Phys. Lett.}
  {\bfseries B735} (2014) 311--313},
  [\href{https://arxiv.org/abs/1405.2219}{{\ttfamily 1405.2219}}].

\bibitem{Grazzini:2015hta}
M.~Grazzini, S.~Kallweit and D.~Rathlev, \emph{{ZZ production at the LHC:
  fiducial cross sections and distributions in NNLO QCD}},
  \href{http://dx.doi.org/10.1016/j.physletb.2015.09.055}{\emph{Phys. Lett.}
  {\bfseries B750} (2015) 407--410},
  [\href{https://arxiv.org/abs/1507.06257}{{\ttfamily 1507.06257}}].

\bibitem{Heinrich:2017bvg}
G.~Heinrich, S.~Jahn, S.~P. Jones, M.~Kerner and J.~Pires, \emph{{NNLO
  predictions for Z-boson pair production at the LHC}},
  \href{http://dx.doi.org/10.1007/JHEP03(2018)142}{\emph{JHEP} {\bfseries 03}
  (2018) 142}, [\href{https://arxiv.org/abs/1710.06294}{{\ttfamily
  1710.06294}}].

\bibitem{Kallweit:2018nyv}
S.~Kallweit and M.~Wiesemann, \emph{{$ZZ$ production at the LHC: NNLO
  predictions for $2\ell2\nu$ and $4\ell$ signatures}},
  \href{http://dx.doi.org/10.1016/j.physletb.2018.10.016}{\emph{Phys. Lett.}
  {\bfseries B786} (2018) 382--389},
  [\href{https://arxiv.org/abs/1806.05941}{{\ttfamily 1806.05941}}].

\bibitem{Gehrmann:2014fva}
T.~Gehrmann, M.~Grazzini, S.~Kallweit, P.~Maierh\"ofer, A.~von Manteuffel,
  S.~Pozzorini, D.~Rathlev and L.~Tancredi, \emph{{$W^+W^-$ Production at
  Hadron Colliders in Next to Next to Leading Order QCD}},
  \href{http://dx.doi.org/10.1103/PhysRevLett.113.212001}{\emph{Phys. Rev.
  Lett.} {\bfseries 113} (2014) 212001},
  [\href{https://arxiv.org/abs/1408.5243}{{\ttfamily 1408.5243}}].

\bibitem{Grazzini:2016ctr}
M.~Grazzini, S.~Kallweit, S.~Pozzorini, D.~Rathlev and M.~Wiesemann,
  \emph{{$W^+W^-$ production at the LHC: fiducial cross sections and
  distributions in NNLO QCD}},
  \href{http://dx.doi.org/10.1007/JHEP08(2016)140}{\emph{JHEP} {\bfseries 08}
  (2016) 140}, [\href{https://arxiv.org/abs/1605.02716}{{\ttfamily
  1605.02716}}].

\bibitem{Grazzini:2016swo}
M.~Grazzini, S.~Kallweit, D.~Rathlev and M.~Wiesemann, \emph{{$W^{\pm}Z$
  production at hadron colliders in NNLO QCD}},
  \href{http://dx.doi.org/10.1016/j.physletb.2016.08.017}{\emph{Phys. Lett.}
  {\bfseries B761} (2016) 179--183},
  [\href{https://arxiv.org/abs/1604.08576}{{\ttfamily 1604.08576}}].

\bibitem{Grazzini:2017ckn}
M.~Grazzini, S.~Kallweit, D.~Rathlev and M.~Wiesemann, \emph{{$W^\pm Z$
  production at the LHC: fiducial cross sections and distributions in NNLO
  QCD}}, \href{http://dx.doi.org/10.1007/JHEP05(2017)139}{\emph{JHEP}
  {\bfseries 05} (2017) 139},
  [\href{https://arxiv.org/abs/1703.09065}{{\ttfamily 1703.09065}}].

\bibitem{Baglio:2012np}
J.~Baglio, A.~Djouadi, R.~Gr\"ober, M.~M\"uhlleitner, J.~Quevillon and
  M.~Spira, \emph{{The measurement of the Higgs self-coupling at the LHC:
  theoretical status}},
  \href{http://dx.doi.org/10.1007/JHEP04(2013)151}{\emph{JHEP} {\bfseries 04}
  (2013) 151}, [\href{https://arxiv.org/abs/1212.5581}{{\ttfamily 1212.5581}}].

\bibitem{Li:2016nrr}
H.~T. Li and J.~Wang, \emph{{Fully Differential Higgs Pair Production in
  Association With a $W$ Boson at Next-to-Next-to-Leading Order in QCD}},
  \href{http://dx.doi.org/10.1016/j.physletb.2016.12.030}{\emph{Phys. Lett. B}
  {\bfseries 765} (2017) 265--271},
  [\href{https://arxiv.org/abs/1607.06382}{{\ttfamily 1607.06382}}].

\bibitem{deFlorian:2019app}
D.~de~Florian, I.~Fabre and J.~Mazzitelli, \emph{{Triple Higgs production at
  hadron colliders at NNLO in QCD}},
  \href{http://dx.doi.org/10.1007/JHEP03(2020)155}{\emph{JHEP} {\bfseries 03}
  (2020) 155}, [\href{https://arxiv.org/abs/1912.02760}{{\ttfamily
  1912.02760}}].

\bibitem{Chawdhry:2019bji}
H.~A. Chawdhry, M.~L. Czakon, A.~Mitov and R.~Poncelet, \emph{{NNLO QCD
  corrections to three-photon production at the LHC}},
  \href{http://dx.doi.org/10.1007/JHEP02(2020)057}{\emph{JHEP} {\bfseries 02}
  (2020) 057}, [\href{https://arxiv.org/abs/1911.00479}{{\ttfamily
  1911.00479}}].

\bibitem{Kallweit:2020gcp}
S.~Kallweit, V.~Sotnikov and M.~Wiesemann, \emph{{Triphoton production at
  hadron colliders in NNLO QCD}},
  \href{http://dx.doi.org/10.1016/j.physletb.2020.136013}{\emph{Phys. Lett. B}
  {\bfseries 812} (2021) 136013},
  [\href{https://arxiv.org/abs/2010.04681}{{\ttfamily 2010.04681}}].

\bibitem{Kallweit:2020gva}
S.~Kallweit, E.~Re, L.~Rottoli and M.~Wiesemann, \emph{{Accurate single- and
  double-differential resummation of colour-singlet processes with
  MATRIX+RADISH: W$^{+}$W$^{-}$ production at the LHC}},
  \href{http://dx.doi.org/10.1007/JHEP12(2020)147}{\emph{JHEP} {\bfseries 12}
  (2020) 147}, [\href{https://arxiv.org/abs/2004.07720}{{\ttfamily
  2004.07720}}].

\bibitem{Frixione:2002ik}
S.~Frixione and B.~R. Webber, \emph{{Matching NLO QCD computations and parton
  shower simulations}},
  \href{http://dx.doi.org/10.1088/1126-6708/2002/06/029}{\emph{JHEP} {\bfseries
  06} (2002) 029}, [\href{https://arxiv.org/abs/hep-ph/0204244}{{\ttfamily
  hep-ph/0204244}}].

\bibitem{Nason:2004rx}
P.~Nason, \emph{{A New method for combining NLO QCD with shower Monte Carlo
  algorithms}},
  \href{http://dx.doi.org/10.1088/1126-6708/2004/11/040}{\emph{JHEP} {\bfseries
  11} (2004) 040}, [\href{https://arxiv.org/abs/hep-ph/0409146}{{\ttfamily
  hep-ph/0409146}}].

\bibitem{Frixione:2007vw}
S.~Frixione, P.~Nason and C.~Oleari, \emph{{Matching NLO QCD computations with
  Parton Shower simulations: the POWHEG method}},
  \href{http://dx.doi.org/10.1088/1126-6708/2007/11/070}{\emph{JHEP} {\bfseries
  11} (2007) 070}, [\href{https://arxiv.org/abs/0709.2092}{{\ttfamily
  0709.2092}}].

\bibitem{Hamilton:2012rf}
K.~Hamilton, P.~Nason, C.~Oleari and G.~Zanderighi, \emph{{Merging H/W/Z + 0
  and 1 jet at NLO with no merging scale: a path to parton shower + NNLO
  matching}}, \href{http://dx.doi.org/10.1007/JHEP05(2013)082}{\emph{JHEP}
  {\bfseries 05} (2013) 082},
  [\href{https://arxiv.org/abs/1212.4504}{{\ttfamily 1212.4504}}].

\bibitem{Alioli:2013hqa}
S.~Alioli, C.~W. Bauer, C.~Berggren, F.~J. Tackmann, J.~R. Walsh and S.~Zuberi,
  \emph{{Matching Fully Differential NNLO Calculations and Parton Showers}},
  \href{http://dx.doi.org/10.1007/JHEP06(2014)089}{\emph{JHEP} {\bfseries 06}
  (2014) 089}, [\href{https://arxiv.org/abs/1311.0286}{{\ttfamily 1311.0286}}].

\bibitem{Hoeche:2014aia}
S.~Höche, Y.~Li and S.~Prestel, \emph{{Drell-Yan lepton pair production at
  NNLO QCD with parton showers}},
  \href{http://dx.doi.org/10.1103/PhysRevD.91.074015}{\emph{Phys. Rev.}
  {\bfseries D91} (2015) 074015},
  [\href{https://arxiv.org/abs/1405.3607}{{\ttfamily 1405.3607}}].

\bibitem{Monni:2019whf}
P.~F. Monni, P.~Nason, E.~Re, M.~Wiesemann and G.~Zanderighi,
  \emph{{MiNNLO$_{\text{PS}}$: A new method to match NNLO QCD to parton
  showers}}, \href{http://dx.doi.org/10.1007/JHEP05(2020)143}{\emph{JHEP}
  {\bfseries 05} (2020) 143},
  [\href{https://arxiv.org/abs/1908.06987}{{\ttfamily 1908.06987}}].

\bibitem{Monni:2020nks}
P.~F. Monni, E.~Re and M.~Wiesemann, \emph{{MiNNLO$_{\text {PS}}$: optimizing
  $2\rightarrow 1$ hadronic processes}},
  \href{http://dx.doi.org/10.1140/epjc/s10052-020-08658-5}{\emph{Eur. Phys. J.
  C} {\bfseries 80} (2020) 1075},
  [\href{https://arxiv.org/abs/2006.04133}{{\ttfamily 2006.04133}}].

\bibitem{Mazzitelli:2020jio}
J.~Mazzitelli, P.~F. Monni, P.~Nason, E.~Re, M.~Wiesemann and G.~Zanderighi,
  \emph{{Next-to-next-to-leading order event generation for top-quark pair
  production}},  \href{https://arxiv.org/abs/2012.14267}{{\ttfamily
  2012.14267}}.

\bibitem{Hamilton:2012np}
K.~Hamilton, P.~Nason and G.~Zanderighi, \emph{{MINLO: Multi-Scale Improved
  NLO}}, \href{http://dx.doi.org/10.1007/JHEP10(2012)155}{\emph{JHEP}
  {\bfseries 10} (2012) 155},
  [\href{https://arxiv.org/abs/1206.3572}{{\ttfamily 1206.3572}}].

\bibitem{Frederix:2015fyz}
R.~Frederix and K.~Hamilton, \emph{{Extending the MINLO method}},
  \href{http://dx.doi.org/10.1007/JHEP05(2016)042}{\emph{JHEP} {\bfseries 05}
  (2016) 042}, [\href{https://arxiv.org/abs/1512.02663}{{\ttfamily
  1512.02663}}].

\bibitem{Hamilton:2013fea}
K.~Hamilton, P.~Nason, E.~Re and G.~Zanderighi, \emph{{NNLOPS simulation of
  Higgs boson production}},
  \href{http://dx.doi.org/10.1007/JHEP10(2013)222}{\emph{JHEP} {\bfseries 10}
  (2013) 222}, [\href{https://arxiv.org/abs/1309.0017}{{\ttfamily 1309.0017}}].

\bibitem{Hoche:2014dla}
S.~Höche, Y.~Li and S.~Prestel, \emph{{Higgs-boson production through gluon
  fusion at NNLO QCD with parton showers}},
  \href{http://dx.doi.org/10.1103/PhysRevD.90.054011}{\emph{Phys. Rev.}
  {\bfseries D90} (2014) 054011},
  [\href{https://arxiv.org/abs/1407.3773}{{\ttfamily 1407.3773}}].

\bibitem{Karlberg:2014qua}
A.~Karlberg, E.~Re and G.~Zanderighi, \emph{{NNLOPS accurate Drell-Yan
  production}}, \href{http://dx.doi.org/10.1007/JHEP09(2014)134}{\emph{JHEP}
  {\bfseries 09} (2014) 134},
  [\href{https://arxiv.org/abs/1407.2940}{{\ttfamily 1407.2940}}].

\bibitem{Alioli:2015toa}
S.~Alioli, C.~W. Bauer, C.~Berggren, F.~J. Tackmann and J.~R. Walsh,
  \emph{{Drell-Yan production at NNLL'+NNLO matched to parton showers}},
  \href{http://dx.doi.org/10.1103/PhysRevD.92.094020}{\emph{Phys. Rev.}
  {\bfseries D92} (2015) 094020},
  [\href{https://arxiv.org/abs/1508.01475}{{\ttfamily 1508.01475}}].

\bibitem{Astill:2016hpa}
W.~Astill, W.~Bizon, E.~Re and G.~Zanderighi, \emph{{NNLOPS accurate associated
  HW production}}, \href{http://dx.doi.org/10.1007/JHEP06(2016)154}{\emph{JHEP}
  {\bfseries 06} (2016) 154},
  [\href{https://arxiv.org/abs/1603.01620}{{\ttfamily 1603.01620}}].

\bibitem{Astill:2018ivh}
W.~Astill, W.~Bizoń, E.~Re and G.~Zanderighi, \emph{{NNLOPS accurate
  associated HZ production with $ H\to b\overline{b} $ decay at NLO}},
  \href{http://dx.doi.org/10.1007/JHEP11(2018)157}{\emph{JHEP} {\bfseries 11}
  (2018) 157}, [\href{https://arxiv.org/abs/1804.08141}{{\ttfamily
  1804.08141}}].

\bibitem{Alioli:2019qzz}
S.~Alioli, A.~Broggio, S.~Kallweit, M.~A. Lim and L.~Rottoli,
  \emph{{Higgsstrahlung at NNLL'$+$NNLO matched to parton showers in GENEVA}},
  \href{http://dx.doi.org/10.1103/PhysRevD.100.096016}{\emph{Phys. Rev.}
  {\bfseries D100} (2019) 096016},
  [\href{https://arxiv.org/abs/1909.02026}{{\ttfamily 1909.02026}}].

\bibitem{Bizon:2019tfo}
W.~Bizo\'n, E.~Re and G.~Zanderighi, \emph{{NNLOPS description of the $H \to
  b\overline{b} $ decay with MiNLO}},
  \href{http://dx.doi.org/10.1007/JHEP06(2020)006}{\emph{JHEP} {\bfseries 06}
  (2020) 006}, [\href{https://arxiv.org/abs/1912.09982}{{\ttfamily
  1912.09982}}].

\bibitem{Alioli:2020fzf}
S.~Alioli, A.~Broggio, A.~Gavardi, S.~Kallweit, M.~A. Lim, R.~Nagar,
  D.~Napoletano and L.~Rottoli, \emph{{Resummed predictions for hadronic Higgs
  boson decays}},  \href{https://arxiv.org/abs/2009.13533}{{\ttfamily
  2009.13533}}.

\bibitem{Re:2018vac}
E.~Re, M.~Wiesemann and G.~Zanderighi, \emph{{NNLOPS accurate predictions for
  $W^+W^-$ production}},
  \href{http://dx.doi.org/10.1007/JHEP12(2018)121}{\emph{JHEP} {\bfseries 12}
  (2018) 121}, [\href{https://arxiv.org/abs/1805.09857}{{\ttfamily
  1805.09857}}].

\bibitem{Lombardi:2020wju}
D.~Lombardi, M.~Wiesemann and G.~Zanderighi, \emph{{Advancing MiNNLO$_{\rm PS}$
  to diboson processes: $Z\gamma$ production at NNLO+PS}},
  \href{https://arxiv.org/abs/2010.10478}{{\ttfamily 2010.10478}}.

\bibitem{Alioli:2020qrd}
S.~Alioli, A.~Broggio, A.~Gavardi, S.~Kallweit, M.~A. Lim, R.~Nagar,
  D.~Napoletano and L.~Rottoli, \emph{{Precise predictions for photon pair
  production matched to parton showers in GENEVA}},
  \href{https://arxiv.org/abs/2010.10498}{{\ttfamily 2010.10498}}.

\bibitem{Alioli:2021egp}
S.~Alioli, A.~Broggio, A.~Gavardi, S.~Kallweit, M.~A. Lim, R.~Nagar and
  D.~Napoletano, \emph{{Next-to-next-to-leading order event generation for $Z$
  boson pair production matched to parton shower}},
  \href{https://arxiv.org/abs/2103.01214}{{\ttfamily 2103.01214}}.

\bibitem{Brown:1978mq}
R.~Brown and K.~Mikaelian, \emph{{$W^+ W^-$ and $Z^0 Z^0$ Pair Production in
  $e^+ e^-$, $p p$, $p \bar{p}$ Colliding Beams}},
  \href{http://dx.doi.org/10.1103/PhysRevD.19.922}{\emph{Phys. Rev.} {\bfseries
  D19} (1979) 922}.

\bibitem{Ohnemus:1991kk}
J.~Ohnemus, \emph{{An Order $\alpha_s$ calculation of hadronic $W^{-} W^{+}$
  production}}, \href{http://dx.doi.org/10.1103/PhysRevD.44.1403}{\emph{Phys.
  Rev.} {\bfseries D44} (1991) 1403--1414}.

\bibitem{Frixione:1993yp}
S.~Frixione, \emph{{A Next-to-leading order calculation of the cross-section
  for the production of $W^+ W^-$ pairs in hadronic collisions}},
  \href{http://dx.doi.org/10.1016/0550-3213(93)90435-R}{\emph{Nucl. Phys.}
  {\bfseries B410} (1993) 280--324}.

\bibitem{Campbell:1999ah}
J.~M. Campbell and R.~K. Ellis, \emph{{An Update on vector boson pair
  production at hadron colliders}},
  \href{http://dx.doi.org/10.1103/PhysRevD.60.113006}{\emph{Phys. Rev.}
  {\bfseries D60} (1999) 113006},
  [\href{https://arxiv.org/abs/hep-ph/9905386}{{\ttfamily hep-ph/9905386}}].

\bibitem{Dixon:1999di}
L.~J. Dixon, Z.~Kunszt and A.~Signer, \emph{{Vector boson pair production in
  hadronic collisions at order $\alpha_s$ : Lepton correlations and anomalous
  couplings}}, \href{http://dx.doi.org/10.1103/PhysRevD.60.114037}{\emph{Phys.
  Rev.} {\bfseries D60} (1999) 114037},
  [\href{https://arxiv.org/abs/hep-ph/9907305}{{\ttfamily hep-ph/9907305}}].

\bibitem{Dixon:1998py}
L.~J. Dixon, Z.~Kunszt and A.~Signer, \emph{{Helicity amplitudes for O(alpha-s)
  production of $W^{+} W^{-}$, $W^\pm Z$, $Z Z$, $W^\pm \gamma$, or $Z \gamma$
  pairs at hadron colliders}},
  \href{http://dx.doi.org/10.1016/S0550-3213(98)00421-0}{\emph{Nucl. Phys. B}
  {\bfseries 531} (1998) 3--23},
  [\href{https://arxiv.org/abs/hep-ph/9803250}{{\ttfamily hep-ph/9803250}}].

\bibitem{Campbell:2011bn}
J.~M. Campbell, R.~K. Ellis and C.~Williams, \emph{{Vector boson pair
  production at the LHC}},
  \href{http://dx.doi.org/10.1007/JHEP07(2011)018}{\emph{JHEP} {\bfseries 07}
  (2011) 018}, [\href{https://arxiv.org/abs/1105.0020}{{\ttfamily 1105.0020}}].

\bibitem{Bierweiler:2012kw}
A.~Bierweiler, T.~Kasprzik, J.~H. K{\"u}hn and S.~Uccirati, \emph{{Electroweak
  corrections to W-boson pair production at the LHC}},
  \href{http://dx.doi.org/10.1007/JHEP11(2012)093}{\emph{JHEP} {\bfseries 1211}
  (2012) 093}, [\href{https://arxiv.org/abs/1208.3147}{{\ttfamily 1208.3147}}].

\bibitem{Baglio:2013toa}
J.~Baglio, L.~D. Ninh and M.~M. Weber, \emph{{Massive gauge boson pair
  production at the LHC: a next-to-leading order story}},
  \href{http://dx.doi.org/10.1103/PhysRevD.88.113005}{\emph{Phys. Rev.}
  {\bfseries D88} (2013) 113005},
  [\href{https://arxiv.org/abs/1307.4331}{{\ttfamily 1307.4331}}].

\bibitem{Billoni:2013aba}
M.~Billoni, S.~Dittmaier, B.~J{\"a}ger and C.~Speckner, \emph{{Next-to-leading
  order electroweak corrections to $pp\rightarrow W^+W^- \rightarrow 4$ leptons
  at the LHC in double-pole approximation}},
  \href{http://dx.doi.org/10.1007/JHEP12(2013)043}{\emph{JHEP} {\bfseries 1312}
  (2013) 043}, [\href{https://arxiv.org/abs/1310.1564}{{\ttfamily 1310.1564}}].

\bibitem{Biedermann:2016guo}
B.~Biedermann, M.~Billoni, A.~Denner, S.~Dittmaier, L.~Hofer, B.~J{\"a}ger and
  L.~Salfelder, \emph{{Next-to-leading-order electroweak corrections to $pp \to
  W^+W^-\to$ 4 leptons at the LHC}},
  \href{http://dx.doi.org/10.1007/JHEP06(2016)065}{\emph{JHEP} {\bfseries 06}
  (2016) 065}, [\href{https://arxiv.org/abs/1605.03419}{{\ttfamily
  1605.03419}}].

\bibitem{Kallweit:2017khh}
S.~Kallweit, J.~M. Lindert, S.~Pozzorini and M.~Schönherr, \emph{{NLO QCD+EW
  predictions for $2\ell2\nu$ diboson signatures at the LHC}},
  \href{http://dx.doi.org/10.1007/JHEP11(2017)120}{\emph{JHEP} {\bfseries 11}
  (2017) 120}, [\href{https://arxiv.org/abs/1705.00598}{{\ttfamily
  1705.00598}}].

\bibitem{Kallweit:2019zez}
M.~Grazzini, S.~Kallweit, J.~M. Lindert, S.~Pozzorini and M.~Wiesemann,
  \emph{{NNLO QCD + NLO EW with Matrix+OpenLoops: precise predictions for
  vector-boson pair production}},
  \href{http://dx.doi.org/10.1007/JHEP02(2020)087}{\emph{JHEP} {\bfseries 02}
  (2020) 087}, [\href{https://arxiv.org/abs/1912.00068}{{\ttfamily
  1912.00068}}].

\bibitem{Glover:1988rg}
E.~W.~N. Glover and J.~J. van~der Bij, \emph{{Z-boson pair production via gluon
  fusion}}, \href{http://dx.doi.org/10.1016/0550-3213(89)90262-9}{\emph{Nucl.
  Phys.} {\bfseries B321} (1989) 561--590}.

\bibitem{Dicus:1987dj}
D.~A. Dicus, C.~Kao and W.~W. Repko, \emph{{Gluon Production of Gauge Bosons}},
  \href{http://dx.doi.org/10.1103/PhysRevD.36.1570}{\emph{Phys. Rev.}
  {\bfseries D36} (1987) 1570}.

\bibitem{Matsuura:1991pj}
T.~Matsuura and J.~van~der Bij, \emph{{Characteristics of leptonic signals for
  Z boson pairs at hadron colliders}},
  \href{http://dx.doi.org/10.1007/BF01475793}{\emph{Z. Phys.} {\bfseries C51}
  (1991) 259--266}.

\bibitem{Zecher:1994kb}
C.~Zecher, T.~Matsuura and J.~van~der Bij, \emph{{Leptonic signals from
  off-shell Z boson pairs at hadron colliders}},
  \href{http://dx.doi.org/10.1007/BF01557393}{\emph{Z. Phys.} {\bfseries C64}
  (1994) 219--226}, [\href{https://arxiv.org/abs/hep-ph/9404295}{{\ttfamily
  hep-ph/9404295}}].

\bibitem{Binoth:2008pr}
T.~Binoth, N.~Kauer and P.~Mertsch, \emph{{Gluon-induced QCD corrections to $pp
  \rightarrow ZZ \rightarrow l\bar{l} l^\prime \bar{l}^\prime$}},
  \href{http://dx.doi.org/10.3360/dis.2008.142}{\emph{Proceedings} {\bfseries
  DIS 2008} (2008) 142}, [\href{https://arxiv.org/abs/0807.0024}{{\ttfamily
  0807.0024}}].

\bibitem{Kauer:2013qba}
N.~Kauer, \emph{{Interference effects for H $\to$ WW/ZZ $\to
  \ell\bar{\nu}_\ell\bar{\ell}\nu_\ell$ searches in gluon fusion at the LHC}},
  \href{http://dx.doi.org/10.1007/JHEP12(2013)082}{\emph{JHEP} {\bfseries 12}
  (2013) 082}, [\href{https://arxiv.org/abs/1310.7011}{{\ttfamily 1310.7011}}].

\bibitem{Cascioli:2013gfa}
F.~Cascioli, S.~H\"oche, F.~Krauss, P.~Maierh\"ofer, S.~Pozzorini and
  F.~Siegert, \emph{{Precise Higgs-background predictions: merging NLO QCD and
  squared quark-loop corrections to four-lepton + 0,1 jet production}},
  \href{http://dx.doi.org/10.1007/JHEP01(2014)046}{\emph{JHEP} {\bfseries 1401}
  (2014) 046}, [\href{https://arxiv.org/abs/1309.0500}{{\ttfamily 1309.0500}}].

\bibitem{Campbell:2013una}
J.~M. Campbell, R.~K. Ellis and C.~Williams, \emph{{Bounding the Higgs width at
  the LHC using full analytic results for $gg \to e^- e^+ \mu^- \mu^+$}},
  \href{http://dx.doi.org/10.1007/JHEP04(2014)060}{\emph{JHEP} {\bfseries 04}
  (2014) 060}, [\href{https://arxiv.org/abs/1311.3589}{{\ttfamily 1311.3589}}].

\bibitem{Ellis:2014yca}
J.~M. Campbell, R.~K. Ellis and C.~Williams, \emph{{Bounding the Higgs Width at
  the LHC}}, {\emph{PoS} {\bfseries LL2014} (2014) 008},
  [\href{https://arxiv.org/abs/1408.1723}{{\ttfamily 1408.1723}}].

\bibitem{Kauer:2015dma}
N.~Kauer, C.~O'Brien and E.~Vryonidou, \emph{{Interference effects for $ H\to
  W\;W\to \ell \nu q{\overline{q}}^{\prime } $ and $ H\to ZZ\to \ell
  \overline{\ell}q\overline{q} $ searches in gluon fusion at the LHC}},
  \href{http://dx.doi.org/10.1007/JHEP10(2015)074}{\emph{JHEP} {\bfseries 10}
  (2015) 074}, [\href{https://arxiv.org/abs/1506.01694}{{\ttfamily
  1506.01694}}].

\bibitem{Gehrmann:2014bfa}
T.~Gehrmann, A.~von Manteuffel, L.~Tancredi and E.~Weihs, \emph{{The two-loop
  master integrals for $q\overline{q} \to VV$}},
  \href{http://dx.doi.org/10.1007/JHEP06(2014)032}{\emph{JHEP} {\bfseries 1406}
  (2014) 032}, [\href{https://arxiv.org/abs/1404.4853}{{\ttfamily 1404.4853}}].

\bibitem{Caola:2014iua}
F.~Caola, J.~M. Henn, K.~Melnikov, A.~V. Smirnov and V.~A. Smirnov,
  \emph{{Two-loop helicity amplitudes for the production of two off-shell
  electroweak bosons in quark-antiquark collisions}},
  \href{http://dx.doi.org/10.1007/JHEP11(2014)041}{\emph{JHEP} {\bfseries 1411}
  (2014) 041}, [\href{https://arxiv.org/abs/1408.6409}{{\ttfamily 1408.6409}}].

\bibitem{Gehrmann:2015ora}
T.~Gehrmann, A.~von Manteuffel and L.~Tancredi, \emph{{The two-loop helicity
  amplitudes for $ q\overline{q}^{\prime}\to {V}_1{V}_2\to 4 $ leptons}},
  \href{http://dx.doi.org/10.1007/JHEP09(2015)128}{\emph{JHEP} {\bfseries 09}
  (2015) 128}, [\href{https://arxiv.org/abs/1503.04812}{{\ttfamily
  1503.04812}}].

\bibitem{Poncelet:2021jmj}
R.~Poncelet and A.~Popescu, \emph{{NNLO QCD study of polarised $W^+ W^-$
  production at the LHC}},  \href{https://arxiv.org/abs/2102.13583}{{\ttfamily
  2102.13583}}.

\bibitem{Caola:2015ila}
F.~Caola, J.~M. Henn, K.~Melnikov, A.~V. Smirnov and V.~A. Smirnov,
  \emph{{Two-loop helicity amplitudes for the production of two off-shell
  electroweak bosons in gluon fusion}},
  \href{http://dx.doi.org/10.1007/JHEP06(2015)129}{\emph{JHEP} {\bfseries 1506}
  (2015) 129}, [\href{https://arxiv.org/abs/1503.08759}{{\ttfamily
  1503.08759}}].

\bibitem{vonManteuffel:2015msa}
A.~von Manteuffel and L.~Tancredi, \emph{{The two-loop helicity amplitudes for
  $gg \to V_1 V_2 \to 4~\mathrm{leptons}$}},
  \href{http://dx.doi.org/10.1007/JHEP06(2015)197}{\emph{JHEP} {\bfseries 1506}
  (2015) 197}, [\href{https://arxiv.org/abs/1503.08835}{{\ttfamily
  1503.08835}}].

\bibitem{Caola:2015rqy}
F.~Caola, K.~Melnikov, R.~R{\"o}ntsch and L.~Tancredi, \emph{{QCD corrections
  to $W^+W^-$ production through gluon fusion}},
  \href{http://dx.doi.org/10.1016/j.physletb.2016.01.046}{\emph{Phys. Lett.}
  {\bfseries B754} (2016) 275--280},
  [\href{https://arxiv.org/abs/1511.08617}{{\ttfamily 1511.08617}}].

\bibitem{Grazzini:2020stb}
M.~Grazzini, S.~Kallweit, M.~Wiesemann and J.~Y. Yook, \emph{{$W^+W^-$
  production at the LHC: NLO QCD corrections to the loop-induced gluon fusion
  channel}},
  \href{http://dx.doi.org/10.1016/j.physletb.2020.135399}{\emph{Phys. Lett. B}
  {\bfseries 804} (2020) 135399},
  [\href{https://arxiv.org/abs/2002.01877}{{\ttfamily 2002.01877}}].

\bibitem{Grazzini:2017mhc}
M.~Grazzini, S.~Kallweit and M.~Wiesemann, \emph{{Fully differential NNLO
  computations with MATRIX}},
  \href{http://dx.doi.org/10.1140/epjc/s10052-018-5771-7}{\emph{Eur. Phys. J.}
  {\bfseries C78} (2018) 537},
  [\href{https://arxiv.org/abs/1711.06631}{{\ttfamily 1711.06631}}].

\bibitem{Cascioli:2011va}
F.~Cascioli, P.~Maierh\"ofer and S.~Pozzorini, \emph{{Scattering Amplitudes
  with Open Loops}},
  \href{http://dx.doi.org/10.1103/PhysRevLett.108.111601}{\emph{Phys. Rev.
  Lett.} {\bfseries 108} (2012) 111601},
  [\href{https://arxiv.org/abs/1111.5206}{{\ttfamily 1111.5206}}].

\bibitem{Buccioni:2017yxi}
F.~Buccioni, S.~Pozzorini and M.~Zoller, \emph{{On-the-fly reduction of open
  loops}}, \href{http://dx.doi.org/10.1140/epjc/s10052-018-5562-1}{\emph{Eur.
  Phys. J.} {\bfseries C78} (2018) 70},
  [\href{https://arxiv.org/abs/1710.11452}{{\ttfamily 1710.11452}}].

\bibitem{Buccioni:2019sur}
F.~Buccioni, J.-N. Lang, J.~M. Lindert, P.~Maierh{\"o}fer, S.~Pozzorini,
  H.~Zhang and M.~F. Zoller, \emph{{OpenLoops 2}},
  \href{http://dx.doi.org/10.1140/epjc/s10052-019-7306-2}{\emph{Eur. Phys. J.
  C} {\bfseries 79} (2019) 866},
  [\href{https://arxiv.org/abs/1907.13071}{{\ttfamily 1907.13071}}].

\bibitem{Dawson:2013lya}
S.~Dawson, I.~M. Lewis and M.~Zeng, \emph{{Threshold resummed and approximate
  next-to-next-to-leading order results for $W^+W^-$ pair production at the
  LHC}}, \href{http://dx.doi.org/10.1103/PhysRevD.88.054028}{\emph{Phys. Rev.}
  {\bfseries D88} (2013) 054028},
  [\href{https://arxiv.org/abs/1307.3249}{{\ttfamily 1307.3249}}].

\bibitem{Grazzini:2015wpa}
M.~Grazzini, S.~Kallweit, D.~Rathlev and M.~Wiesemann,
  \emph{{Transverse-momentum resummation for vector-boson pair production at
  NNLL+NNLO}}, \href{http://dx.doi.org/10.1007/JHEP08(2015)154}{\emph{JHEP}
  {\bfseries 08} (2015) 154},
  [\href{https://arxiv.org/abs/1507.02565}{{\ttfamily 1507.02565}}].

\bibitem{Dawson:2016ysj}
S.~Dawson, P.~Jaiswal, Y.~Li, H.~Ramani and M.~Zeng, \emph{{Resummation of jet
  veto logarithms at N$^3$LL$_a$ + NNLO for $W^+ W^-$ production at the LHC}},
  \href{http://dx.doi.org/10.1103/PhysRevD.94.114014}{\emph{Phys. Rev.}
  {\bfseries D94} (2016) 114014},
  [\href{https://arxiv.org/abs/1606.01034}{{\ttfamily 1606.01034}}].

\bibitem{Wiesemann:2020gbm}
M.~Wiesemann, L.~Rottoli and P.~Torrielli, \emph{{The Z$\gamma$
  transverse-momentum spectrum at NNLO+N$^3$LL}},
  \href{http://dx.doi.org/10.1016/j.physletb.2020.135718}{\emph{Phys. Lett. B}
  {\bfseries 809} (2020) 135718},
  [\href{https://arxiv.org/abs/2006.09338}{{\ttfamily 2006.09338}}].

\bibitem{MatrixRadishurl}
\emph{{\textsc{Matrix+RadISH} is an interface to \textsc{RadISH} within
  \textsc{Matrix} by S. Kallweit, E. Re, L. Rottoli, M. Wiesemann}},
  \href{https://arxiv.org/abs/https://matrix.hepforge.org/matrix+radish.html}{{\ttfamily
  https://matrix.hepforge.org/matrix+radish.html}}.

\bibitem{Monni:2016ktx}
P.~F. Monni, E.~Re and P.~Torrielli, \emph{{Higgs Transverse-Momentum
  Resummation in Direct Space}},
  \href{http://dx.doi.org/10.1103/PhysRevLett.116.242001}{\emph{Phys. Rev.
  Lett.} {\bfseries 116} (2016) 242001},
  [\href{https://arxiv.org/abs/1604.02191}{{\ttfamily 1604.02191}}].

\bibitem{Bizon:2017rah}
W.~Bizon, P.~F. Monni, E.~Re, L.~Rottoli and P.~Torrielli,
  \emph{{Momentum-space resummation for transverse observables and the Higgs
  $p_\perp$ at N$^3$LL+NNLO}},
  \href{https://arxiv.org/abs/1705.09127}{{\ttfamily 1705.09127}}.

\bibitem{Monni:2019yyr}
P.~F. Monni, L.~Rottoli and P.~Torrielli, \emph{{Higgs transverse momentum with
  a jet veto: a double-differential resummation}}, vol.~124.
\newblock 2020,
  \href{http://dx.doi.org/10.1103/PhysRevLett.124.252001}{10.1103/PhysRevLett.124.252001}.

\bibitem{Jaiswal:2014yba}
P.~Jaiswal and T.~Okui, \emph{{Explanation of the $WW$ excess at the LHC by
  jet-veto resummation}},
  \href{http://dx.doi.org/10.1103/PhysRevD.90.073009}{\emph{Phys. Rev.}
  {\bfseries D90} (2014) 073009},
  [\href{https://arxiv.org/abs/1407.4537}{{\ttfamily 1407.4537}}].

\bibitem{Meade:2014fca}
P.~Meade, H.~Ramani and M.~Zeng, \emph{{Transverse momentum resummation effects
  in $W^+W^-$ measurements}},
  \href{http://dx.doi.org/10.1103/PhysRevD.90.114006}{\emph{Phys. Rev.}
  {\bfseries D90} (2014) 114006},
  [\href{https://arxiv.org/abs/1407.4481}{{\ttfamily 1407.4481}}].

\bibitem{Becher:2014aya}
T.~Becher, R.~Frederix, M.~Neubert and L.~Rothen, \emph{{Automated NNLL $+$ NLO
  resummation for jet-veto cross sections}},
  \href{http://dx.doi.org/10.1140/epjc/s10052-015-3368-y}{\emph{Eur.Phys.J.}
  {\bfseries C75} (2015) 154},
  [\href{https://arxiv.org/abs/1412.8408}{{\ttfamily 1412.8408}}].

\bibitem{Monni:2014zra}
P.~F. Monni and G.~Zanderighi, \emph{{On the excess in the inclusive
  $W^+W^-\rightarrow l^+l^-\nu\bar\nu$ cross section}},
  \href{http://dx.doi.org/10.1007/JHEP05(2015)013}{\emph{JHEP} {\bfseries 1505}
  (2015) 013}, [\href{https://arxiv.org/abs/1410.4745}{{\ttfamily 1410.4745}}].

\bibitem{ATLAS:2014xea}
{\scshape ATLAS} collaboration, \emph{{Measurement of the $W^+W^-$ production
  cross section in proton-proton collisions at $\sqrt{s} =8$ TeV with the ATLAS
  detector}},  \href{https://arxiv.org/abs/ATLAS-CONF-2014-033}{{\ttfamily
  ATLAS-CONF-2014-033}}.

\bibitem{Arpino:2019fmo}
L.~Arpino, A.~Banfi, S.~J{\"a}ger and N.~Kauer, \emph{{BSM $WW$ production with
  a jet veto}}, \href{http://dx.doi.org/10.1007/JHEP08(2019)076}{\emph{JHEP}
  {\bfseries 08} (2019) 076},
  [\href{https://arxiv.org/abs/1905.06646}{{\ttfamily 1905.06646}}].

\bibitem{Hamilton:2010mb}
K.~Hamilton, \emph{{A positive-weight next-to-leading order simulation of weak
  boson pair production}},
  \href{http://dx.doi.org/10.1007/JHEP01(2011)009}{\emph{JHEP} {\bfseries 01}
  (2011) 009}, [\href{https://arxiv.org/abs/1009.5391}{{\ttfamily 1009.5391}}].

\bibitem{Bellm:2016cks}
J.~Bellm, S.~Gieseke, N.~Greiner, G.~Heinrich, S.~Pl{\"a}tzer, C.~Reuschle and
  J.~F. von Soden-Fraunhofen, \emph{{Anomalous coupling, top-mass and
  parton-shower effects in ${W^+W^-}$ production}},
  \href{http://dx.doi.org/10.1007/JHEP05(2016)106}{\emph{JHEP} {\bfseries 05}
  (2016) 106}, [\href{https://arxiv.org/abs/1602.05141}{{\ttfamily
  1602.05141}}].

\bibitem{Bellm:2015jjp}
J.~Bellm et~al., \emph{{Herwig 7.0/Herwig++ 3.0 release note}},
  \href{http://dx.doi.org/10.1140/epjc/s10052-016-4018-8}{\emph{Eur. Phys. J.}
  {\bfseries C76} (2016) 196},
  [\href{https://arxiv.org/abs/1512.01178}{{\ttfamily 1512.01178}}].

\bibitem{Hoche:2010pf}
S.~Hoche, F.~Krauss, M.~Schonherr and F.~Siegert, \emph{{Automating the POWHEG
  method in Sherpa}},
  \href{http://dx.doi.org/10.1007/JHEP04(2011)024}{\emph{JHEP} {\bfseries 04}
  (2011) 024}, [\href{https://arxiv.org/abs/1008.5399}{{\ttfamily 1008.5399}}].

\bibitem{Nason:2013ydw}
P.~Nason and G.~Zanderighi, \emph{{$W^+ W^-$ , $W Z$ and $Z Z$ production in
  the POWHEG-BOX-V2}},
  \href{http://dx.doi.org/10.1140/epjc/s10052-013-2702-5}{\emph{Eur. Phys. J.}
  {\bfseries C74} (2014) 2702},
  [\href{https://arxiv.org/abs/1311.1365}{{\ttfamily 1311.1365}}].

\bibitem{Melia:2011tj}
T.~Melia, P.~Nason, R.~Röntsch and G.~Zanderighi, \emph{{W+W-, WZ and ZZ
  production in the POWHEG BOX}},
  \href{http://dx.doi.org/10.1007/JHEP11(2011)078}{\emph{JHEP} {\bfseries 11}
  (2011) 078}, [\href{https://arxiv.org/abs/1107.5051}{{\ttfamily 1107.5051}}].

\bibitem{Gehrmann:2012yg}
T.~Gehrmann, S.~H{\"o}che, F.~Krauss, M.~Schonherr and F.~Siegert, \emph{{NLO
  QCD matrix elements + parton showers in $e^+e^- \rightarrow$ hadrons}},
  \href{http://dx.doi.org/10.1007/JHEP01(2013)144}{\emph{JHEP} {\bfseries 01}
  (2013) 144}, [\href{https://arxiv.org/abs/1207.5031}{{\ttfamily 1207.5031}}].

\bibitem{Hoeche:2012yf}
S.~H{\"o}eche, F.~Krauss, M.~Schonherr and F.~Siegert, \emph{{QCD matrix
  elements + parton showers: The NLO case}},
  \href{http://dx.doi.org/10.1007/JHEP04(2013)027}{\emph{JHEP} {\bfseries 04}
  (2013) 027}, [\href{https://arxiv.org/abs/1207.5030}{{\ttfamily 1207.5030}}].

\bibitem{Frederix:2012ps}
R.~Frederix and S.~Frixione, \emph{{Merging meets matching in MC@NLO}},
  \href{http://dx.doi.org/10.1007/JHEP12(2012)061}{\emph{JHEP} {\bfseries 12}
  (2012) 061}, [\href{https://arxiv.org/abs/1209.6215}{{\ttfamily 1209.6215}}].

\bibitem{Alwall:2014hca}
J.~Alwall, R.~Frederix, S.~Frixione, V.~Hirschi, F.~Maltoni, O.~Mattelaer,
  H.~S. Shao, T.~Stelzer, P.~Torrielli and M.~Zaro, \emph{{The automated
  computation of tree-level and next-to-leading order differential cross
  sections, and their matching to parton shower simulations}},
  \href{http://dx.doi.org/10.1007/JHEP07(2014)079}{\emph{JHEP} {\bfseries 07}
  (2014) 079}, [\href{https://arxiv.org/abs/1405.0301}{{\ttfamily 1405.0301}}].

\bibitem{Alioli:2010xd}
S.~Alioli, P.~Nason, C.~Oleari and E.~Re, \emph{{A general framework for
  implementing NLO calculations in shower Monte Carlo programs: the POWHEG
  BOX}}, \href{http://dx.doi.org/10.1007/JHEP06(2010)043}{\emph{JHEP}
  {\bfseries 06} (2010) 043},
  [\href{https://arxiv.org/abs/1002.2581}{{\ttfamily 1002.2581}}].

\bibitem{Hamilton:2016bfu}
K.~Hamilton, T.~Melia, P.~F. Monni, E.~Re and G.~Zanderighi, \emph{{Merging WW
  and WW+jet with MINLO}},
  \href{http://dx.doi.org/10.1007/JHEP09(2016)057}{\emph{JHEP} {\bfseries 09}
  (2016) 057}, [\href{https://arxiv.org/abs/1606.07062}{{\ttfamily
  1606.07062}}].

\bibitem{Brauer:2020kfv}
S.~Br\"auer, A.~Denner, M.~Pellen, M.~Sch\"onherr and S.~Schumann,
  \emph{{Fixed-order and merged parton-shower predictions for WW and WWj
  production at the LHC including NLO QCD and EW corrections}},
  \href{http://dx.doi.org/10.1007/JHEP10(2020)159}{\emph{JHEP} {\bfseries 10}
  (2020) 159}, [\href{https://arxiv.org/abs/2005.12128}{{\ttfamily
  2005.12128}}].

\bibitem{Chiesa:2020ttl}
M.~Chiesa, C.~Oleari and E.~Re, \emph{{NLO QCD+NLO EW corrections to diboson
  production matched to parton shower}},
  \href{http://dx.doi.org/10.1140/epjc/s10052-020-8419-3}{\emph{Eur. Phys. J.
  C} {\bfseries 80} (2020) 849},
  [\href{https://arxiv.org/abs/2005.12146}{{\ttfamily 2005.12146}}].

\bibitem{Collins:1977iv}
J.~C. Collins and D.~E. Soper, \emph{{Angular Distribution of Dileptons in
  High-Energy Hadron Collisions}},
  \href{http://dx.doi.org/10.1103/PhysRevD.16.2219}{\emph{Phys. Rev. D}
  {\bfseries 16} (1977) 2219}.

\bibitem{hepforge:VVamp}


\bibitem{Matrixurl}
\emph{{\textsc{Matrix} is the abbreviation of Munich Automates qT subtraction
  and Resummation to Integrate X-sections by M. Grazzini, S. Kallweit, M.
  Wiesemann}},
  \href{https://arxiv.org/abs/http://matrix.hepforge.org}{{\ttfamily
  http://matrix.hepforge.org}}.

\bibitem{Jezo:2015aia}
T.~Je\v{z}o and P.~Nason, \emph{{On the Treatment of Resonances in
  Next-to-Leading Order Calculations Matched to a Parton Shower}},
  \href{http://dx.doi.org/10.1007/JHEP12(2015)065}{\emph{JHEP} {\bfseries 12}
  (2015) 065}, [\href{https://arxiv.org/abs/1509.09071}{{\ttfamily
  1509.09071}}].

\bibitem{Binoth:2006mf}
T.~Binoth, M.~Ciccolini, N.~Kauer and M.~Kr\"amer, \emph{{Gluon-induced W-boson
  pair production at the LHC}},
  \href{http://dx.doi.org/10.1088/1126-6708/2006/12/046}{\emph{JHEP} {\bfseries
  0612} (2006) 046}, [\href{https://arxiv.org/abs/hep-ph/0611170}{{\ttfamily
  hep-ph/0611170}}].

\bibitem{Kauer:2012hd}
N.~Kauer and G.~Passarino, \emph{{Inadequacy of zero-width approximation for a
  light Higgs boson signal}},
  \href{http://dx.doi.org/10.1007/JHEP08(2012)116}{\emph{JHEP} {\bfseries 1208}
  (2012) 116}, [\href{https://arxiv.org/abs/1206.4803}{{\ttfamily 1206.4803}}].

\bibitem{Alioli:2021wpn}
S.~Alioli, S.~Ferrario~Ravasio, J.~M. Lindert and R.~R\"ontsch,
  \emph{{Four-lepton production in gluon fusion at NLO matched to parton
  showers}},  \href{https://arxiv.org/abs/2102.07783}{{\ttfamily 2102.07783}}.

\bibitem{Alwall:2007st}
J.~Alwall, P.~Demin, S.~de~Visscher, R.~Frederix, M.~Herquet, F.~Maltoni,
  T.~Plehn, D.~L. Rainwater and T.~Stelzer, \emph{{MadGraph/MadEvent v4: The
  New Web Generation}},
  \href{http://dx.doi.org/10.1088/1126-6708/2007/09/028}{\emph{JHEP} {\bfseries
  09} (2007) 028}, [\href{https://arxiv.org/abs/0706.2334}{{\ttfamily
  0706.2334}}].

\bibitem{Campbell:2012am}
J.~M. Campbell, R.~K. Ellis, R.~Frederix, P.~Nason, C.~Oleari and C.~Williams,
  \emph{{NLO Higgs Boson Production Plus One and Two Jets Using the POWHEG BOX,
  MadGraph4 and MCFM}},
  \href{http://dx.doi.org/10.1007/JHEP07(2012)092}{\emph{JHEP} {\bfseries 07}
  (2012) 092}, [\href{https://arxiv.org/abs/1202.5475}{{\ttfamily 1202.5475}}].

\bibitem{Cullen:2014yla}
G.~Cullen et~al., \emph{{G$\scriptsize{O}$S$\scriptsize{AM}$-2.0: a tool for
  automated one-loop calculations within the Standard Model and beyond}},
  \href{http://dx.doi.org/10.1140/epjc/s10052-014-3001-5}{\emph{Eur. Phys. J.
  C} {\bfseries 74} (2014) 3001},
  [\href{https://arxiv.org/abs/1404.7096}{{\ttfamily 1404.7096}}].

\bibitem{Campbell:2019dru}
J.~Campbell and T.~Neumann, \emph{{Precision Phenomenology with MCFM}},
  \href{http://dx.doi.org/10.1007/JHEP12(2019)034}{\emph{JHEP} {\bfseries 12}
  (2019) 034}, [\href{https://arxiv.org/abs/1909.09117}{{\ttfamily
  1909.09117}}].

\bibitem{Salam:2008qg}
G.~P. Salam and J.~Rojo, \emph{{A Higher Order Perturbative Parton Evolution
  Toolkit (HOPPET)}},
  \href{http://dx.doi.org/10.1016/j.cpc.2008.08.010}{\emph{Comput. Phys.
  Commun.} {\bfseries 180} (2009) 120--156},
  [\href{https://arxiv.org/abs/0804.3755}{{\ttfamily 0804.3755}}].

\bibitem{Gehrmann:2001pz}
T.~Gehrmann and E.~Remiddi, \emph{{Numerical evaluation of harmonic
  polylogarithms}},
  \href{http://dx.doi.org/10.1016/S0010-4655(01)00411-8}{\emph{Comput. Phys.
  Commun.} {\bfseries 141} (2001) 296--312},
  [\href{https://arxiv.org/abs/hep-ph/0107173}{{\ttfamily hep-ph/0107173}}].

\bibitem{Nason:2013uba}
P.~Nason and C.~Oleari, \emph{{Generation cuts and Born suppression in
  POWHEG}},  \href{https://arxiv.org/abs/1303.3922}{{\ttfamily 1303.3922}}.

\bibitem{Gehrmann:2005pd}
T.~Gehrmann, T.~Huber and D.~Maitre, \emph{{Two-loop quark and gluon
  form-factors in dimensional regularisation}},
  \href{http://dx.doi.org/10.1016/j.physletb.2005.07.019}{\emph{Phys. Lett.}
  {\bfseries B622} (2005) 295--302},
  [\href{https://arxiv.org/abs/hep-ph/0507061}{{\ttfamily hep-ph/0507061}}].

\bibitem{Gonsalves:1983nq}
R.~J. Gonsalves, \emph{{Dimensionally Regularized Two Loop On-shell Quark Form
  Factor}}, \href{http://dx.doi.org/10.1103/PhysRevD.28.1542}{\emph{Phys. Rev.
  D} {\bfseries 28} (1983) 1542}.

\bibitem{vanNeerven:1985xr}
W.~van Neerven, \emph{{Dimensional Regularization of Mass and Infrared
  Singularities in Two Loop On-shell Vertex Functions}},
  \href{http://dx.doi.org/10.1016/0550-3213(86)90165-3}{\emph{Nucl. Phys. B}
  {\bfseries 268} (1986) 453--488}.

\bibitem{Kramer:1986sr}
G.~Kramer and B.~Lampe, \emph{{Integrals for Two Loop Calculations in Massless
  QCD}}, \href{http://dx.doi.org/10.1063/1.527586}{\emph{J. Math. Phys.}
  {\bfseries 28} (1987) 945}.

\bibitem{btwxturl}
\emph{{The {\sc Btwxt} general-purpose, N-dimensional interpolation library, by
  N.~Kruis, T.~Scimone and P.~Sullivan}},
  \href{https://arxiv.org/abs/https://github.com/bigladder/btwxt}{{\ttfamily
  https://github.com/bigladder/btwxt}}.

\bibitem{doi:10.1002/sapm1960391258}
G.~Birkhoff and H.~L. Garabedian, \emph{Smooth surface interpolation},
  \href{http://dx.doi.org/10.1002/sapm1960391258}{\emph{Journal of Mathematics
  and Physics} {\bfseries 39} (1960) 258--268}.

\bibitem{CATMULL1974317}
E.~Catmull and R.~Rom, \emph{A class of local interpolating splines},  in
  \emph{Computer Aided Geometric Design} (R.~E. Barnhill and R.~F. Riesenfiled,
  eds.), pp.~317 -- 326.
\newblock Academic Press, 1974.
\newblock
  \href{http://dx.doi.org/https://doi.org/10.1016/B978-0-12-079050-0.50020-5}{DOI}.

\bibitem{Catani:2013tia}
S.~Catani, L.~Cieri, D.~de~Florian, G.~Ferrera and M.~Grazzini,
  \emph{{Universality of transverse-momentum resummation and hard factors at
  the NNLO}},
  \href{http://dx.doi.org/10.1016/j.nuclphysb.2014.02.011}{\emph{Nucl. Phys.}
  {\bfseries B881} (2014) 414--443},
  [\href{https://arxiv.org/abs/1311.1654}{{\ttfamily 1311.1654}}].

\bibitem{Patrignani:2016xqp}
{\scshape Particle Data Group} collaboration, C.~Patrignani et~al.,
  \emph{{Review of Particle Physics}},
  \href{http://dx.doi.org/10.1088/1674-1137/40/10/100001}{\emph{Chin. Phys.}
  {\bfseries C40} (2016) 100001}.

\bibitem{Ball:2014uwa}
{\scshape NNPDF} collaboration, R.~D. Ball et~al., \emph{{Parton distributions
  for the LHC Run II}},
  \href{http://dx.doi.org/10.1007/JHEP04(2015)040}{\emph{JHEP} {\bfseries 04}
  (2015) 040}, [\href{https://arxiv.org/abs/1410.8849}{{\ttfamily 1410.8849}}].

\bibitem{Buckley:2014ana}
A.~Buckley, J.~Ferrando, S.~Lloyd, K.~Nordström, B.~Page, M.~Rüfenacht,
  M.~Schönherr and G.~Watt, \emph{{LHAPDF6: parton density access in the LHC
  precision era}},
  \href{http://dx.doi.org/10.1140/epjc/s10052-015-3318-8}{\emph{Eur. Phys. J.}
  {\bfseries C75} (2015) 132},
  [\href{https://arxiv.org/abs/1412.7420}{{\ttfamily 1412.7420}}].

\bibitem{Sjostrand:2014zea}
T.~Sjöstrand, S.~Ask, J.~R. Christiansen, R.~Corke, N.~Desai, P.~Ilten,
  S.~Mrenna, S.~Prestel, C.~O. Rasmussen and P.~Z. Skands, \emph{{An
  Introduction to PYTHIA 8.2}},
  \href{http://dx.doi.org/10.1016/j.cpc.2015.01.024}{\emph{Comput. Phys.
  Commun.} {\bfseries 191} (2015) 159--177},
  [\href{https://arxiv.org/abs/1410.3012}{{\ttfamily 1410.3012}}].

\bibitem{TheATLAScollaboration:2014rfk}
\emph{{ATLAS Pythia 8 tunes to 7 TeV datas, ATL-PHYS-PUB-2014-021}},  11, 2014.

\bibitem{Stewart:2011cf}
I.~W. Stewart and F.~J. Tackmann, \emph{{Theory Uncertainties for Higgs and
  Other Searches Using Jet Bins}},
  \href{http://dx.doi.org/10.1103/PhysRevD.85.034011}{\emph{Phys. Rev. D}
  {\bfseries 85} (2012) 034011},
  [\href{https://arxiv.org/abs/1107.2117}{{\ttfamily 1107.2117}}].

\bibitem{Banfi:2012jm}
A.~Banfi, P.~F. Monni, G.~P. Salam and G.~Zanderighi, \emph{{Higgs and Z-boson
  production with a jet veto}},
  \href{http://dx.doi.org/10.1103/PhysRevLett.109.202001}{\emph{Phys. Rev.
  Lett.} {\bfseries 109} (2012) 202001},
  [\href{https://arxiv.org/abs/1206.4998}{{\ttfamily 1206.4998}}].

\bibitem{Catani:1997xc}
S.~Catani and B.~R. Webber, \emph{{Infrared safe but infinite: Soft gluon
  divergences inside the physical region}},
  \href{http://dx.doi.org/10.1088/1126-6708/1997/10/005}{\emph{JHEP} {\bfseries
  10} (1997) 005}, [\href{https://arxiv.org/abs/hep-ph/9710333}{{\ttfamily
  hep-ph/9710333}}].

\end{thebibliography}\endgroup
\bibliographystyle{JHEP}

\end{document}